\documentclass[iop,apj,twocolappendix,numberedappendix,revtex4]{emulateapj}

\newbox\grsign \setbox\grsign=\hbox{$>$}
\newdimen\grdimen \grdimen=\ht\grsign
\newbox\laxbox \newbox\gaxbox
\setbox\gaxbox=\hbox{\raise.5ex\hbox{$>$}\llap
     {\lower.5ex\hbox{$\sim$}}}\ht1=\grdimen\dp1=0pt
\setbox\laxbox=\hbox{\raise.5ex\hbox{$<$}\llap
     {\lower.5ex\hbox{$\sim$}}}\ht2=\grdimen\dp2=0pt

\newcommand{\lax}{$\mathrel{\copy\laxbox}$}

\newcommand{\rot}{ \color{red} }

\newcommand{\rg}{R_{\rm g}}
\newcommand{\rs}{R_{\rm S}}
\newcommand{\tg}{t_{\rm g}}

\usepackage[usenames,dvipsnames]{color}
\usepackage{amsmath}
\usepackage{amssymb}
\usepackage{ctable} 
\usepackage{graphicx}
\usepackage{natbib}
\usepackage[titletoc,title]{appendix}
\usepackage{lipsum}
\usepackage{mwe}
\usepackage[caption=false]{subfig}

\begin{document}

\title{GR-MHD disk winds and jets from black holes and resistive accretion disks}
\shorttitle{BH disk \& jets}
\shortauthors{Vourellis et al.}
\author{Christos Vourellis\altaffilmark{1,*},
        Christian Fendt\altaffilmark{1}, 
        Qian Qian\altaffilmark{1,*},
        Scott Noble\altaffilmark{2}
}
\altaffiltext{1}{Max Planck Institute for Astronomy, Heidelberg, Germany}
\altaffiltext{2}{Department of Physics and Engineering Physics, University of Tulsa, Tulsa, USA}
\altaffiltext{*}{Member of the International Max Planck Research School for Astronomy \& Cosmic Physics at the University of Heidelberg}
\email{ vourellis@mpia.de, fendt@mpia.de  }                                   
\begin{abstract}
We perform GR-MHD simulations of outflow launching from thin accretion disks.
As in the non-relativistic case, resistivity is essential for the mass loading of the disk wind.
We implemented resistivity in the ideal GR-MHD code \texttt{HARM3D}, extending previous works 
\citep{QQ1,QQ2} for larger physical grids, higher spatial resolution, and longer simulation time.\\
We consider an initially thin, resistive disk orbiting the black hole, threaded by a large-scale magnetic flux.
As the system evolves, outflows are launched from the black hole magnetosphere and the disk surface.
We mainly focus on disk outflows, investigating their MHD structure and energy output in comparison with 
the Poynting-dominated black hole jet.
The disk wind encloses two components -- a fast component dominated by the toroidal magnetic field and a slower 
component dominated by the poloidal field.
The disk wind transitions from sub to super-Alfv\'enic speed, reaching velocities $\simeq 0.1c$. 
We provide parameter studies varying spin parameter and resistivity level, and measure the respective mass and 
energy fluxes.
A higher spin strengthens the $B_{\phi}$-dominated disk wind along the inner jet.\\
We disentangle a critical resistivity level that leads to a maximum matter and energy output for both, resulting 
from the interplay between re-connection and diffusion, which in combination govern the magnetic flux and 
the mass loading.
For counter-rotating black holes the outflow structure shows a magnetic field reversal.
We estimate the opacity of the inner-most accretion stream and the outflow structure around it.
This stream may be critically opaque for a lensed signal, while the axial jet funnel remains optically thin. 
\end{abstract}

%
\section{Introduction}
Astrophysical jets appear as linearly collimated structures of high speed that are typically found in young stellar objects,
X-Ray binaries, gamma-ray bursts, or active galactic nuclei (AGN).
The physical mechanisms which produce these jets (jet launching) have been studied extensively.
A consensus has been achieved, that launching of relativistic jets requires the existence of an accretion disk around 
a black hole and a strong magnetic field.

\citet{BP1982} have proposed that jets can be formed as a result of magneto-centrifugal acceleration of matter from 
the surface of an accretion disk (thereafter BP mechanism). 
On the other hand, \citet{BZ1977} suggested that relativistic jets can be launched from the magnetosphere of a 
black hole by extracting rotational energy (thereafter BZ mechanism).
An interesting question for AGN jets is which of these mechanisms is responsible for generating the jets
we observe.
One way to investigate and compare the efficiency of these mechanisms is to use magneto-hydrodynamic (MHD) simulations. 
In the case of the BZ mechanism, the equations of MHD need to be solved in a general relativistic (GR, GR-MHD) 
context \citep{Einstein1915}. 

Despite the abundance of observational data, it is almost impossible to resolve the jet launching area for 
more than a few sources.
Applying VLBI, \citet{Doeleman2012} determined a jet base of M87 of approximately $5.5\rs$ (Schwarzschild radii), which may 
imply that the BZ mechanism is responsible for feeding energy into the jet.
On the other hand, \citet{Boccardi2016} find for the launching region of the Cygnus A jet a scale of $227\,\rs$, 
thus suggesting that at least part of the jet may result from a disk wind (BP mechanism).
Only high resolution radio observations of the jet base and the accretion disk may
discriminate which of the two mechanisms is more involved in the launching of jets.
Another unresolved problem connected to the launching question is the matter content of relativistic jets.
BZ-driven jets are expected to be leptonic and mass loaded by pair-production in the
strong radiation field of the black hole-disk corona, 
while jets launched as disk winds would be fed with hadronic disk material.

Most recently, the long-lasting search for a direct proof for the existence of supermassive black holes 
succeeded when the Event Horizon Telescope Collaboration (EHTC) released the first striking pictures of the shadow 
of the central black hole in M87, observed with short wavelength $1.3\;$mm \citep{EHT2019a}. 
Radiation from an asymmetric ring around the black hole was detected and identified as signature of the photon sphere 
around a Kerr black hole \citep{EHT2019e}.

As for the launching area, the jet propagation has been extensively investigated as well.
For the example of M87m 15\,GHz VLA observations find that the jet knots are moving (outwards)
with apparent velocities of about 0.5\,c \citep{Biretta1995}.
More recently, for the same source radio observations by \citet{AsadaNakamura2012} find indication of a change in 
the jet opening angle at $10^5$ Schwarzschild radii distance from the central black hole such that the jet shape 
changes from parabolic to conical.
A similar behavior was detected for jet and counter jet of NGC\,4261 \citep{Nakahara2018}.

A physical complete theory that will fully connect the AGN jet launching mechanism with the observed behavior of the 
jet is still under development. 
The general approach is to perform GR-MHD simulations of the close environment of the central black hole and the accretion 
disk to investigate and compare the launching mechanisms of relativistic extragalactic jets.
In the past twenty years a significant number of GR-MHD codes have been developed and used to simulate rotating disks around 
black holes and their resulting outflows 
\citep{Koide1999, Gammie2003, DeVilliers2003a, Noble2006, DelZanna2007, Noble2009, BdZ2013, McKinneyTchekh2014, ZanottiDumbser2015,  Porth2017}.

\citet{Koide1999} studied the development of a relativistic jet in a Schwarzschild space-time and identified a magnetic 
driven and a gas-pressure driven component. 
\citet{DeVilliers2003a} focused in the accretion process between the disk and the black hole for different black hole spins.
\citet{McKinney2004} examined the energy flux in the black hole horizon in an attempt to detect the BZ mechanism.
\citet{McKinneyTchekh2012} tested magnetically choked accretion flows and detected quasi periodic oscillations between the 
accreting inflow and the jet magnetosphere.
In \citet{TchekhNarayan2010, TchekhNarayan2011} the authors simulated accretion flows into extreme Kerr black holes to
measure the energy extracted by the BZ mechanism.
Radiative transfer in combination with GR-MHD codes 
allows the study of spectrum of GR accretion disks \citep{Noble2011} or their evolution in the super-Eddington 
limit \citep{SadowskiNarayan2014, McKinneyTchekh2014}.
Recently, \citet{Nakamura2018} compared the jet funnel seen in GR-MHD and force-free electrodynamic simulations 
with VLBI data of M87, finding good agreement concerning a parabolic jet shape.

With a number of codes available, it is possible to perform comparison studies, as analytical test problems in GR-MHD
do not  exist.
A major breakthrough along these lines has been achieved as an integral part of the EHTC studies, comparing a set of 
GR-MHD codes (including \texttt{HARM3D}) in the ideal-MHD limit, simulating a torus around a black hole \citep{Porth2019}.
All codes produce very similar results confirming the robustness of the methods used.

In contrast to most of the GR-MHD simulations including the above-mentioned code-comparison studies,
one of the specific features of our present study is that we follow the evolution of a {\em thin disk} right from 
the start of the simulation.
Thin disks were first studied in a purely hydrodynamic approach by \citet{SS1973} in the non-relativistic limit 
and by \citet{NT1973} for the general relativistic case.
As a seminal step forward, the $\alpha$-viscosity was invented as a mean driver of angular momentum exchange in disks
\citep{SS1973}.
Launching simulations of jets out of thin disks using non-relativistic resistive MHD were pioneered by \citet{CK2002}.
Those simulations and many follow-up studies essentially apply resistivity or {\em magnetic diffusivity} to allow matter to be
accreted through the magnetic field that threads the disk, and also disk material to be loaded on the jet magnetic field, 
eventually leading the system into an inflow-outflow structure in quasi-stationary state.
We further refer to \citet{Zanni2007} who studied the efficiency of the magneto-centrifugal acceleration mechanism
for different levels of resistivity (see also \citealt{Somayeh2012}).

It thus seems essential to apply resistive MHD for disk-jet launching also for the relativistic case.
Magnetically diffusive MHD codes for the relativistic case have been developed only rather recently.
Resistive MHD for special relativistic simulations was pioneered by \citet{Komissarov2007}.
\citet{Palenzuela2009} applied an implicit-explicit solver for the resistive GR-MHD equations in order to deal with the 
stiff part of the electric field, allowing them to model magnetized rotating neutron stars \citep{Palenzuela2013}.
A similar scheme was used by \citet{Dionysopoulou2013} for the resistive version of the \texttt{WHISKY} code
\citep{Whisky2005}, which was then to study collisions of binary neutron stars \citep{Dionysopoulou2015}. 
The ideal MHD code \texttt{ECHO} \citet{DelZanna2007} was also extended to the resistive regime considering as a fully covariant 
mean-field dynamo closure \citep{BdZ2013}.
Subsequently, \citet{Bugli2014} investigated the evolution of a kinematic mean-field dynamo in thick accretion disks.
\citet{Porth2017} presented a GR-MHD code particularly suited for black hole accretion and \citet{Ripperda2019} evolved it further including resistivity and a new inverse solver for the electric field.

In the present paper we have expanded the physics of the parallel, 3D, conservative, GR-MHD code \texttt{HARM3D} 
\citep{Gammie2003, Noble2006, Noble2009} by implementing resistivity in the form of a magnetic diffusivity, 
following \citet{BdZ2013} and \citet{QQ1,QQ2}
This allows us to run axisymmetric (so-called 2.5D) simulations of thin accretion disks around black holes in order 
to investigate the detailed launching conditions that favour the generation of relativistic jets.
In particular we are interested in comparing the energy budget of the jet launched from the black hole magnetosphere 
to the jets launched from the disk and to compare the outflow mass fluxes to that of the disk accretion.
Compared to our previous works \citep{QQ1,QQ2}, we can now take advantage of the parallelization of the code and can
aim for long lasting simulation runs on larger domains and with better grid resolution.

Our paper is structured as follows.
Section~\ref{sec:theory} introduces the basic theory of (resistive) GR-MHD.
Section~\ref{sec:setup} includes a summary of the initial setup as well as the characteristic properties of the simulations.
Section~\ref{sec:refsim} discusses our reference simulation and the outflows it develops.
Section~\ref{sec:comparison} compares the reference simulation with simulations of different black hole rotation and levels 
of magnetic diffusivity.
In Section~\ref{sec:EHT} we briefly discuss our results in the light of the recently detected black hole shadow in the center 
of M87.
Finally, Section~\ref{sec:summary} summarizes our work.
In the Appendix we present test simulations for the implementation of resistivity and a test simulation of the thin disk
setup using the GR-MHD code in the mildly-relativistic limit.

\section{Theoretical background}
\label{sec:theory}
Here we review the basic equations of resistive GR-MHD as a basis for our implementation of resistivity (in the form 
of magnetic diffusivity) in the formerly ideal GR-MHD code \texttt{HARM3D} \citep{Gammie2003, Noble2006, Noble2009}.

We adopt the signature of \citet{MTW1973} for the metric ($-,+,+,+$) and use geometrized units where $G=c=1$. 
Greek letters run for 0,1,2,3 ($t,r,\theta,\phi$ )while Latin letters run for 1,2,3 ($r,\theta,\phi$). 
Radii are expressed in units of the gravitational radius, $\rg = GM/c^2$, while time is in units of light travel time $\tg = GM/c^3$.
Vector quantities are denoted with bold letters while the vector and tensor components are indicated with their respective indices.

Our code uses the {"}3+1{"} decomposition of the GR-MHD equations where the time component is separated from the spacial 
components which are expressed as 3-dimensional manifolds. 
The space-time is described by the metric $g_{\mu \nu}$ in Kerr-Schilds coordinates with $g \equiv det(g_{\mu \nu})$.
A zero angular momentum observer frame (ZAMO) exists in the spacelike manifolds moving only in time with velocity 
$n_{\mu} = (-\alpha,0,0,0)$ where $\alpha = 1/ \sqrt{-g^{tt}}$ is the lapse function. The gravitational shift is 
$\beta^i = \alpha^2 g^{ti}$.
The four velocity of the fluid in the co-moving frame is $u^{\mu} = (u^t, 0, 0, u^{\phi})$ .
The code solves the equations of resistive GR-MHD using a conservative scheme based on the previous works of 
\citet{Gammie2003} and \citet{Noble2006}.

We extended the physics simulated by the code to the resistive GR-MHD regime by implementing magnetic diffusivity in the ideal 
GR-MHD version of the code, following the work of \citet{BdZ2013} and \citet{QQ1}.
As a result we were required to increase the number of variables from 8 to 11 adding the three components of the electric field.
We denote our new resistive GR-MHD code with \texttt{rHARM3D}. 

\subsection{Basic GR-MHD equations}
A magnetized fluid in a general relativistic environment is described by the Maxwell equations \citep{Maxwell1865} in covariant form
\begin{equation}
     \nabla_{\nu} ^{*}F^{\mu \nu} = 0, \quad\quad
     \nabla_{\nu} F^{\mu \nu} = J^{\mu},
\label{eq:Maxwell}
\end{equation}
where
\begin{equation}
\begin{split}
F^{\mu \nu} & =  u^{\mu}e^{\nu} - e^{\mu}u^{\nu} + \epsilon^{\mu \nu \alpha \beta} u_{\alpha} b_{\beta}
 \\
^{*}F^{\mu \nu} & =  - u^{\mu}b^{\nu} + b^{\mu}u^{\nu} + \epsilon^{\mu \nu \alpha \beta} u_{\alpha} e_{\beta}
\end{split}
\label{eq:Faraday}
\end{equation}
are the anti-symmetric Faraday and Maxwell tensors, $e^{\mu}$ and $b^{\mu}$ are the electric and magnetic field in the fluid 
rest frame, and $\epsilon^{\mu \nu \kappa \lambda}$ is the Levi-Civita symbol
\begin{equation}
\epsilon_{\alpha \beta \gamma \delta} =             \sqrt{-g}[\alpha \beta \gamma \delta], \quad
\epsilon^{\alpha \beta \gamma \delta} = - \frac{1}{\sqrt{-g}}[\alpha \beta \gamma \delta].
\label{eq:LevCiv}
\end{equation}
The magnetic and electric field as measured by the normal observer are defined as $\mathcal B^i = n_{\mu} ^{*}F^{\mu \nu} = \alpha ^{*} F^{it}$ and $\mathcal E^i = n_{\nu} F^{\mu \nu} = - \alpha F^{it}$.
The equations of motion for the magnetized 
fluid are
\begin{eqnarray}
   \nabla _{\nu} T^{\mu \nu} = 0,
\label{eq:motion}
\end{eqnarray}
where $T^{\mu \nu} = T^{\mu \nu}_{fluid} + T^{\mu \nu}_{EM}$ is the stress-energy tensor which can be split into a fluid part and an electromagnetic part.
The fluid component can be written as
\begin{equation}
    T^{\mu \nu}_{fluid} = (\rho + u + p)u^{\mu}u^{\nu} + pg^{\mu \nu}, 
\label{eq:Tfluid}
\end{equation}
where $\rho$ is the mass density, $u$ is the internal energy and $p$ is the thermal pressure. 
Pressure and internal energy are connected through
\begin{equation}
u = \frac{p}{\Gamma -1},
\label{eq:eos}
\end{equation}
where $\Gamma$ is the polytropic exponent. 
The electromagnetic component can be written as
\begin{equation}
\begin{split}
    T^{\mu \nu}_{EM} = & (b^2 + e^2) \left(u^{\mu}u^{\nu} + \frac{g^{\mu \nu}}{2} \right) - b^{\mu}b^{\nu} - e^{\mu}e^{\nu} \\ & - u_{\alpha}e_{\beta}b_{\gamma} \left( u^{\mu}\epsilon^{\nu \alpha \beta \gamma } + u^{\nu}\epsilon^{\mu \alpha \beta \gamma } \right)
\label{eq:Tem}
\end{split}
\end{equation}
These two components can be combined into the total stress energy tensor
\begin{eqnarray}
    T^{\mu \nu} &=& (\rho + u + p + b^2 + e^2)u^{\mu}u^{\nu} \nonumber \\
    & +& \left(p + \frac{1}{2}(b^2 + e^2) \right)g^{\mu \nu} \nonumber \\
    & -& b^{\mu}b^{\nu} - e^{\mu}e^{\nu} - u_{\alpha}e_{\beta}b_{\gamma} \left( u^{\mu}\epsilon^{\nu \alpha \beta \gamma } + u^{\nu}\epsilon^{\mu \alpha \beta \gamma } \right)
\label{eq:Ttot}
\end{eqnarray}

\subsection{From ideal to resistive GR-MHD}
Resistivity enters the equations in the form of an (anomalous) magnetic diffusivity $\eta = \eta(r,\theta)$
that is believed to be of turbulent nature.
In ideal MHD, the electric field is given by Ohm's Law $\pmb{E}+\pmb{v}\times \pmb{B} = 0$. 
In the resistive regime Ohm's Law becomes
\begin{equation}
    \pmb{E}+\pmb{v}\times \pmb{B} = \eta \pmb{J},
\label{eq:Ohm}
\end{equation}
or in covariant form in the fluid frame 
\begin{equation}
    e^{\mu} = \eta j^{\mu},
\label{eq:Ohm2}
\end{equation}
where $J$ is the electric current density. 
In the resistive environment, the electric field can no longer be calculated by the cross product of fluid velocity and magnetic field and 
new equations need to be formulated.
By setting $\eta=0$ we get back into the ideal case $e^{\mu}=0$.

Furthermore, magnetic diffusivity puts a restriction in the time-step of a numerical simulation, as the diffusive time step goes as
$dt_{\eta} = \left( \Delta x_i \right)^2 / \eta$, where $\Delta x_i$ is the smallest cell size in any dimension of the grid. 
Thus, for high values of magnetic diffusivity we expect the diffusive time step to become lower than the MHD step and
to effectively determine the evolution of the simulation. 

\begin{figure}
    \centering
    \includegraphics[width=0.95\columnwidth]{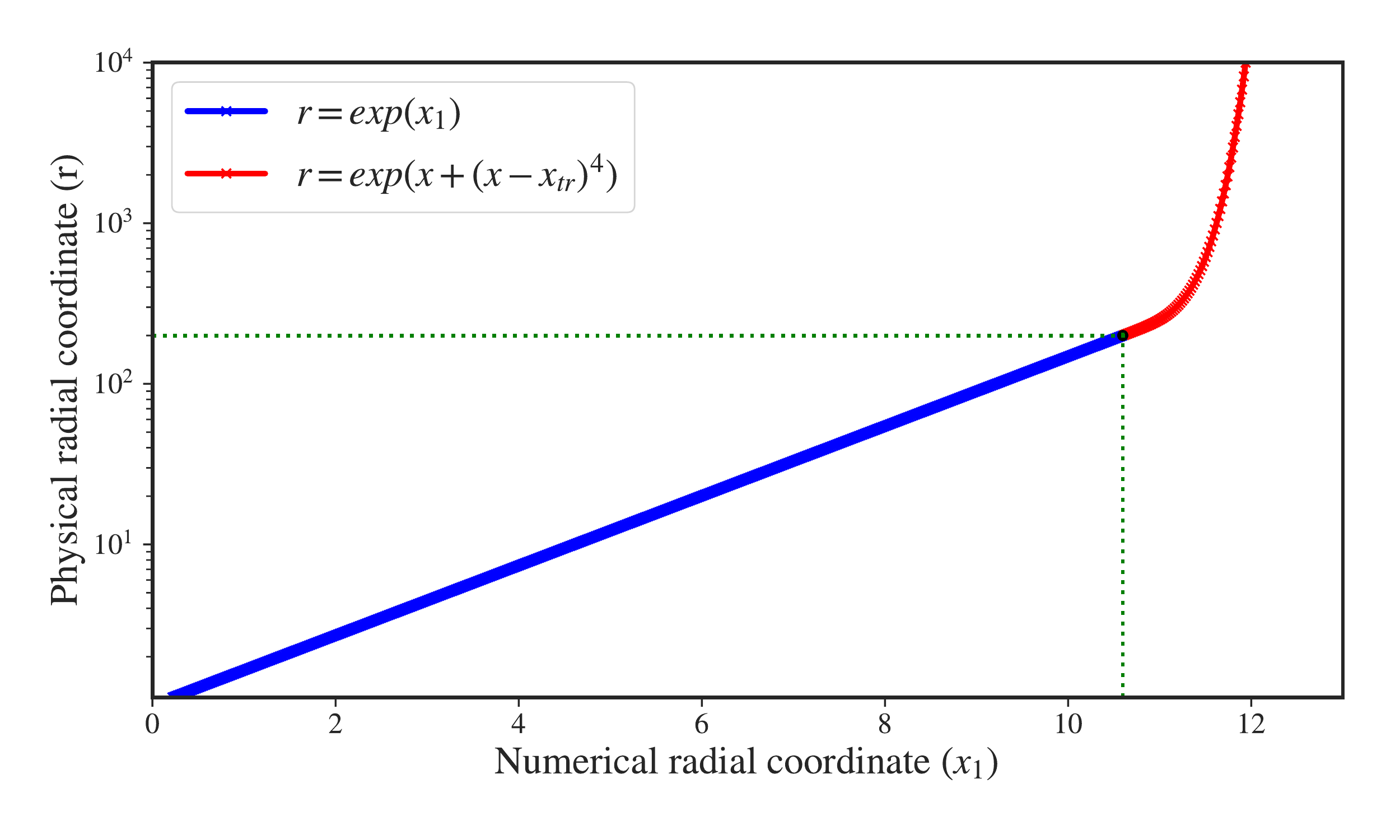}
    \caption{Interrelation between the numerical and physical radial coordinates for the stretched grid. 
    Up to a transition radius $\left(x_{\text{tr}}, R_{\text{tr}}\right)$, we use a simple logarithmic grid (blue), 
    beyond which the grid transits into a hyper-logarithmic scaling (red).}
    \label{fig:coords}
\end{figure}

\section{Simulation setup}
\label{sec:setup}
This paper considers GR-MHD simulations of thin accretion disks that
rotate differentially around a (rotating) black hole and are threaded by a poloidal magnetic field. 
Here we describe the initial conditions we use for our models, the boundary conditions and 
other numerical details of the simulation.

\begin{figure}
    \begin{tabular}[b]{cc}
        \subfloat{%
            \includegraphics[width=0.42\columnwidth]{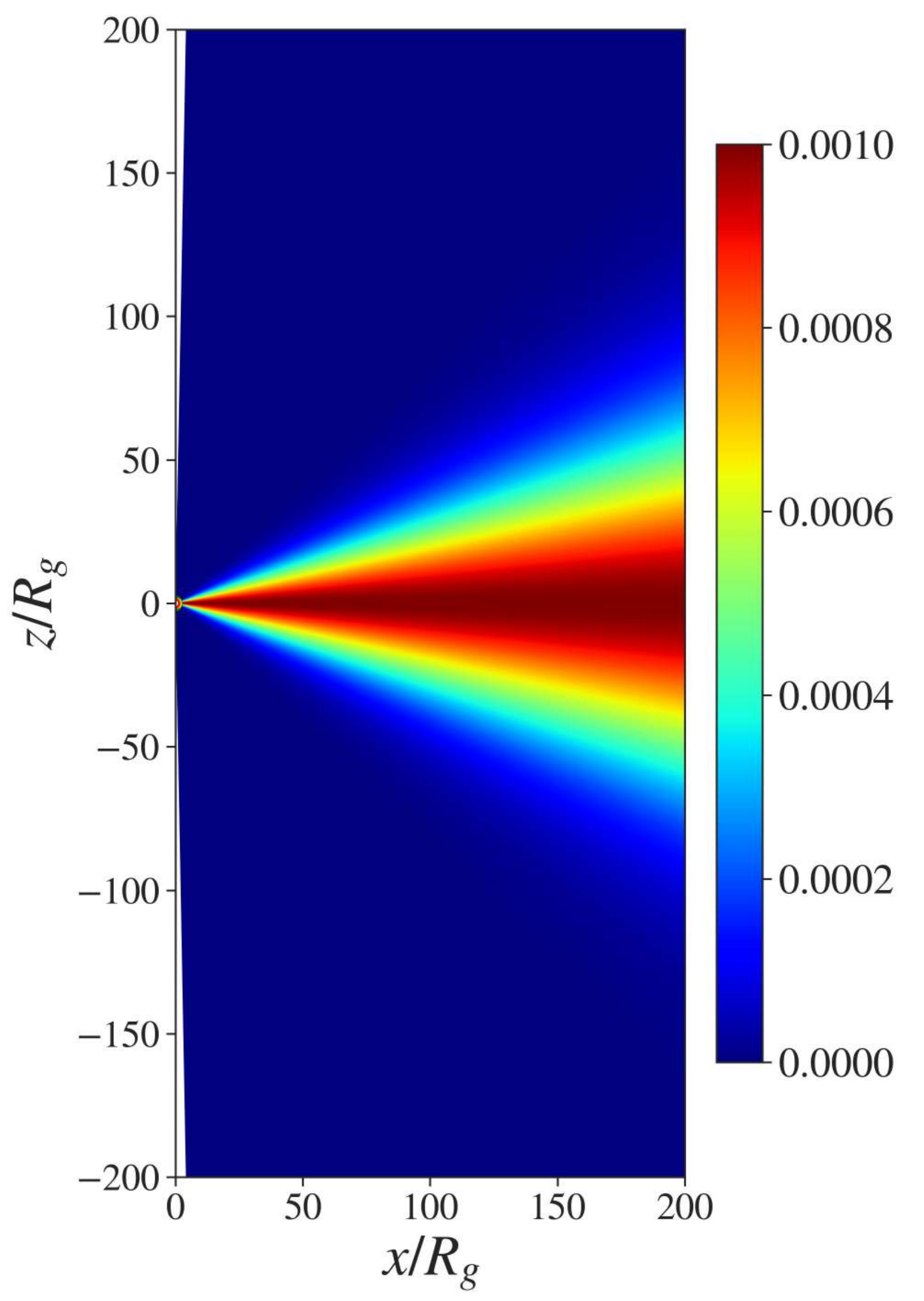}
        } 
    &
        \begin{tabular}[b]{c}
            \subfloat{%
                \includegraphics[width=0.48\columnwidth]{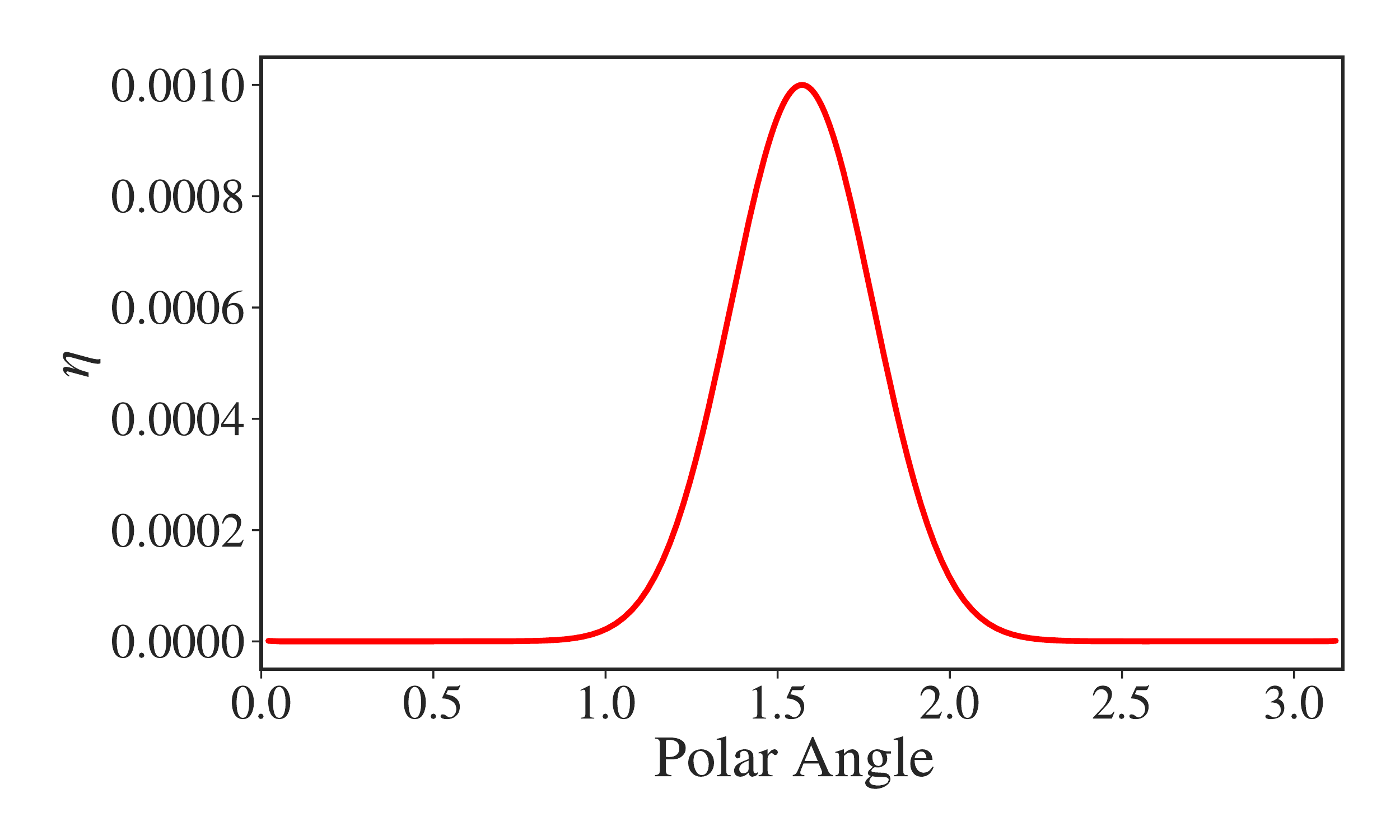}
            } \\
            \subfloat{%
                \includegraphics[width=0.48\columnwidth]{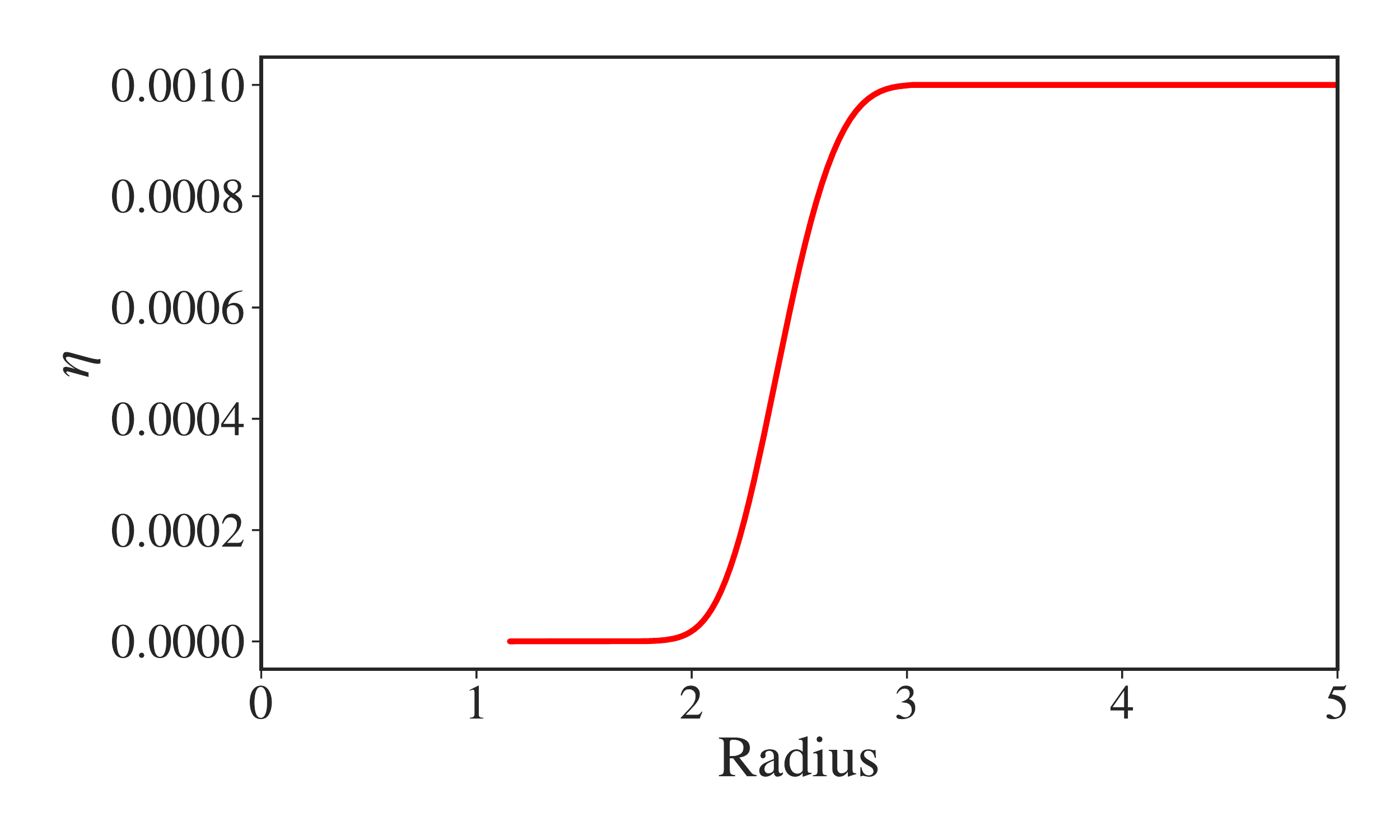}
            } 
        \end{tabular}
    \end{tabular}
    \caption{The distribution of the magnetic diffusivity. 
    We define a diffusivity for $r>3$.
    Along the equatorial plane, the diffusivity saturates to a constant value $\eta_0$,
    while in both (spherically) radial and angular direction it follows a Gaussian profile.}   
     \label{fig:diffusivity}
\end{figure}

\subsection{Numerical grid}
\label{sec:grid}
Depending on our problem setup we apply a different numerical grid.
The original grid of \texttt{HARM} applying modified Kerr-Schild coordinates is used for our test simulations of diffusivity
and for the simulation in the mildly-relativistic limit (see Appendix). 

For our science applications we decided to construct a stretched grid in order to shift the outer boundary condition 
as far out as possible.
This grid is an extension of the original \texttt{HARM} grid and is based on the hyper-logarithmic grid of \citet{TchekhMcKinney2009}.
With that we may concentrate cells close to the black hole in radial direction, and concentrate cells close to the 
equatorial plane or the polar axis in polar direction, allowing us
to resolve the turbulent disk and the polar jet at the same time. 
Furthermore, with such a scheme the outer boundary is causally disconnected from the inner simulated area of interest 
close to the black hole or the disk.

In the hyper-logarithmic grid the radial coordinate is split in two parts. 
The first part follows a logarithmic scaling as in the original \texttt{HARM} code \citep{Gammie2003}.
Beyond a transition radius $R_{\text{tr}}$, the grid becomes substantially more scarce, up to the outer radius $R_{\text{out}}$. 

Physical and numerical radial coordinates translate as
\begin{equation}
    r(x_1) = \text{exp}\left[ \frac{1}{2} x_1 + 4H\,\left(x_1 - x_{1_{\text{tr}}}\right)^4 \right],
\end{equation}
where $x_1$ is the uniformly spaced numerical radial coordinate and $x_{1_{\text{tr}}}$ is the transition radius 
(corresponding to $R_{\text{tr}}$). 
The function $H = H(x_1 - x_{1_{\text{tr}}})$ is a step function that is equal to unity for $x_1 > x_{1_{\text{tr}}}$ and 
vanishes otherwise. 
In Figure \ref{fig:coords} we show the relation between the numerical and the physical radial coordinates.

The physical and numerical polar coordinates are connected by 
\begin{equation}
    \theta(x_2) = \theta_{\text{start}} + x_2 \theta_{\text{length}} - h_{\text{slope}} \sin (4\pi x_2),
\end{equation}
where $\theta$ and $x_2$ are the physical and numerical polar coordinates respectively, while $\theta_{\text{start}}$ 
denotes the starting angle and $\theta_{\text{length}} = \pi - 2\theta_{\text{start}}$ the angular length of the coordinate
in radians.
The factor $h_{\text{slope}}$ governs how many grid cells are focused towards the equatorial plane and towards 
the symmetry axis.
We note that these coordinates are slightly different from the original \texttt{HARM} code, where the choice of focusing 
coordinates and the increase of resolution for the polar coordinate is only possible towards the equatorial plane.

Our typical maximum resolution in the polar coordinate is $\Delta\theta = 0.00625$ along the polar axis and in the equatorial 
plane while the minimum resolution is $\Delta\theta = 0.025$ at $45^{\circ}$. 
The radial coordinate is best resolved close to the horizon where $\Delta r_{r=2} = 0.02$, and is radially decreasing with 
$\Delta r_{r=10} = 0.1$, $\Delta r_{r=50} = 0.5$ and $\Delta r_{r=100} = 1$. 

\begin{figure}[t]
    \centering
    \includegraphics[width=0.9\columnwidth]{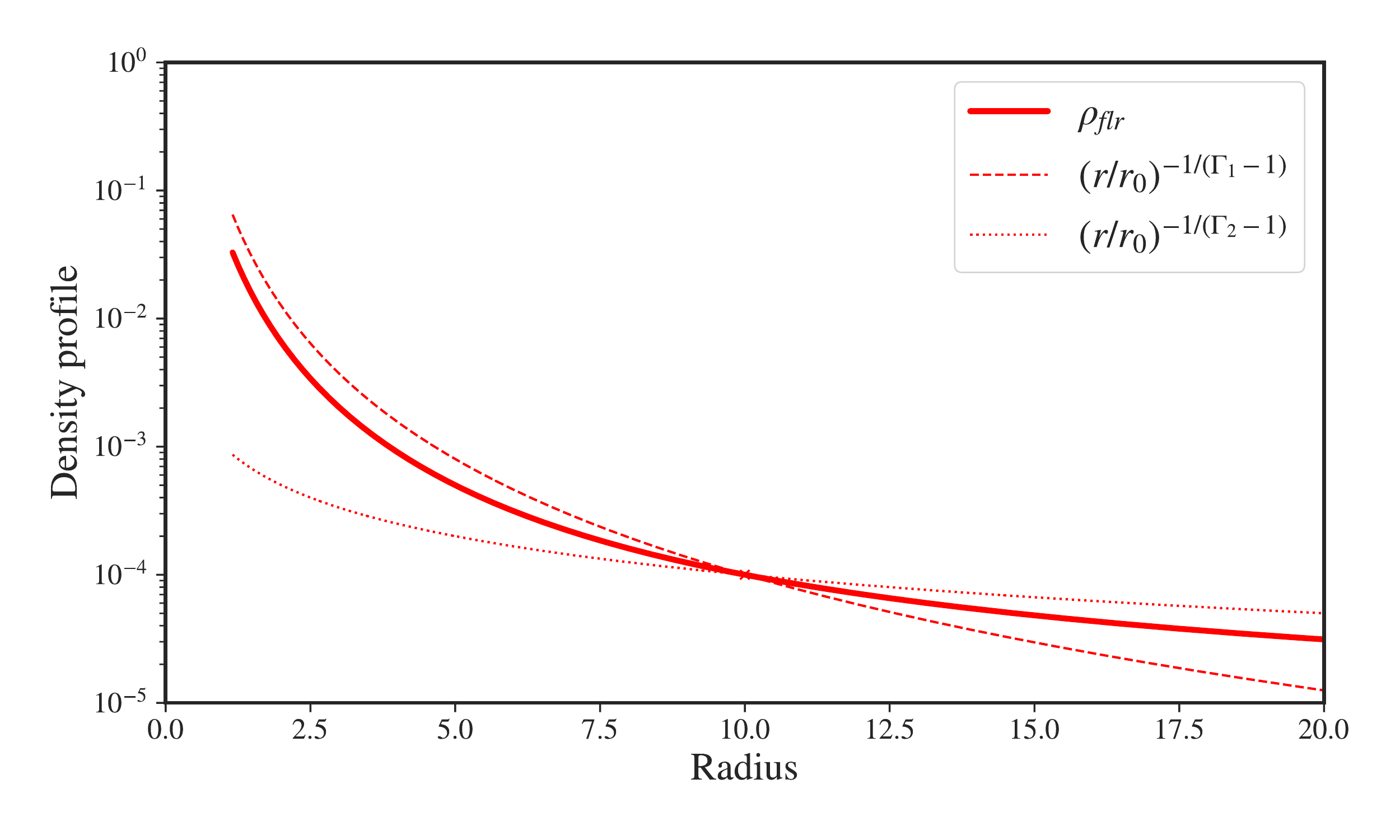}
    \includegraphics[width=0.9\columnwidth]{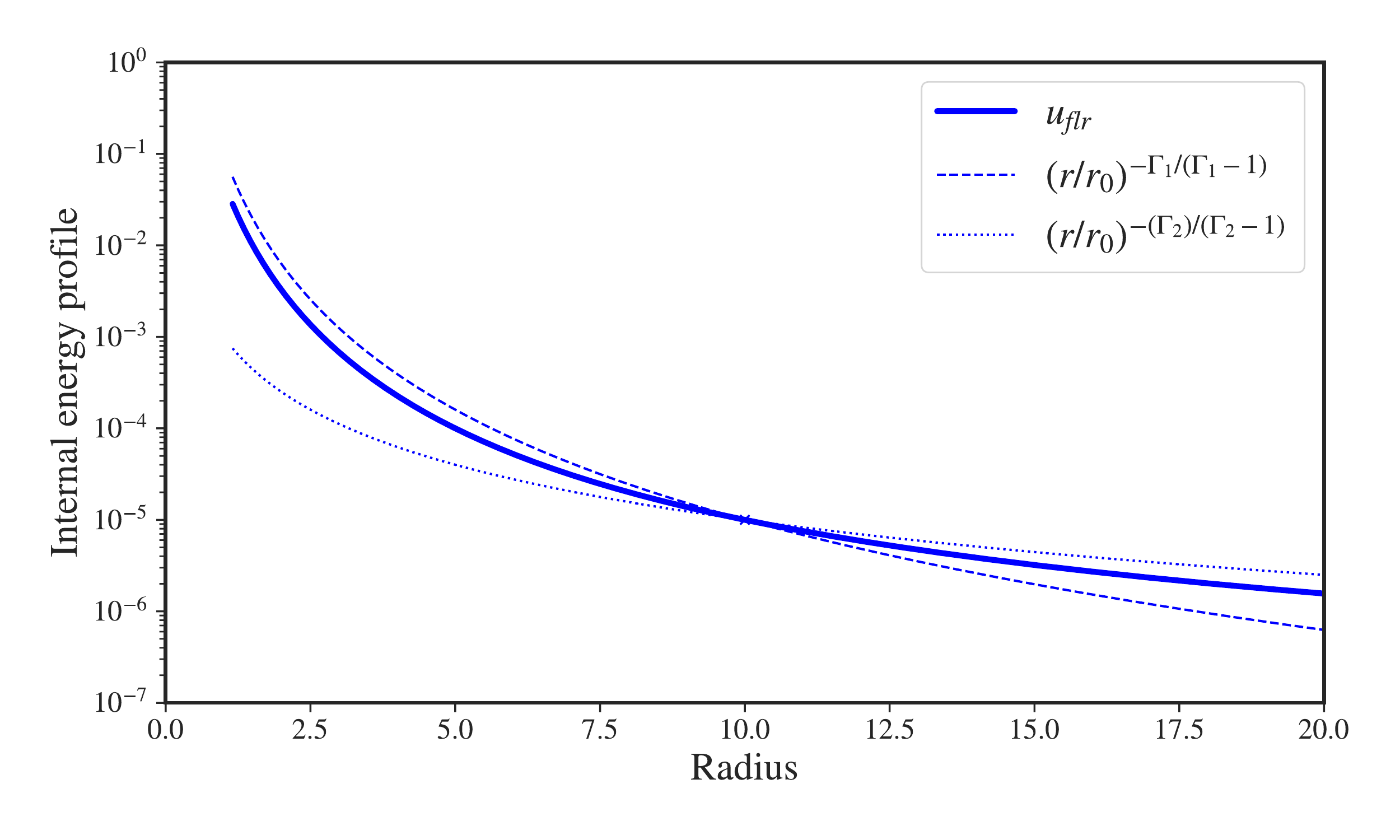}
    \caption{The floor model used in the science simulations. The radial density (top) and internal energy (bottom) distributions are calculated as broken power laws.}
    \label{fig:floors}
\end{figure}

{
\subsection{Boundary conditions}
For our simulations we use outflow boundary conditions in the inner and outer radial boundary.
The values of the primitive variables are copied from the boundary cells to the ghost cells.
At the same time we make sure there is no inflow from the boundaries by checking the velocity is pointing outwards 
at each boundary cell.
As an extra measure in the inner boundary, we make sure we have 10 cells of our grid inside the black hole event 
horizon in order to prevent numerical effects from propagating outside of it.

Furthermore, one of the reasons we modified our numerical grid into the hyper-logarithmic version we described 
before is because we wanted to have the outer boundary as far away from the disk as possible.
Before adopting the hyper-logarithmic grid we had noticed a collimation effect in the magnetic field lines which we
had deemed as artificial (see Appendix B in \citet{QQ2}).
By selecting an outer radius of $R_{\text{out}} = 10000$ we make sure that the outer boundary stays causally 
disconnected from the disk.
In the axial boundary we impose axisymmetric boundary conditions where the vector values are being reflected 
along the small cutout in both axes.
}

\subsection{Initial conditions}
The initial disk density distribution is described by a non-relativistic vertical equilibrium profile,
such as applied in \citet{Somayeh2012}, 
\begin{equation}
    \rho(r,\theta) = \left[ \frac{\Gamma -1}{\Gamma} \frac{r_{in}}{r} \frac{1}{\epsilon^2} \left( \sin \theta + \epsilon^2 \frac{\Gamma}{\Gamma -1} \right) \right]^{1 / (\Gamma -1)},
    \label{eq:Somayeh}
\end{equation}
slightly modified to fit into our code. 
Here, $r_{\rm in}$ is the initial inner disk, and $\epsilon = H/r$ is the initial disk aspect ratio as is defined 
by the vertical equilibrium of a disk with a local pressure scale height $H(r)$.
The pressure and internal energy are given by the polytropic equation of state $p = K \rho^{\Gamma}$ and Equation~\ref{eq:eos}, where $K$ is the 
polytropic constant.
For the polytropic exponent we will use different values for different simulations as specified in the sections below.

Around the disk we prescribe an initial ''corona".
For the choice of a polytropic index of $\Gamma = 4/3$, the disk has a finite outer radius much smaller than the 
outer radius of the stretched grid.
Furthermore, the upper and lower disk surfaces do not follow lines of constant polar angle as implied 
by Eq.~\eqref{eq:Somayeh}.
The initial coronal density and pressure are given by
\begin{equation}
\rho_{\text{cor}} \propto r^{1 / (1-\Gamma)}, \quad\quad p_{\text{cor}} = K_{\text{cor}} \rho_{\text{cor}}^{\Gamma}.
    \label{eq:corona}
\end{equation}
The coronal temperature is chosen to be much higher than the disk temperature, $K_{\text{cor}} >> K$.
This implies a density jump between disk and corona, but a pressure equilibrium along the disk surface.
More specifically, for our simulation we chose $K=0.001$ for the disk initial condition and $K=1$ for 
the initial corona. 
{ The corona collapses instantly the moment the simulation starts and part of it is also expelled by 
the initial ejections from the disk, meaning that the values are quickly replaced by the floor values of 
the simulation (see Sect.~\ref{sec:floor}).
However, the polytropic equation $P = K \: \rho^{\Gamma}$ is not enforced in any step of the code except
the initial condition. 
The code uses Equation~\ref{eq:eos} to connect pressure and internal energy, which means that 
entropy and temperature are free to change.}

The disk is given an initial orbital velocity following \citet{PW},
\begin{equation}
    \Tilde{u}^{\phi} = r^{-3/2} \left(\frac{r}{r-R_{\text{PW}}} \right).
    \label{eq:PW}
\end{equation}
where $R_{\text{PW}}$ is a constant of choice, here equal to the gravitational radius $\rg$.
This approximation is applied in the $\phi$-component of the fluid velocity $u^{\phi}$.

\begin{figure*}
    \centering
    \includegraphics[width=0.66\columnwidth]{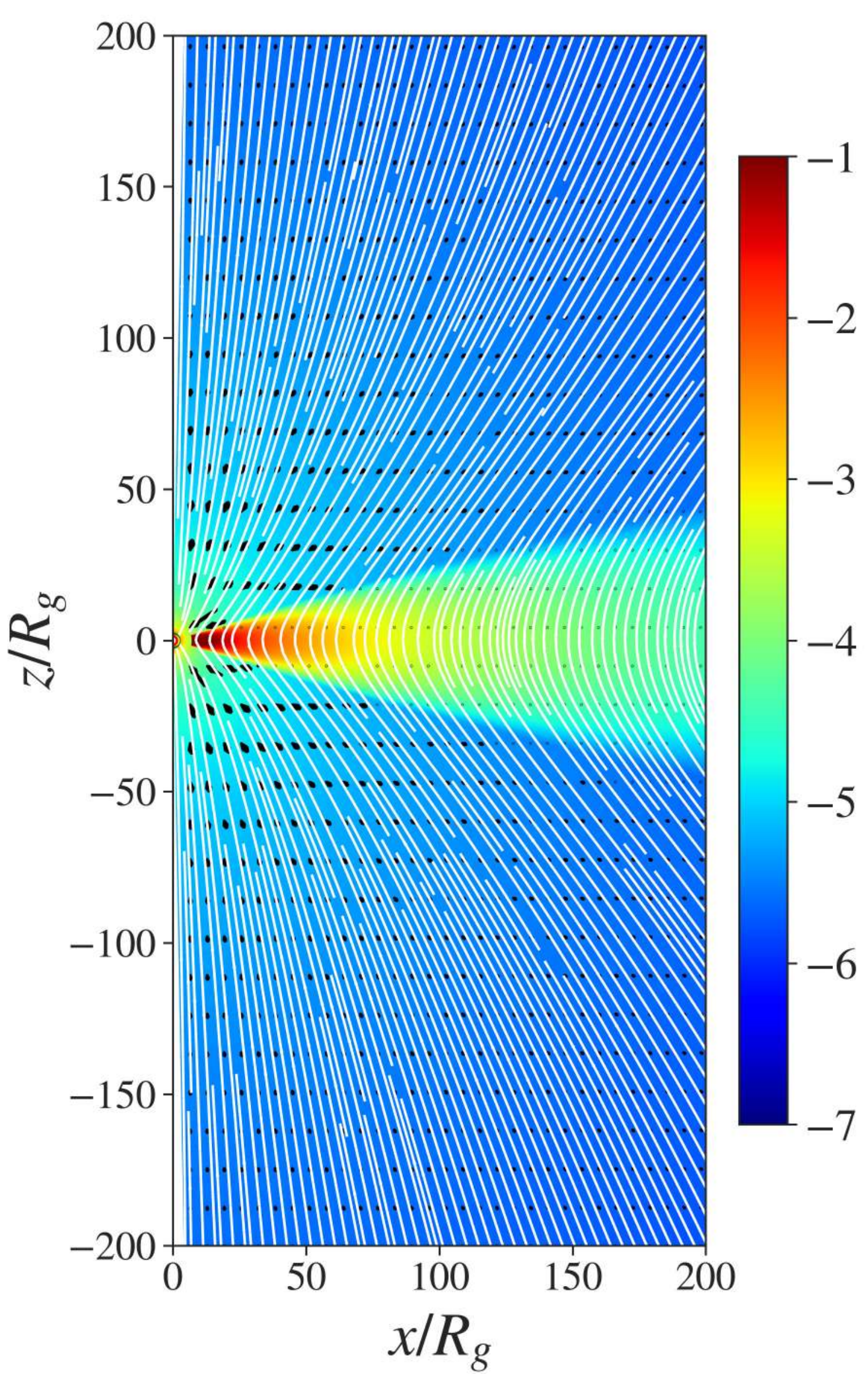}
    \includegraphics[width=0.66\columnwidth]{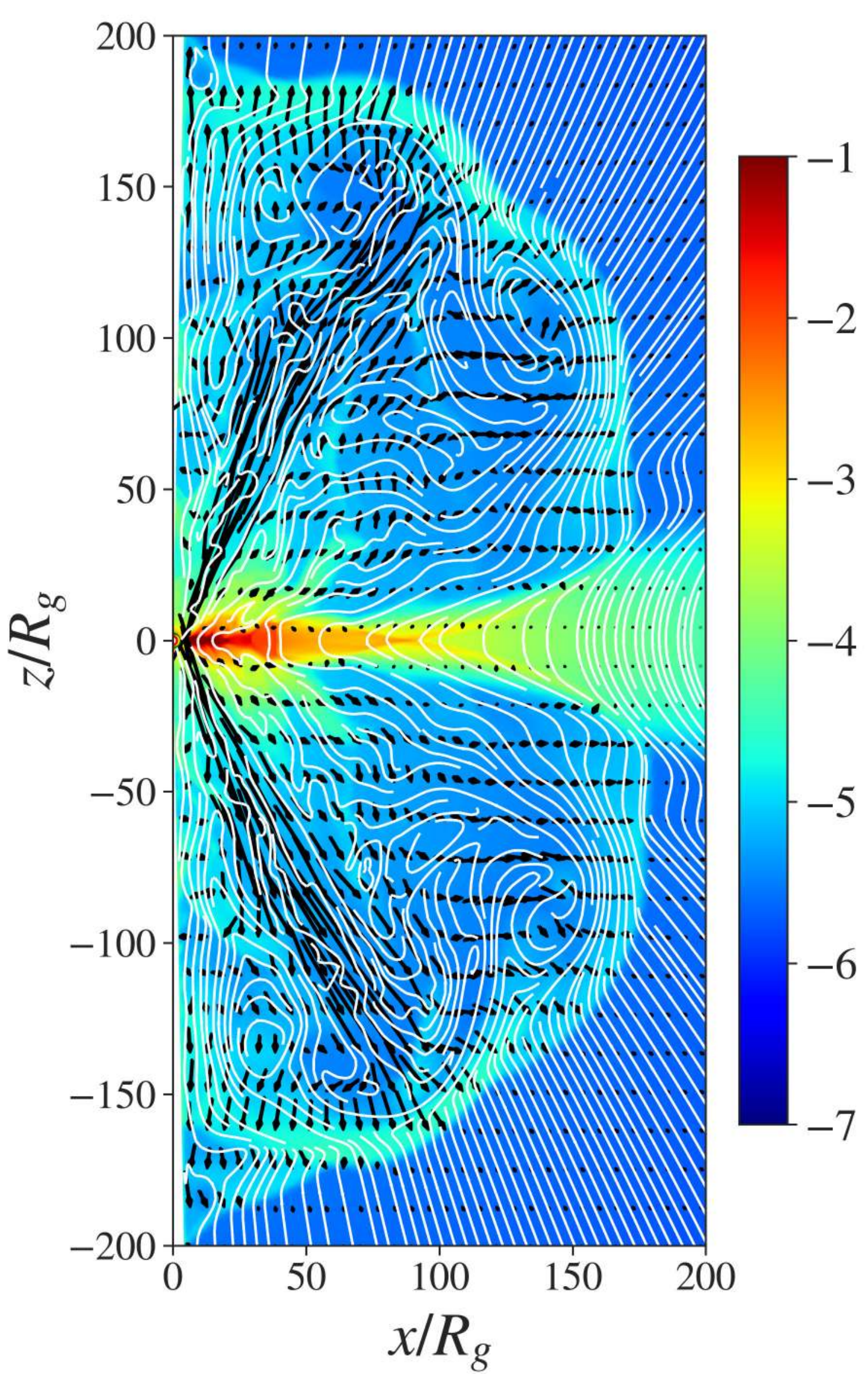}
    \includegraphics[width=0.66\columnwidth]{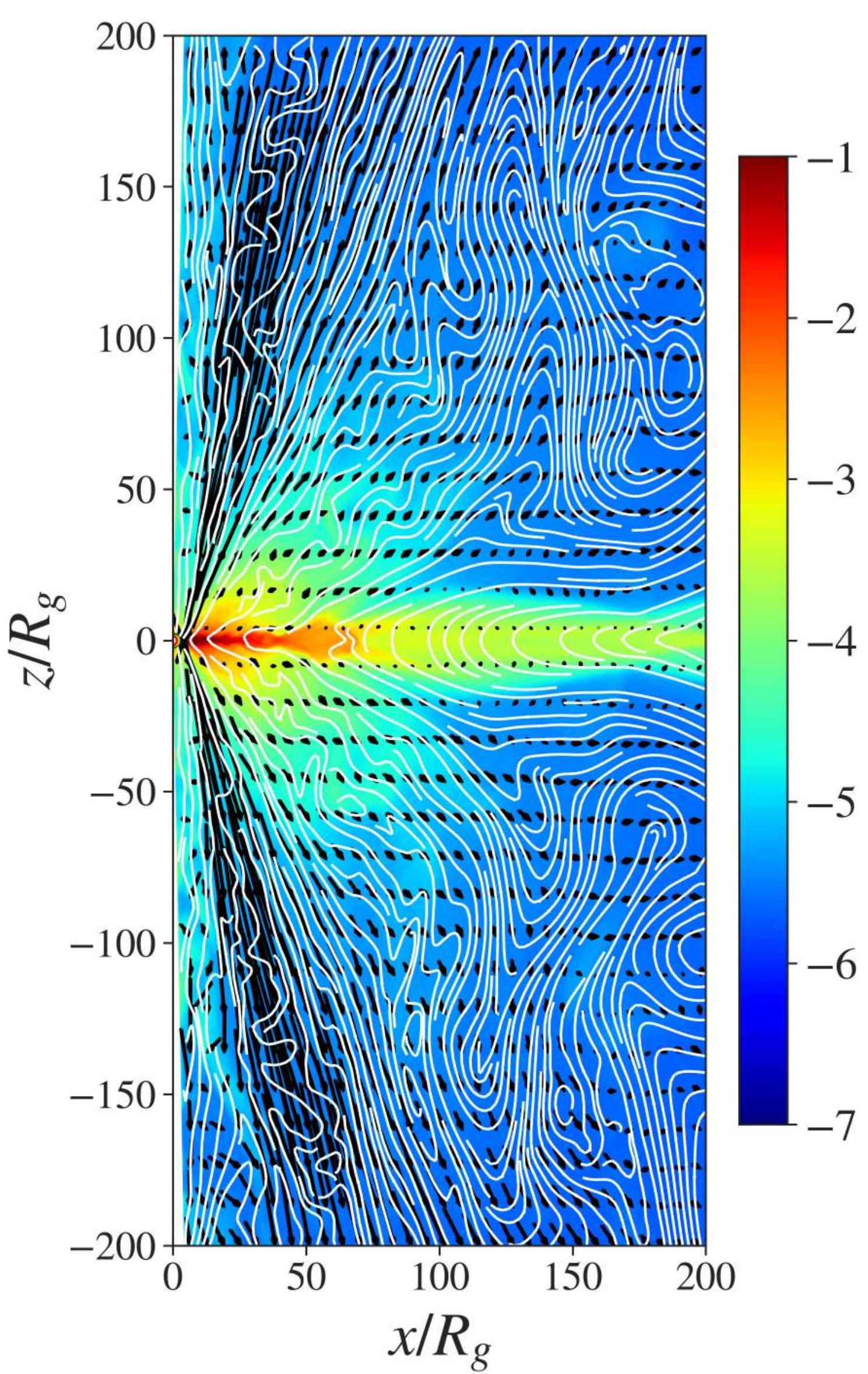}
    \includegraphics[width=0.66\columnwidth]{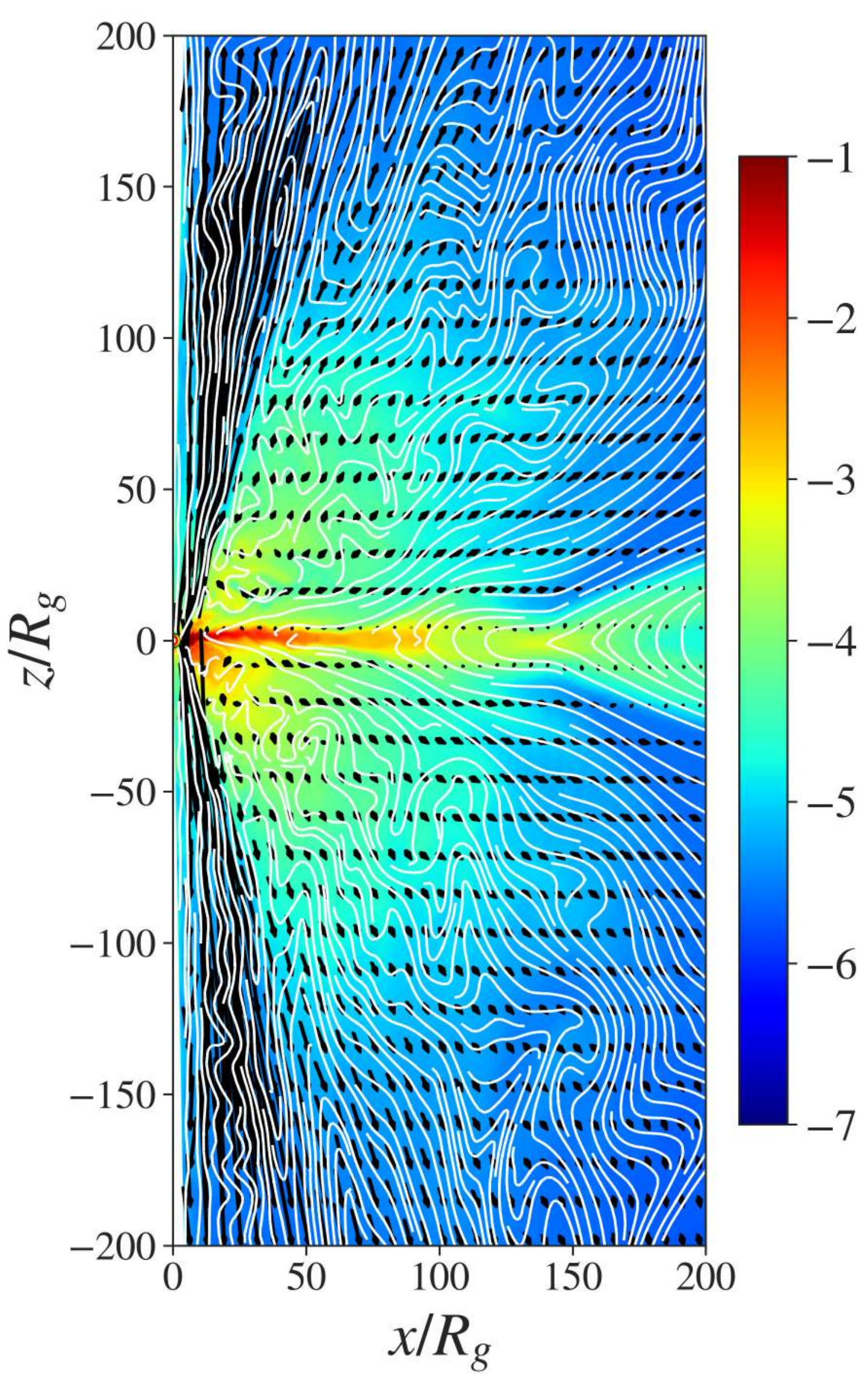}
    \includegraphics[width=0.66\columnwidth]{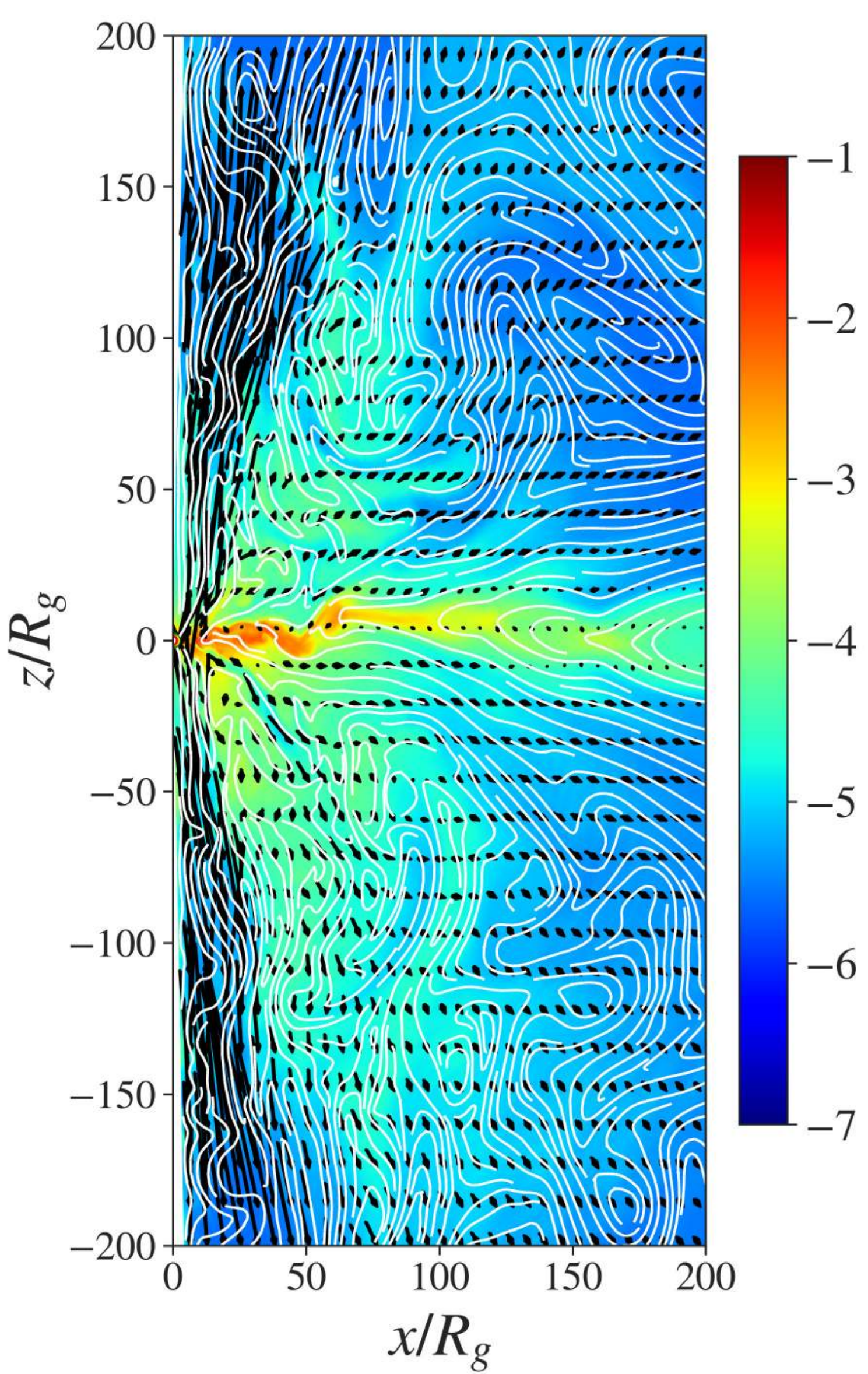}
    \includegraphics[width=0.66\columnwidth]{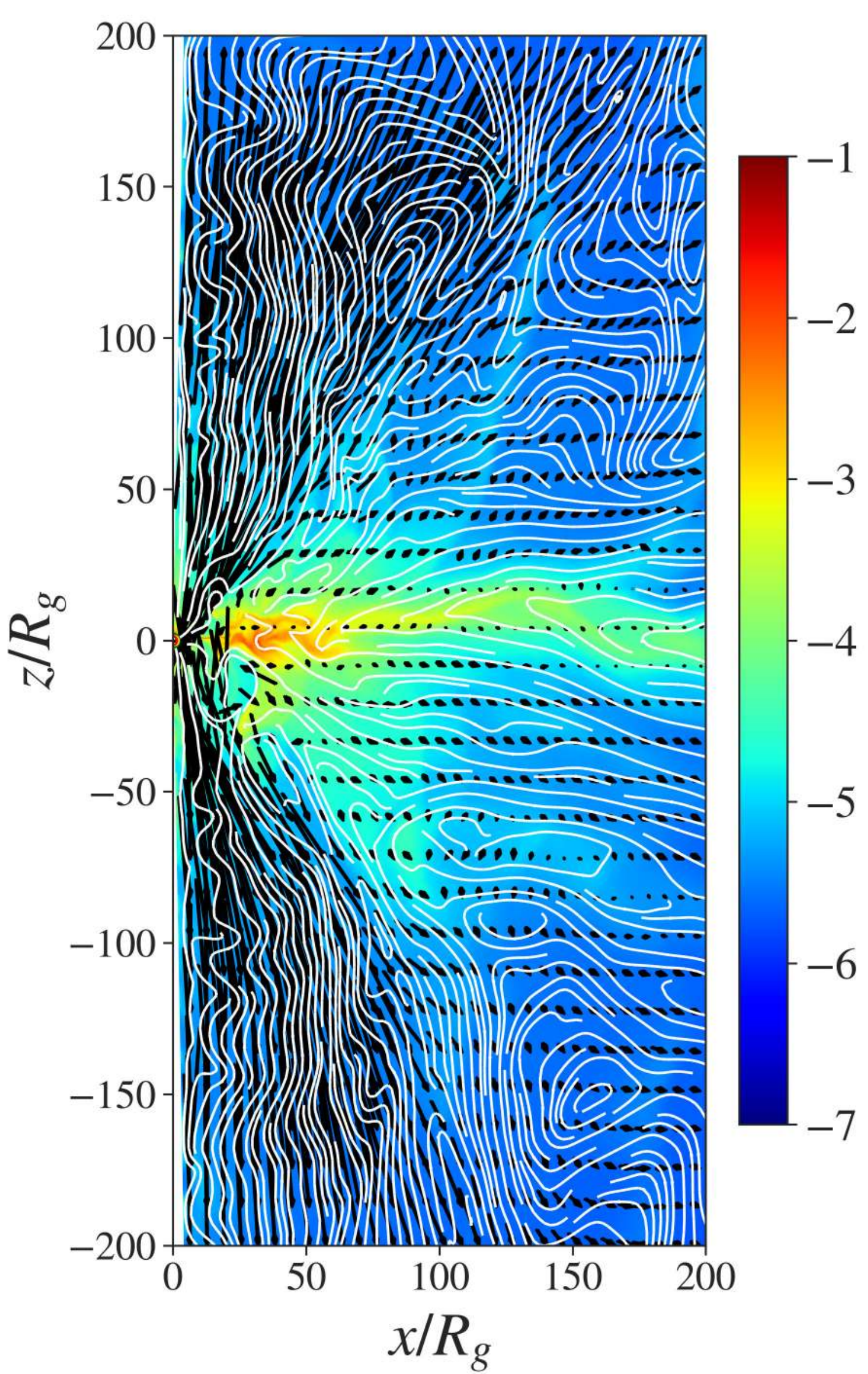}
    \caption{Snapshots of our reference simulation. Shown is the density distribution (in log scale) at times
    $t =$ 0, 1000, 2000, 4000, 6000, and 8000. 
    The poloidal magnetic field lines are shown (white lines), while the poloidal velocity field is represented by the black arrows.}
    \label{fig:h20.3}
\end{figure*}


The disk is initially threaded by a large scale poloidal magnetic field, implemented via the magnetic vector potential
$A_{\phi}$ following $B = \nabla \times A_{\phi}$. 
In most cases we use the inclined field profile suggested by \cite{Zanni2007},
\begin{equation}
    A_{\phi}(r,\theta) \propto ( r \sin\theta )^{3/4} \frac{ m^{5/4} }{ \left(m^2 + \tan^{-2}\theta \right)^{5/8} }.
    \label{eq:VecPot}
\end{equation}
The parameter $m$ determines the initial inclination of the field, which plays an important role for the magneto-centrifugal
launching of disk winds \citep{BP1982}.
The magnetic field strength is then normalized by choice of the plasma-$\beta = p_{\text{gas}} / p_{\text{magn}}$.

\subsection{The magnetic diffusivity}
\label{sec:diffusivity}
The simulations presented in this paper apply a scalar function for the magnetic diffusivity that is constant in time.
The diffusivity is assumed to be of turbulent nature, thus much larger than the microscopic resistivity
and thought to be generated by the magneto-rotational instability (MRI, \citealt{BH1991}).
In general, the magnetic diffusivity distribution is chosen such that it resembles a magnetized diffusive disk
within a ideal-MHD wind and jet area. 

We apply a magnetic diffusivity profile as it is typically used in jet launching simulations
(see e.g. \citealt{Zanni2007, Somayeh2012, QQ2}), i.e. a Gaussian profile along the polar angle with a maximum
at the initial disk mid-plane,
\begin{equation}
    \eta(r,\theta) = \eta_0 \exp \left[-2 \left( \frac{\alpha}{\alpha_{\eta}} \right)^2 \right],
    \label{eq:diffusivity}
\end{equation}
where $\eta_0$ is the level of diffusivity along the equatorial plane, $\alpha = \pi/2 -\theta$ is the angle towards the 
disk mid-plane and $\alpha_{\eta} = \arctan (\chi \times \epsilon)$ is
the angle that measures the scale height of the diffusivity profile. 
The parameter $\chi$ compares the scale height of the diffusivity profile with the disk pressure scale height.
This profile -- as artificial as it might seem -- focuses the high diffusivity values in the equatorial plane, allowing for a highly 
resistive material and for a lowly resistive to asymptotically ideal-MHD disk wind and jet.
Since we take resistivity as a result of turbulence, we expect higher diffusivity in the highly turbulent areas of the interior of the disk. 

Initially, we also considered an anisotropic resistivity profile with different values of $\eta$ affecting the 
poloidal and toroidal components of the magnetic field.
According to \citet{Ferreira1997} such a profile would help stabilize the disk evolution reaching a stationary state.
In contrast to \citealt{Zanni2007, MurphyFerreiraZanni2010, Somayeh2012} who applied an anisotropic diffusivity
in their simulations, 
in our case the disk loses its mass rather quickly, mainly due to the disk wind.
This rapid mass loss is actually minimizing the stabilization effect by an anisotropic magnetic diffusivity.
Furthermore, the initial ejections created by the absence of equilibrium between the disk and the black hole delay the
reach of a stationary condition even more.
Based on that, we decided that the introduction of anisotropic diffusivity would not contribute much in the stability 
of the the disk.

When testing the performance of our code, we found that the simulations become more stable when we apply a low 
background diffusivity (1000 times lower than in the disk) along the rotational axis.
We thus apply an exponentially decreasing profile along the axial boundary within 6 grid cells.
As this axial diffusivity is confined within an opening angle of $<3.5^{\circ}$, it does not affect the 
physics of the jet launching.
We also apply an exponential decrease in the radial diffusivity profile from radius $r=3$ towards the horizon, 
resulting in a smooth transition from the high disk diffusivity to the ideal MHD black hole environment.
Figure~\ref{fig:diffusivity} shows the 2D distribution as well as the radial and angular profiles
of $\eta$.

For the magnitude of the magnetic diffusivity $\eta_0$ we apply a range of values, $\eta_0 = 10^{-10} \ldots 10^{-2}$
(in code units).
These values correspond to some kind of standard parameters applied in the literature in diffusive MHD simulation 
in GR \citep{BdZ2013,Bugli2014,QQ1,QQ2}, 
in non-relativistic simulations \citep{CK2002,Zanni2007,Somayeh2012,Stepanovs1}, 
but have been modeled concerning strength and spatial distribution also by direct simulations, e.g. 
by \citet{Gressel2010}.
Concerning the diffusive numerical time stepping and the strength of the numerical diffusivity 
we refer to the discussion in our previous works \citep{QQ1,QQ2}.

Here we emphasize another important impact of physical resistivity: 
{\em It suppresses the magneto-rotational instability, MRI,} \citep{Fleming2000,LongarettiLesur2010}.
Overall, we thus do not expect to detect any MRI being resolve in our disk structure. 

As discussed in \citet{QQ1} the diffusion rate will be of order $k^2\eta$ \citep{Fleming2000}, with the wave number $k$.
From \citet{BH1991} we know that the MRI grows only in a certain range of wave numbers 
$k \in [0,k_{\rm max}]$, in the linear MRI regime -- depending on whether the numerical grid may resolve certain wave lengths
and whether certain wave lengths will fit into the the disk pressure scale height.
Furthermore, there exists a wave number $k_{\rm MRI}$ for which the MRI growth rate reaches a maximum (see \citealt{HawleyBalbus1992} 
for the case of a Keplerian disk). 
A certain number of MRI modes can therefore be damped out when $k_{\rm MRI}^{2}\eta$ is comparable to the maximum growth 
rate of MRI. 
Moreover, for a large enough $\eta$, it is even possible to damp out most of the MRI modes in the linear evolution of MRI.

In \citet{QQ1} a thorough investigation of resistive effects on the accretion rate of an initial Fishbone \& Moncrief \citep{FishboneMoncrief1976} 
torus was presented. 
They could show that for this setup for $\eta$ \lax $10^{-3}$ the MRI seemed to be completed damped, while for lower $\eta$ the onset of the
MRI and thus of massive accretion was substantially delayed.
This result was claimed to be consistent with \citet{LongarettiLesur2010}, demonstrating that the growth rate of the MRI 
substantially decreases with $1/\eta$ beyond a critical diffusivity.

In addition to the point that we do not expect the MRI to play a role in our simulations due to the disk resistivity,
we also note that we consider a thin disk that is thread by a strong magnetic field.
Thus, angular momentum transport is dominated by the torque of the magnetic lever arm.

Another consequences of considering a magnetic diffusivity are physical reconnection of the magnetic field
and also physical ohmic heating.
Both effects are present in our simulations and we will discuss their impact on the accretion-ejection system
accordingly. 

\subsection{The density floor model}
\label{sec:floor}
As typical for any MHD code, \texttt{rHARM3D} cannot work in vacuum.
This is a problem also for relativistic MHD codes, in particular for their inversion schemes, 
so usually a floor model is applied to circumvent numerical problems when the initial 
disk corona collapses.

Depending on the model setup, we apply a different floor model for the density and pressure.
Note that in particular in our approach that applies a large scale initial disk magnetic flux, we potentially
deal with a high magnetization $\propto B^2/\rho$ and / or low plasma-$beta$ $\propto P/B^2$ at large radii.
Thus, for simulations on a large grid, we choose a floor profile following a broken power law.
For the density we apply
\begin{equation}
    \rho_{\rm flr}(r) \propto \left[ \left( \frac{r}{r_0} \right)^{1/(1-\Gamma_1)}
                                   + \left( \frac{r}{r_0} \right)^{1/(1-\Gamma_2)}
                              \right],
\end{equation}
while the internal energy follows
\begin{equation}
      u_{\rm flr}(r) \propto \left[ \left( \frac{r}{r_0} \right)^{\Gamma_1/(1-\Gamma_1)}
                                  + \left( \frac{r}{r_0} \right)^{\Gamma_2/(1-\Gamma_2)} 
                              \right],
\end{equation}
with $\Gamma_1 = 4/3$ and $\Gamma_2 = 2$, and
where $r_0$ marks the transition radius between the two power laws with typically $r_0=10\,\rg$
(see Figure \ref{fig:floors}).
With that we implement higher floor values for large radii in order to avoid a too high magnetization.
Close to the black hole we apply the same floor profile as in the original \texttt{HARM} code.

\begin{figure}
    \centering
    \includegraphics[width=0.95\columnwidth]{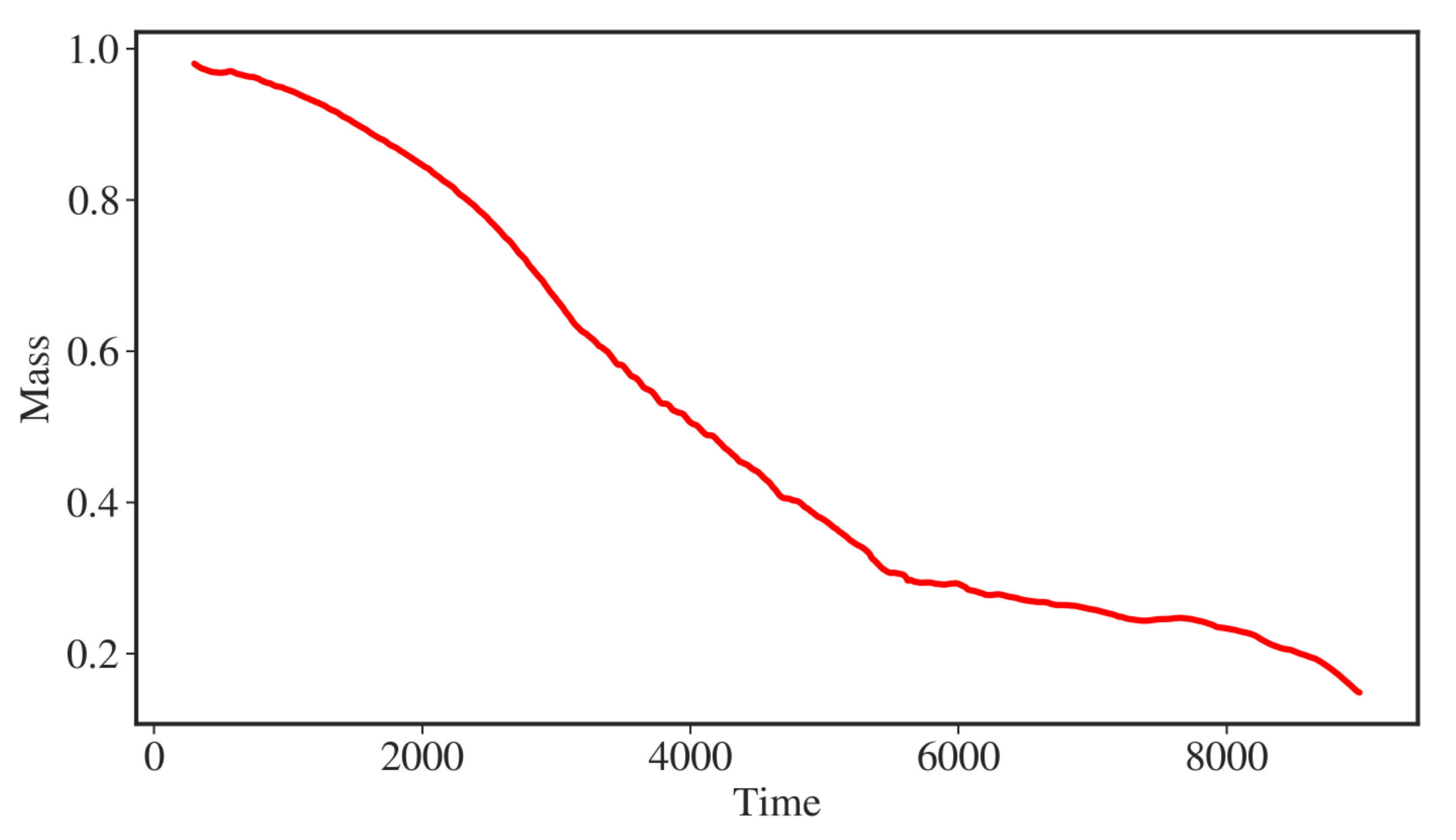}
     \caption{The evolution of the disk mass in our reference simulation measured in a reference area as described 
     in the text. The mass is normalized to the initial disk mass.}
    \label{fig:h20.3_disk_time_massr1r5}
\end{figure}

\subsection{Characteristic quantities of the simulations}
Here we define a number of physical quantities, that will later be used to characterize the evolution in different
simulations.
The mass contained in a disk-shaped area between radii $r_1$ and $r_2$ and between surfaces of constant angle 
$\theta_1$ and $\theta_2$ is calculated as
\begin{equation}
    M_{\rm disk} = 2 \pi \int_{r_1}^{r_2} \int_{\theta_1}^{\theta_2} \rho(r,\theta) \: \sqrt{-g(r,\theta)}\: d\theta\: dr,
    \label{eq:Mass}
\end{equation}
where $\sqrt{-g}$ is the square root of the determinant of the metric.
The mass flux through a sphere of radius $R$ between angles $(\theta_1, \theta_2)$ is
\begin{equation}
    \Dot{M}(R) = 2 \pi \int_{\theta_1}^{\theta_2} \rho(R,\theta) \: u^r(R,\theta) \: \sqrt{-g(R,\theta)}\: d\theta.
    \label{eq:MassFlux_R}
\end{equation}
Similarly we calculate the mass flux in $\theta$-direction considering the $u^{\theta}$ component and the area element 
$\sqrt{-g(R,\theta)}\: d\phi dr$.
This is in particular used to calculate the disk wind mass flux from the disk surface, considering two surfaces with a constant 
opening angle $\Theta$ that is chosen to be similar to the initial disk density distribution.
We thus obtain
\begin{equation}
    \Dot{M}(\Theta) = 2 \pi \int_{r_1}^{r_2} \rho(r,\Theta) \: u^{\theta}(r,\Theta) \: \sqrt{-g(r,\Theta)}\: dr.
    \label{eq:MassFlux_TH}
\end{equation}
The Poynting flux per solid angle is defined as 
\begin{equation}
 \begin{split}
    F_{\text{EM}}(r,\theta) &= -T^r_t                 \\
            & = - \left[ \left(b^2 + e^2 \right) \left(u^r u_t + \frac{1}{2} g^r_t \right) - b^r b_t - e^r e_t \right. \\
            & \left. - u_{\beta} e_{\gamma} b_{\delta} \left(u^r \epsilon_t^{\beta \gamma \delta} + u_t \epsilon^{r \beta \gamma \delta} \right) \vphantom{\frac{1}{2}} \right].
       \label{eq:Poynting_angle}      
 \end{split}
\end{equation}
By integration along the polar angle we obtain the flux through a sphere of radius $R$,
\begin{equation}
    \dot{E}_{\text{EM}} (R) = 2\pi \int_0^{\pi} \sqrt{-g(R,\theta)} \: F_{\text{EM}}(R,\theta) \: d\theta.
\end{equation}
The corresponding electromagnetic flux is
\begin{equation}
 \begin{split}
    F_{\text{EM}}(r,\theta) &= -T^{\theta}_t                 \\
            & = - \left[ \left(b^2 + e^2 \right) \left(u^{\theta} u_t + \frac{1}{2} g^{\theta}_t \right) - b^{\theta} b_t - e^{\theta} e_t \right. \\
            & \left. - u_{\beta} e_{\gamma} b_{\delta} \left(u^{\theta} \epsilon_t^{\beta \gamma \delta} + u_t \epsilon^{\theta \beta \gamma \delta} \right) \vphantom{\frac{1}{2}} \right].
            \label{eq:Poynting_tot}    
 \end{split}
\end{equation}
By integration along the radius we obtain the flux through a surface of constant angle $\Theta$,
\begin{equation}
    \dot{E}_{\text{EM}} (\Theta) = 2\pi \int_{r_1}^{r_2} \sqrt{-g(r,\Theta)} \: F_{\text{EM}}(r,\Theta) \: dr.
\end{equation}
The poloidal Alfv\'en Mach number is 
$M_{\rm A,p} = \left[ h \rho u_{\text{P}}^2 B_{\text{P}}^{-2} \right]^{1/2} $
(see also \citealt{QQ2}),
where $h\equiv \Gamma/(\Gamma-1)(P/\rho) +1 $ is the specific enthalpy of the fluid.
Alfv\'enic Mach numbers $M_{\rm A,p} < 1$ imply that the magnetic energy is dominating the kinetic energy of the fluid
and that the dynamics of the outflow is most likely governed by the strong magnetic field in that area.

\section{A reference simulation}
\label{sec:refsim}
In the following we will first describe the details of our reference simulation which will be used to
compare our parameter runs for characteristic properties of the source.
The reference simulation runs for 9000 $\tg$ corresponding to approximately 67 disk orbits at the initial inner disk radius.
In Figure \ref{fig:h20.3} we show the evolution of the density distribution, the poloidal magnetic field lines,
and the poloidal velocity field up to time $t=8000$.

\subsection{Initial conditions}
The distributions for the initial density, pressure, angular velocity and magnetic vector potential
are given by Equations \eqref{eq:Somayeh}, \eqref{eq:eos}, \eqref{eq:PW} and \eqref{eq:VecPot}.
For the disk rotation \eqref{eq:PW} we impose a factor of 0.95 in order to treat a sub-Keplerian disk.
For the disk gas law we apply $\Gamma = 4/3$ and $K = 0.001$. 
For a Kerr parameter of $a = 0.9$, the horizon is located at $r = 1.4358$ and the innermost stable 
circular orbit (ISCO) at $r = 2.32088$.

For the numerical grid we choose a transition radius $R_{\text{tr}} = 200$ and an outer grid radius $R_{\text{out}} = 10^4$.
The initial inner disk radius $r_{\text{in}} = 7$ is outside the ISCO in order to avoid possible initial ejections of 
gas as the initial disk is not in force-equilibrium within GR.
At this radius the initial angular velocity of the disk is $\Omega_{\text{r}_{\text{in}}} \approx 0.047$, thus slightly 
lower than the Keplerian value $\Omega_{K} \approx 0.052$, and corresponding to an orbital period of 
$T_{\text{r}_{\text{in}}} \approx 135$.

The initial corona is given by Eq.~\eqref{eq:corona} with $K_{\text{cor}} = 1$, resulting in a higher coronal entropy.
The initial magnetic field structure follows Eq.~\eqref{eq:VecPot} with $m=0.6$. 
The magnetic field strength is fixed by the choice of the plasma-$\beta = 10$ at the initial inner disk radius.
The magnetic diffusivity profile is given by Eq.~\eqref{eq:diffusivity} with $\chi = 3$ and  $\eta_0 = 0.001$ 
(see Figure~\ref{fig:diffusivity}).

\subsection{Evolution of disk mass and disk accretion}
\label{sec:evo}
As the disk evolves, accretion sets in and the inner disk radius changes to lower values, 
extending down to $r \approx 3$ right outside of the ISCO after having completed more than 200 rotations
at this radius.
Since the shape of the disk changes constantly, it is difficult to measure the total disk mass.
One option is to measure the mass within a disk area defined by the inner surface located at $r = 3$, an outer 
surface at $r = 100$ and the surfaces of constant opening angle of $\theta \approx 80^{\circ}$ and $\theta \approx 100^{\circ}$ 
degrees.
The disk mass is then obtained by integrating the mass density in  the disk area as specified above
\footnote{The disk mass and mass flux are normalized by the mass of the {\em initial} disk material included in the 
disk area as specified above}.

The evolution of the disk mass is shown in Figure \ref{fig:h20.3_disk_time_massr1r5}.
Since the disk is not in equilibrium, there is a rapid change in the innermost part of the disk that causes a small initial increase.
We understand that the extra mass for the disk arises from the initial corona which immediately starts to collapse, and by
that squeezes and relaxes the inner disk until a quasi equilibrium is reached at $t \approx 300$.
After that, the disk mass decreases steadily until $t \approx 5500$ when the slope of the disk mass evolution changes.
This is mainly due to changes in the disk outflow.
Note that by the end of the simulation the disk has lost more than 80\% of its initial mass.

\begin{figure}
    \centering
    \includegraphics[width=0.95\columnwidth]{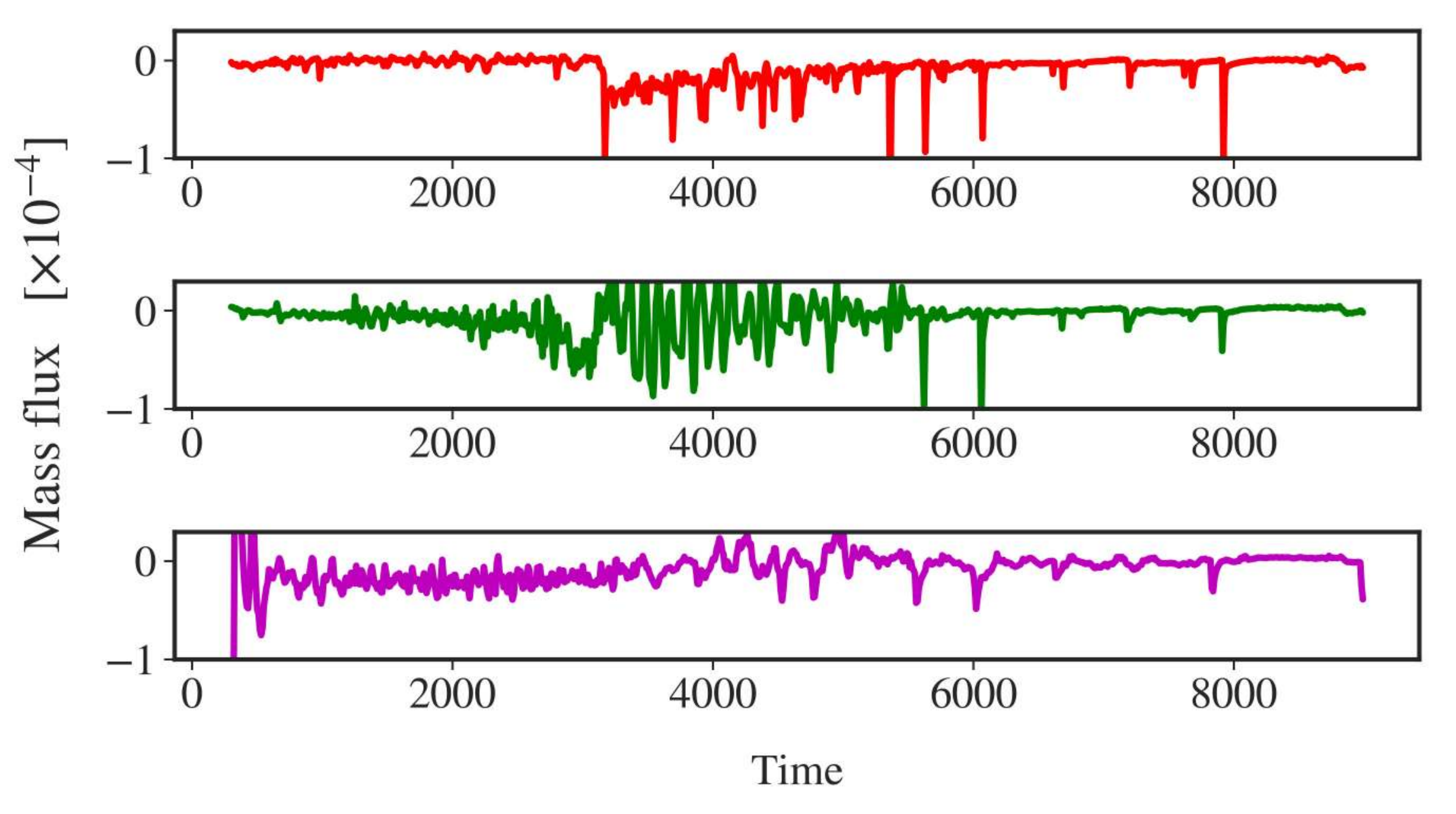}
     \caption{The accretion rate measured in three different radii ($r=2,\;4,\;13$, top to bottom) for the duration of the simulation. 
     The mass flux was integrated between $80^\circ$ and $100^\circ$ using the negative values of radial velocity. }
    \label{fig:h20.3_disk_time_totalaccretion}
\end{figure}

Figure \ref{fig:h20.3_disk_time_totalaccretion} shows the normalized accretion rates measured through three different 
radii, $r=2,\;4,\;13$ and integrated over the disk scale height.
Close to the horizon, measured at radius $r=2$, the accretion rate is first negligible,
mainly because of the absence of disk material in that radius.
After $t=3000$ accretion rate increases. 
Note that by now the inner disk radius, located initially at $r=7$, has moved closer to the black hole, populating that area 
with dense disk material.
The enhanced accretion level is accompanied by substantial accretion spikes.
However, the underlying base accretion rate seems to decrease as the disk loses mass.
The accretion mass flux in the inner area is of the order of $10^{-4}$.

After $t=6000$ and until the end of the simulation, the innermost area around $r=2$ becomes almost empty again with 
the exception of a thin stream of material that is connecting the disk with the black hole.
It looks like that at this point in time all material close to the black hole has fallen into it, but has
not been replenished by disk material from larger radii. 
As a result, accretion at $r=2$ is halted completely\footnote{with the exception of the floor density accretion}
for a substantial period of time until it is temporarily restarted by disk material that has newly arrived (accreted) from 
larger radii.
This relaunch of accretion is indicated by the spikes in the accretion rate at late times.

\begin{figure}
    \centering
    \includegraphics[width=4cm]{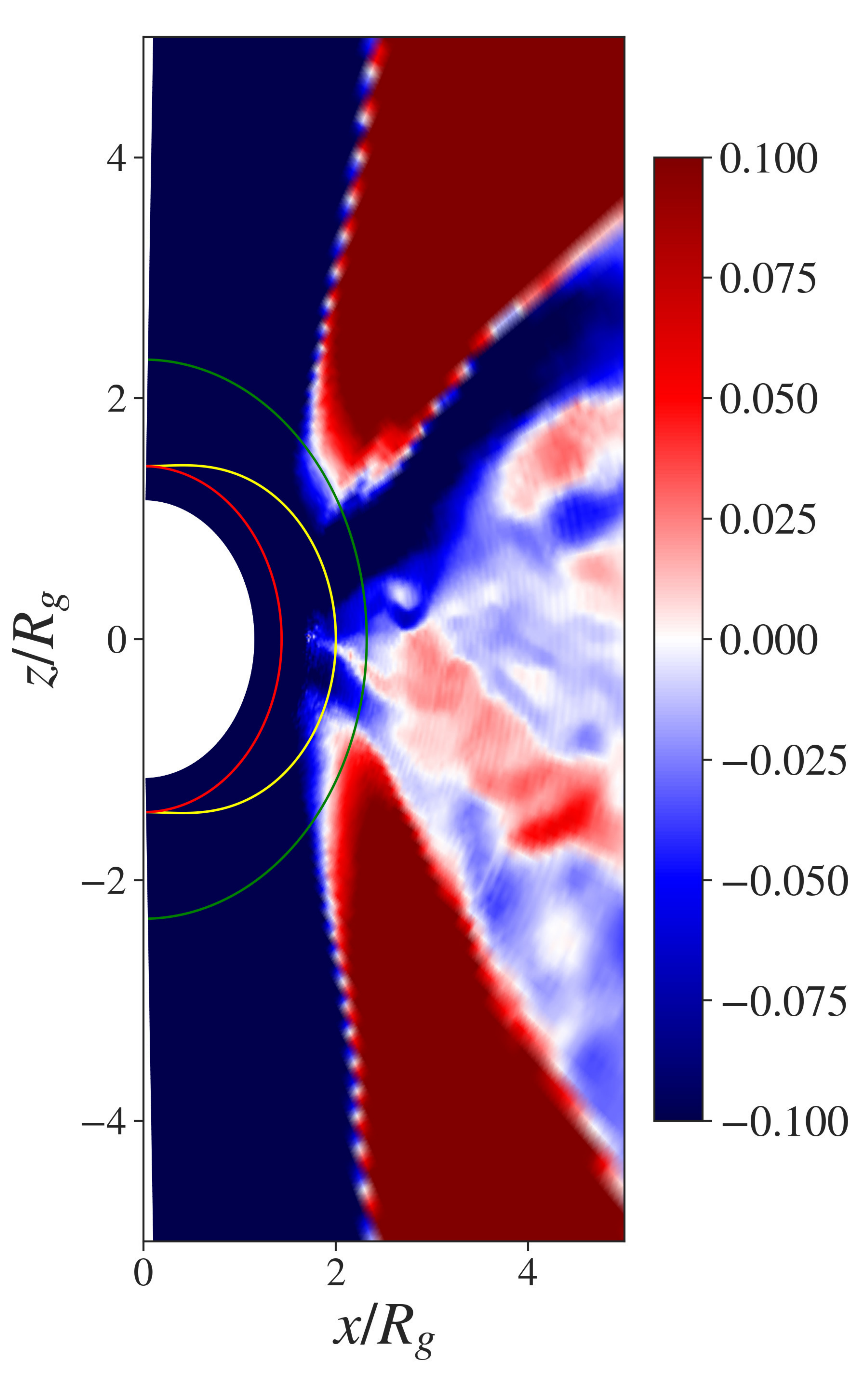}
    \includegraphics[width=4cm]{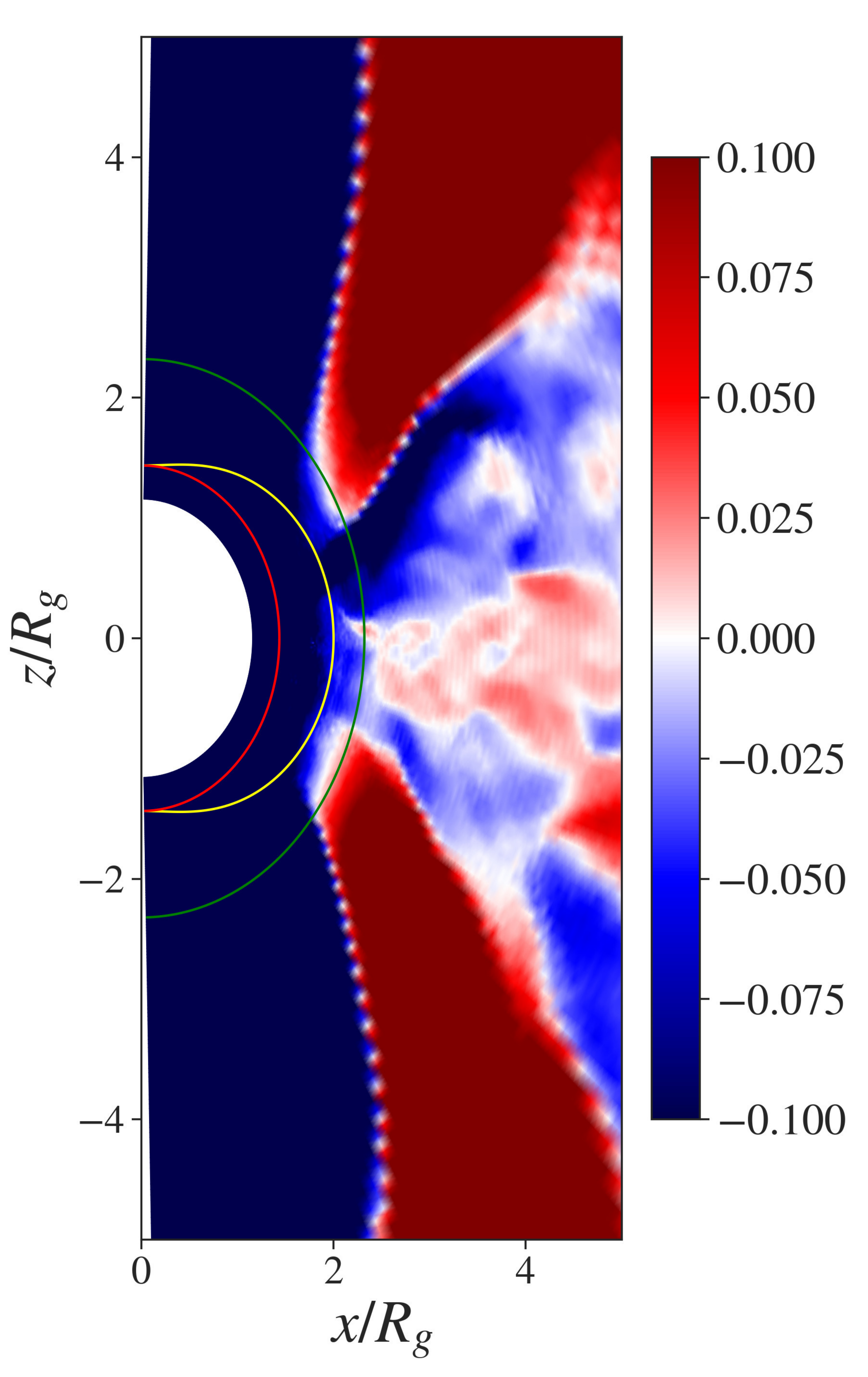}
     \caption{The turbulent dynamic structure of the inner disk. 
     Shown is the radial velocity distribution at a short time interval, at t=3800 (left) and at t=3850 (right). 
     Positive and negative velocities indicate the turbulent nature of the inner disk area. 
     The red semicircle marks the horizon, the yellow line marks the ergosphere, and the green line indicates the 
     radius of the ISCO.
     }
    \label{fig:tmp1}
\end{figure}

Similar to $r=2$, at radius $r=4$ accretion is not significant until $t=1500$, while it gradually increases
afterwards till $t=3000$.
In the following strong accretion phase, $\left( t \in [3000,6000] \right)$, there is also a significant amount of material 
moving outwards.
In the inner part of the disk, just outside the ISCO, the gas is actually moving in both directions, radially inwards and 
outwards, thus indicating the turbulent character of the motion. 
The highly turbulent nature of the inner accretion flow is shown in Figure~\ref{fig:tmp1}.
The figure demonstrates the rapid change in density and velocity within short time.
Note the strong gradient in velocity at the ergosphere (yellow line; dark blue indicates high infall speed).

Since the average accretion rate at radius $r=4$ is similar to that measured at  $r=2$, we conclude that the accretion mass flux is conserved, and, thus, no outflow is ejected 
from this area close to the black hole.
Even further out, at radius $r=13$ the accretion process looks quite different.
The accretion rate is again of an order of magnitude similar to smaller radii.
The accretion spikes that are seen at lower radii now are replaced with much broader time periods of high mass accretion 
indicating a slower change to the accretion rate.

However, we still detect a few accretion spikes during the third phase of evolution.
In fact, the accretion spikes that are observed at $r=13$ are subsequently followed by spikes at $r=4$ and $r=2$.
We measure a time delay between the spikes at $r=13$ and $r=4$ varying between $\Delta t=75$ and $\Delta t=40$.
The time delay between the spikes at $r=4$ and $r=2$ is $\Delta t \simeq 10$. \footnote{This is also the time sequence for our 
data dumps, so we cannot provide a higher time resolution for the pattern speed of the spikes.}
An approximate average accretion velocity can be defined by dividing the distance traveled by the fluid by the time delay 
of the spikes.
For the three major spikes appearing at radius $r=2$ at $t=[5630,6070,7920]$ we measure a similar velocity from radius 
$r=4$ to radius $r=2$ of 0.2 for all three spikes.
For the spikes pattern speed from $r=13$ to $r=4$ we measure velocities of 0.12, 0.225 and 0.16 for the three spikes,
respectively.
These values derived for the pattern speed agree well with the radial velocity that we observe in this area of the disk.

\begin{figure*}
    \centering
    \includegraphics[width=5cm]{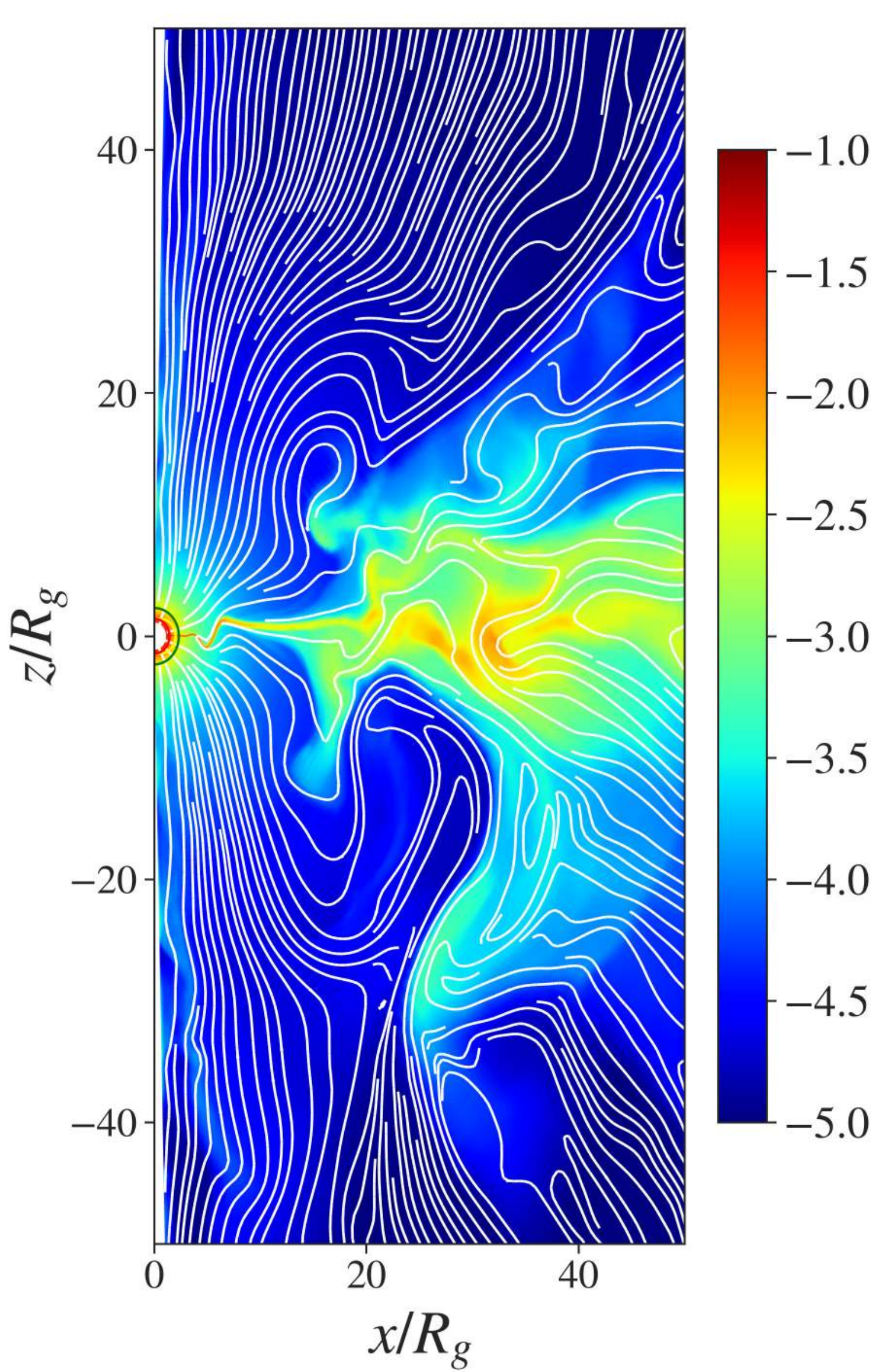}
    \includegraphics[width=5cm]{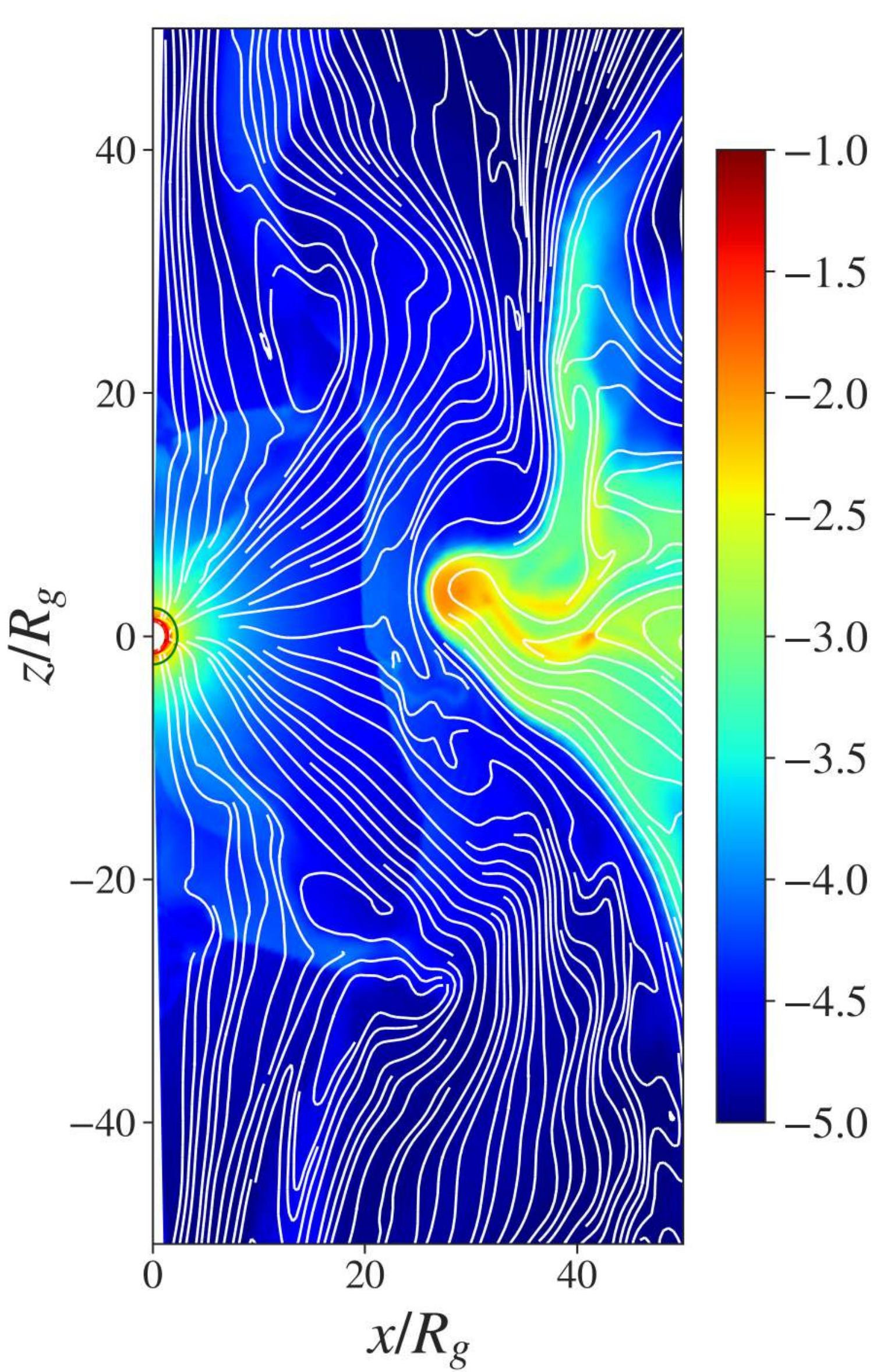}
    \includegraphics[width=5cm]{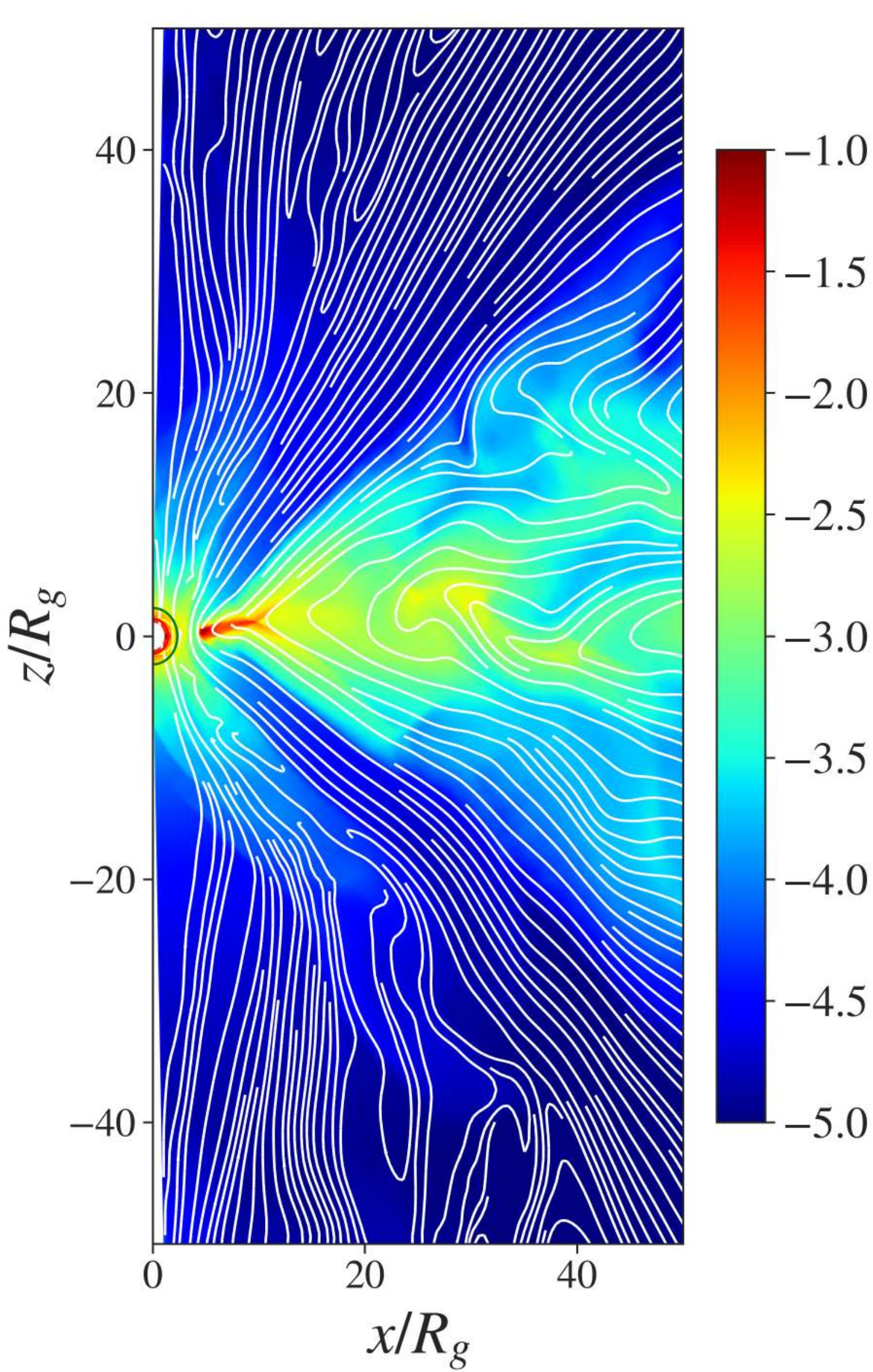}
    \caption{Snapshots of of our reference simulation. Shown is the density distribution (log scale) at times
    $\left(t = 8000,\; 8500, \; 9000 \right)$. The poloidal magnetic field lines are shown (white lines). 
    At late times the accretion disk disconnects from the black hole which results in 
    halting of mass accretion and ejection of the BZ jet.
    }
    \label{fig:h20.3_r50}
\end{figure*}

At late stages of the simulation (between $t=6000$ and $t=9000$) we notice a decline in the accretion rate at
all three radii.
This is accompanied with the opening of a larger gap between the horizon and the inner disk, meaning that the inner
disk radius moves out.
At this time, the disk has already lost 70\% of its mass.
During this period, the disk accretion becomes disconnected from the black hole.
We interpret this as follows.
Due to the decrease of density and pressure (following accretion and ejection of disk material), this area
becomes magnetically dominated.
The strong magnetization leads to the structure of an magnetically arrested disk (MAD, see \citet{Narayan2003}).
When the magnetic flux is advected to the black hole respectively to the rotational axis, the magnetization in this 
area decreases again, and accretion restarts (see Fig.\ref{fig:h20.3_r50}).  

\begin{figure}
    \centering
    \includegraphics[width=0.95\columnwidth]{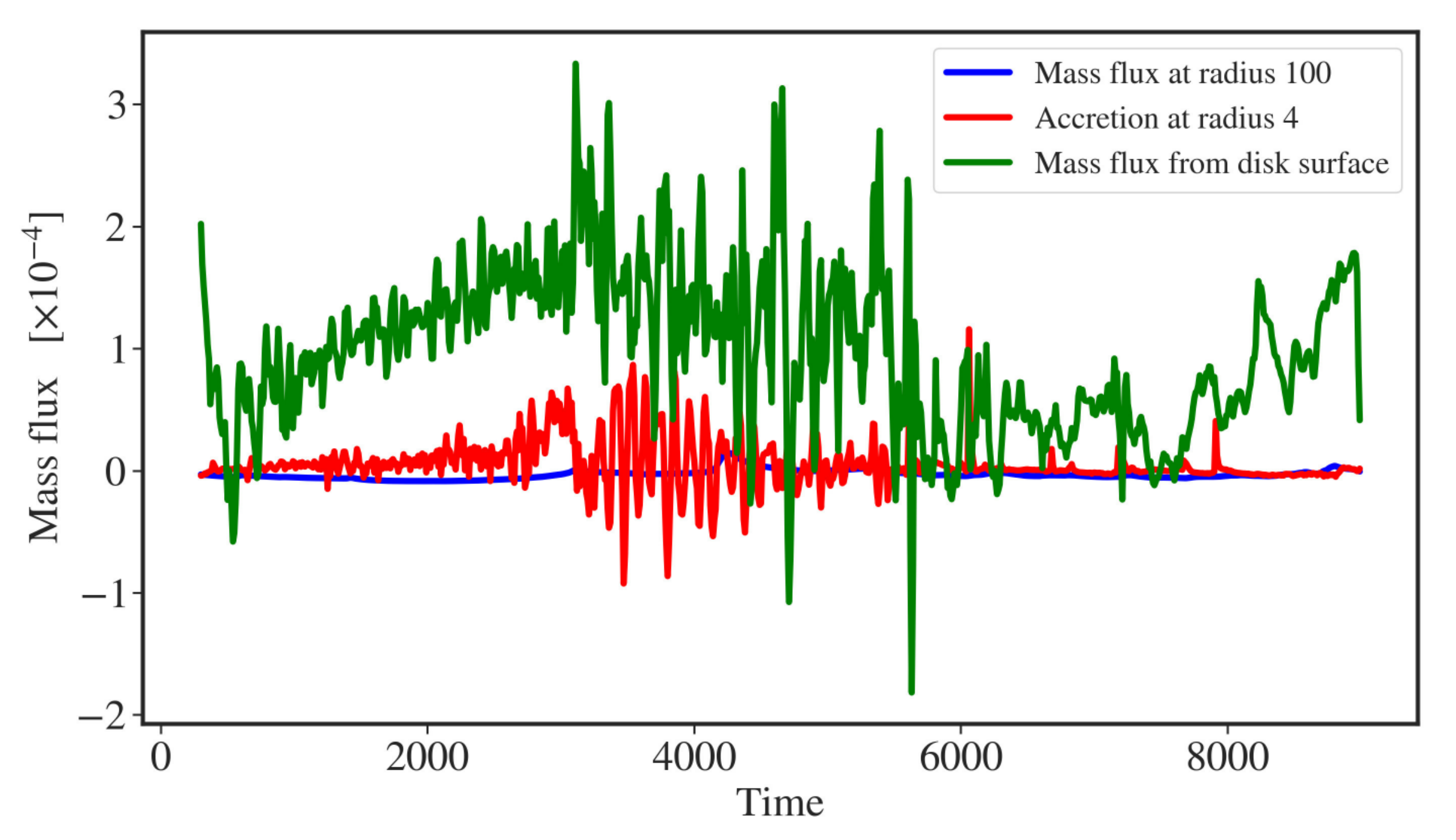}
     \caption{The integrated mass flux through the outer disk at $r=100$ (blue), through the inner disk radius at $r=3$
     (red) and through the disk surfaces (green).}
    \label{fig:h20.3_disk_time_radialmassloss}
\end{figure}

Finally, in Figure~\ref{fig:h20.3_disk_time_radialmassloss} we display the mass fluxes vertical to the surfaces of 
constant opening angles ($\theta= 80^\circ, 100^\circ$) that approximate the surfaces of the initial disk density distribution.
We see that the mass fluxes of accretion and radial outflow along the disk are comparable.
However, both are dominated by the vertical mass loss from the disk surface.
The low accretion rate is comparable to a MAD structure \citep{TchekhNarayan2011} and due to the strong disk magnetic field a strong outflow is launched, but at the same time the accretion rate decreases.
Obviously, also the strength of magnetic diffusivity plays a role (see our comparison study below).
We may conclude that most of the mass that the disk is losing is due to the strong disk wind that is launched.

We note that since a substantial disk wind is present during the whole simulation, the wind mass loss rate is changing.
The wind mass flux increases until about $t=3000$ and then decreases again until $t=6000$.
In the late stages of the simulation the wind mass flux is highly variable.
These two different phases of wind ejection seem to correspond to similar phases in the disk accretion,
visible in Figure~\ref{fig:h20.3_disk_time_totalaccretion} (middle panel) that shows large variations in the mass accretion rate,
or in Figure~\ref{fig:h20.3_disk_time_massr1r5} that indicates a change in the disk mass evolution at $t=3000$.

\begin{figure}
    \centering
    \includegraphics[width=0.95\columnwidth]{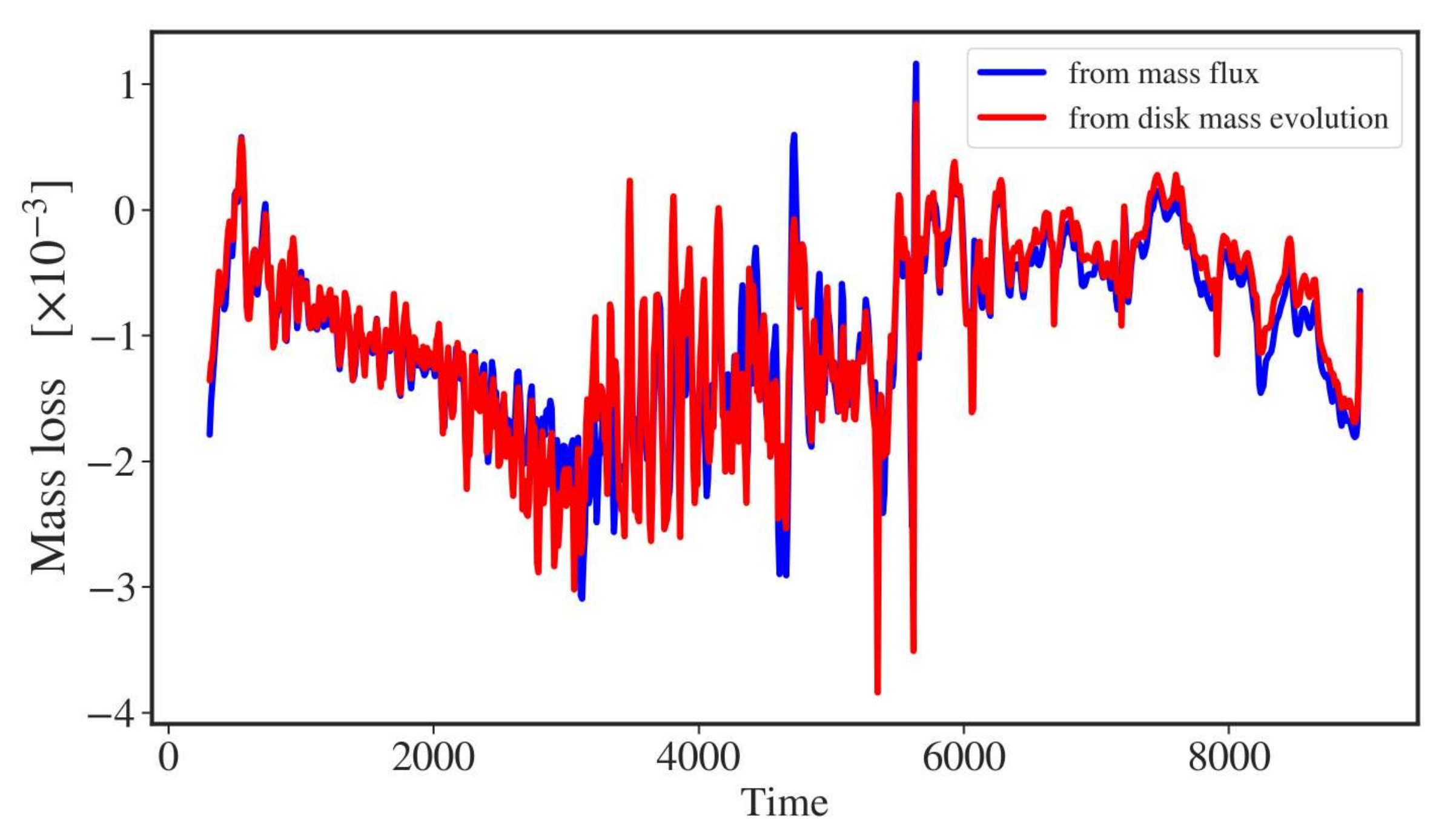}
     \caption{Comparison of the disk mass loss as calculated directly from the disk mass evolution (red) and 
     from the outflow mass flux (blue). 
     Both curves coincide demonstrating the robustness of our integration tools.
     }
    \label{fig:h20.3_disk_time_totalmassloss}
\end{figure}

In order to double-check our mass flux integration, we have measured the total mass loss of the disk with two different methods. 
Firstly, we integrate all mass flux leaving the surfaces of the disk area as specified previously.
Secondly, we calculate the mass loss from the mass evolution of the disk (see Figure~\ref{fig:h20.3_disk_time_massr1r5}).
Figure~\ref{fig:h20.3_disk_time_totalmassloss} compares the time evolution of the two measurements.
Essentially, both show excellent agreement, confirming our methods to determine the evolution of the disk.

The mass loss remains negative for the majority of the simulation with small exceptions of momentarily mass increase 
especially in the later stages.
On average, we have a mass loss rate of the order of $10^{-3} / \tg$.
The rate of mass loss however changes a lot, following a repeating pattern similar to the one appearing in the vertical 
mass flux from the disk surface, demonstrating that the large mass loss is due to the disk wind.
Based on Figure \ref{fig:h20.3_disk_time_radialmassloss}, if we integrate over time we find that out of a total of 85\% 
of the disk mass lost during the simulation, approximately 73\% is from the disk wind, 10\% is from accretion to the black 
hole and 2\% is across the outer disk radius.

\subsection{Outflow from the black hole magnetosphere}

\begin{figure}
    \centering    
    \includegraphics[width=4.1cm]{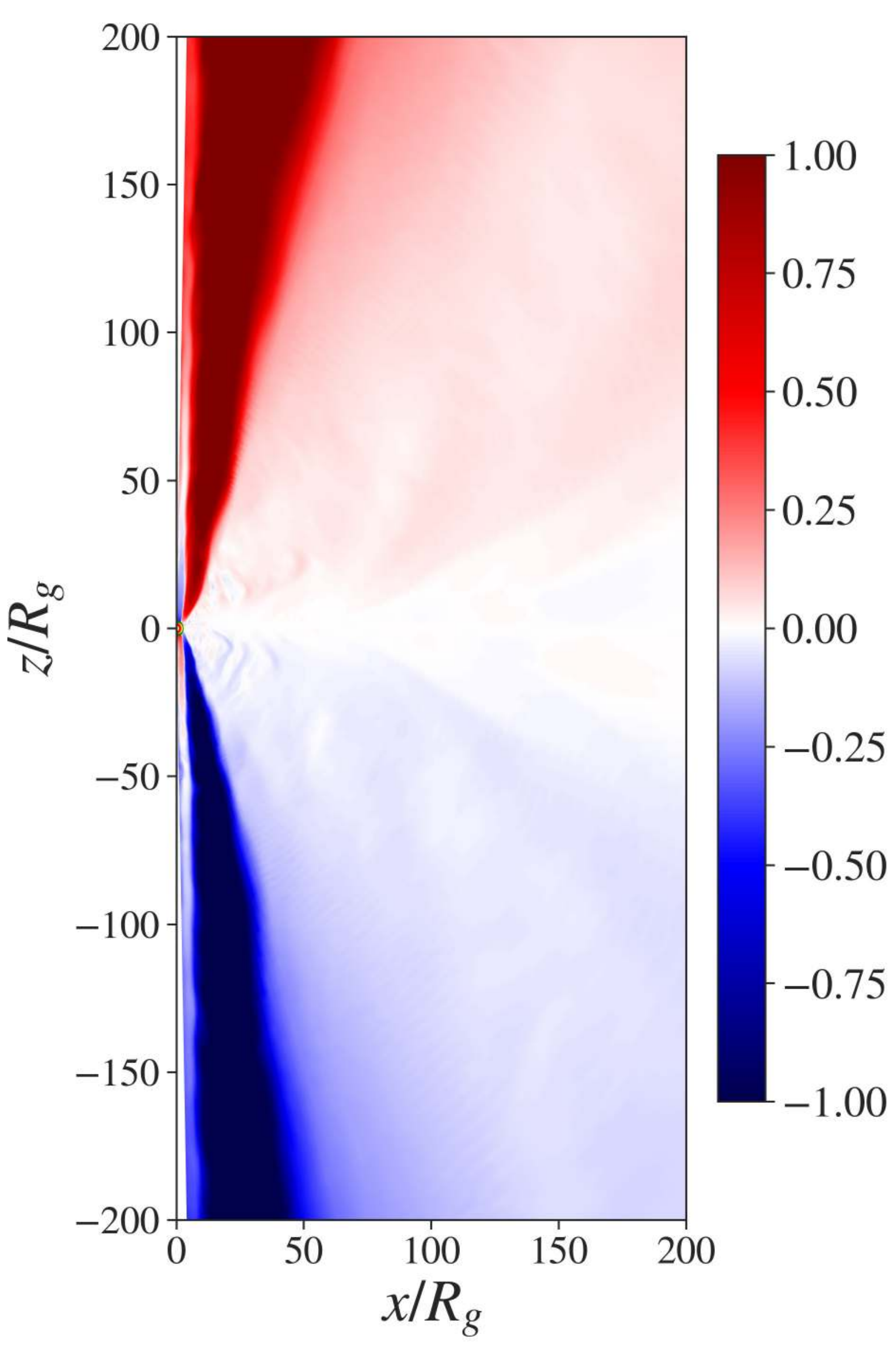}
    \includegraphics[width=4.1cm]{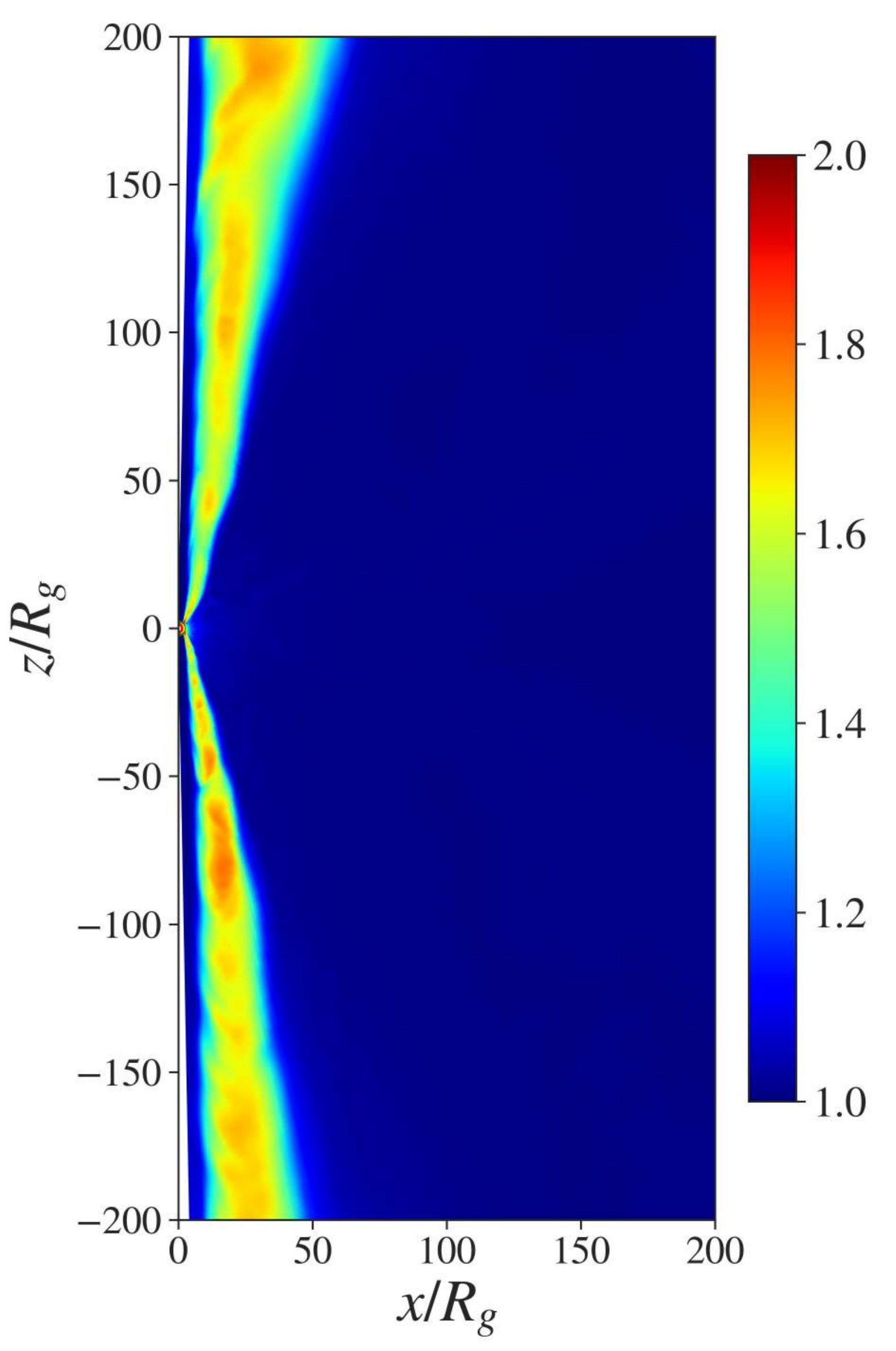}
    \caption{The vertical component of the velocity and the Lorentz factor on a sub-grid 
    of 200 $\rg$.}
    \label{fig:h20.3_Vz_gamma_4000}
\end{figure}

\begin{figure*}
    \centering
    \includegraphics[width=4.1cm]{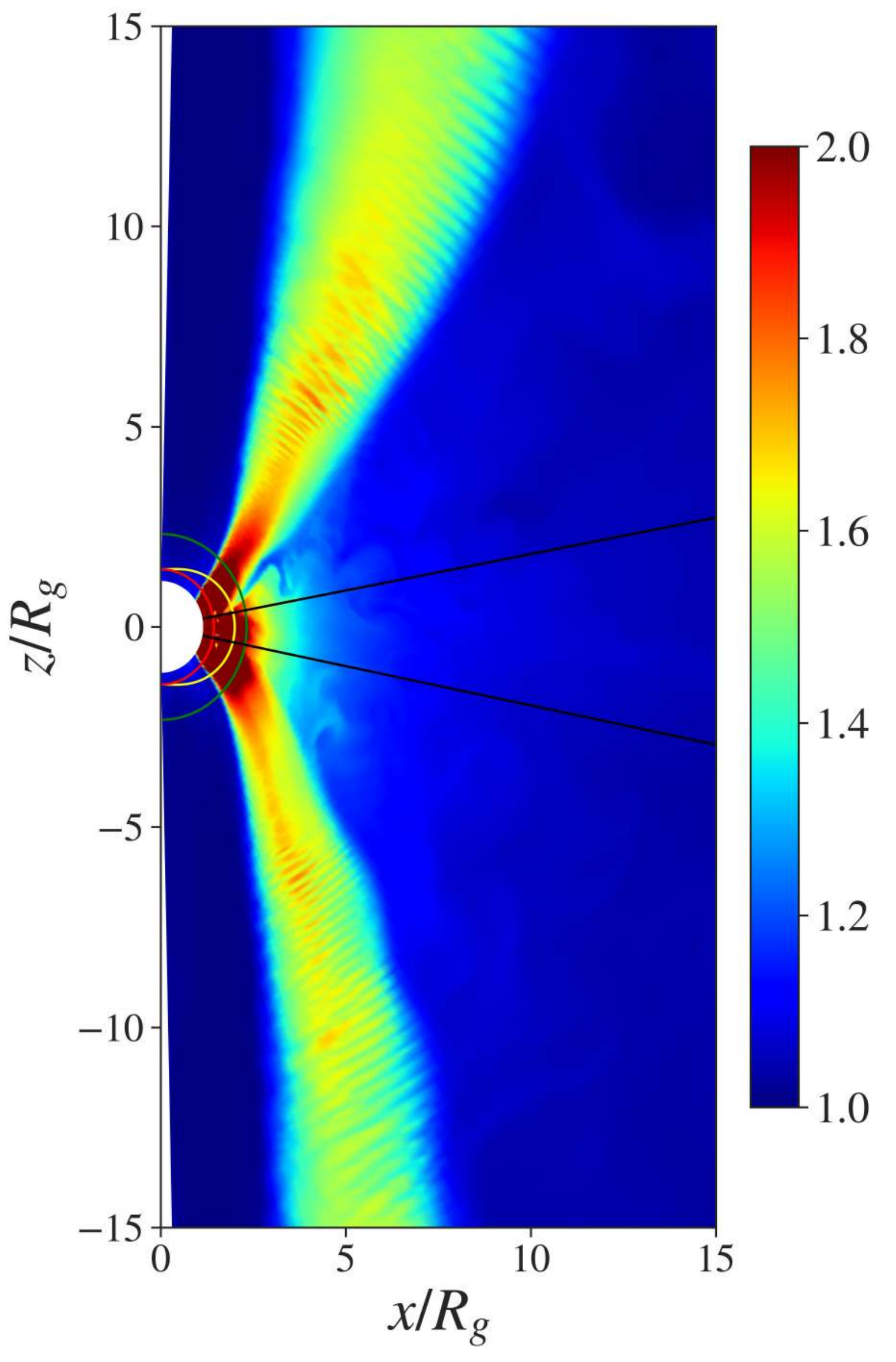}
    \includegraphics[width=4.1cm]{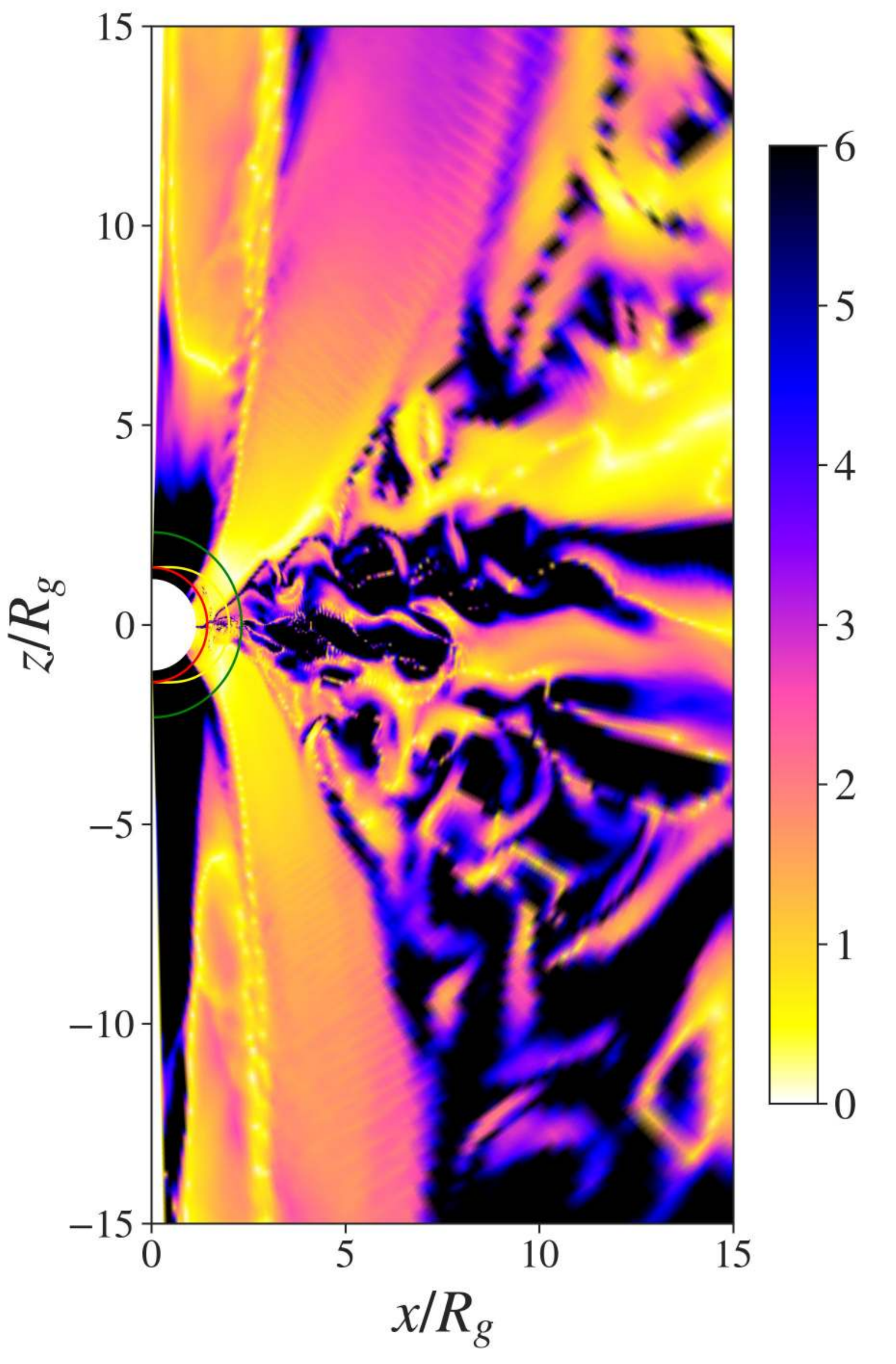}
    \includegraphics[width=4.1cm]{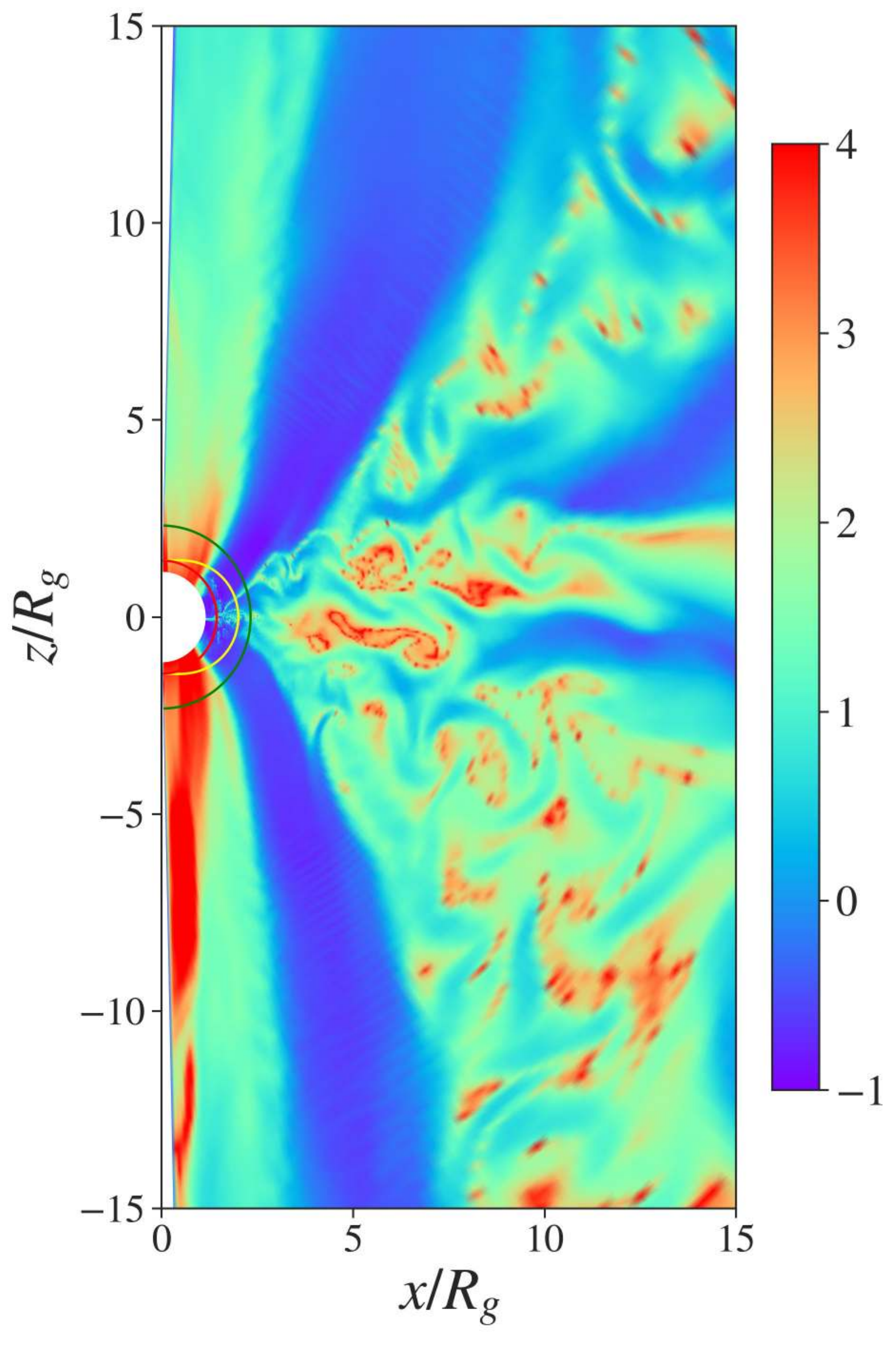}
    \includegraphics[width=4.1cm]{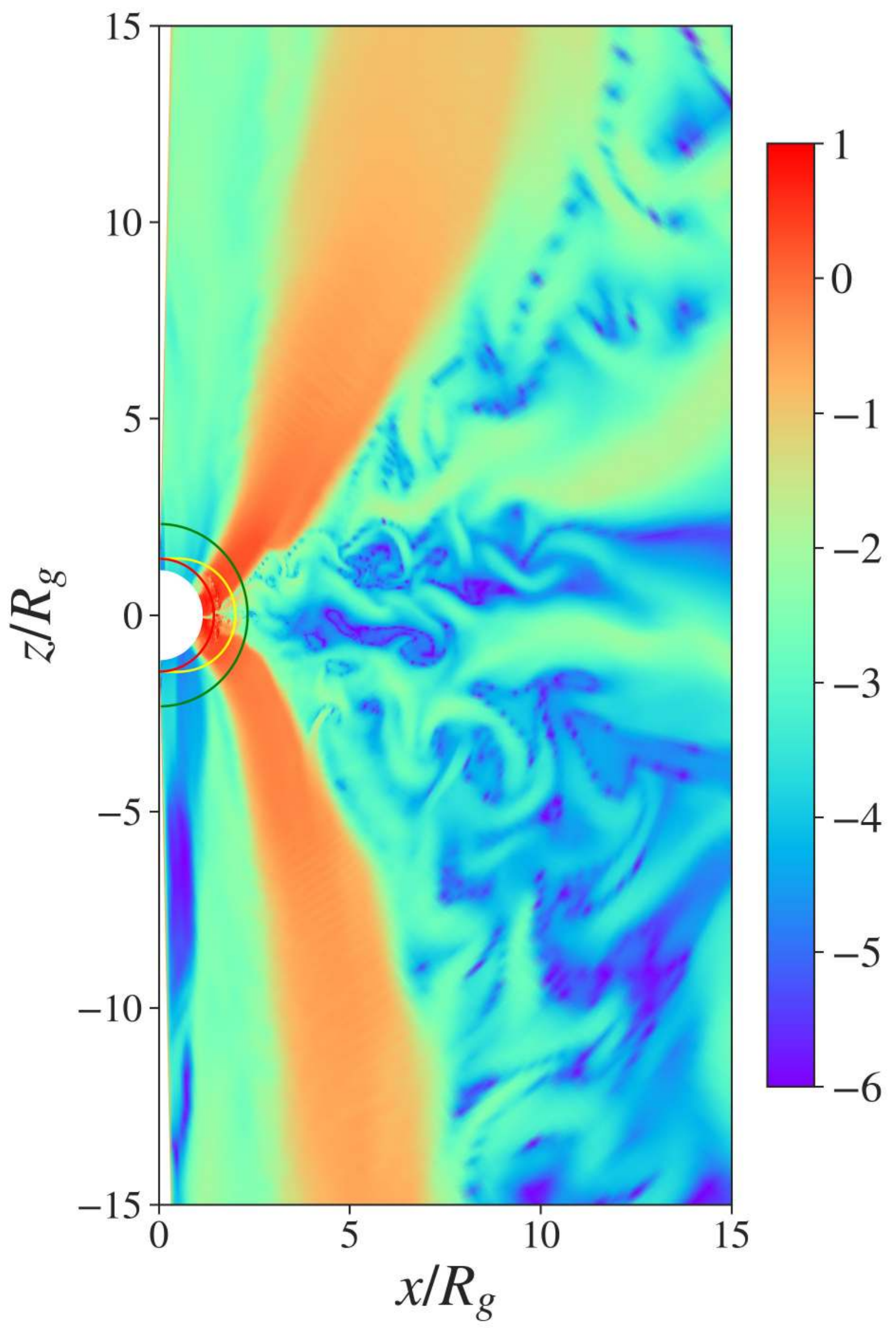}
    \caption{Magnetohydrodynamic accretion-ejection structure close to the black hole, $r<15$ at $t=4000$.
    Shown is the Lorentz factor (left),  the poloidal Alfv\'en Mach number (log scale, second left), 
    the plasma-$\beta$ (log scale, second right), and the magnetization $\rho/B^2$ (log scale, right).
    The high Alfv\'en Mach number indicates flows that are dominated by kinetic energy, regardless of the highly 
    magnetized area. 
    The red semicircle marks the horizon, the yellow line marks the ergosphere, 
    and the green line indicates the radius of the ISCO.}
    \label{fig:h20.3_4000_r15}
\end{figure*}

\begin{figure*}
    \centering
    \includegraphics[width=0.95\columnwidth]{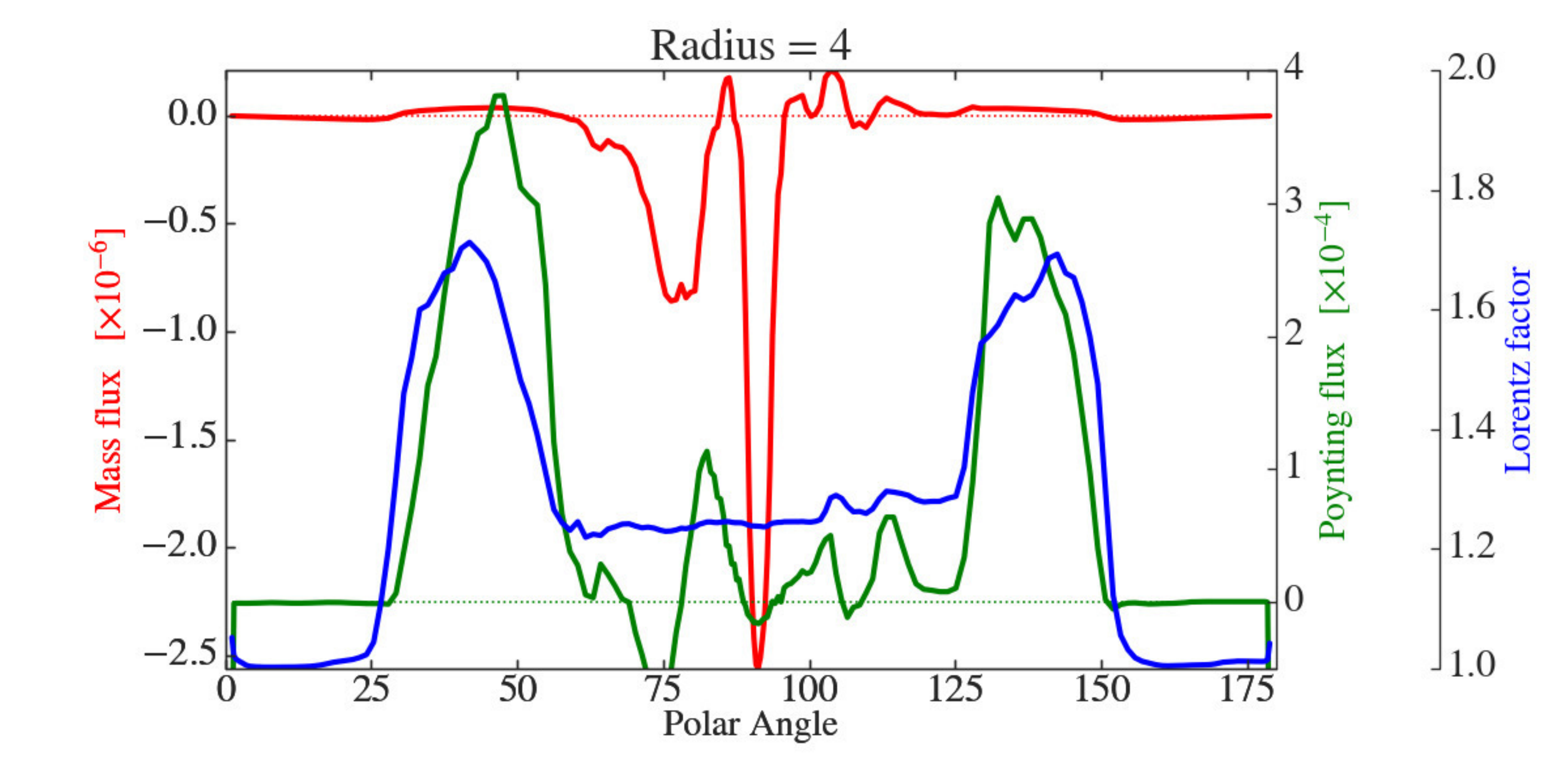}
    \includegraphics[width=0.95\columnwidth]{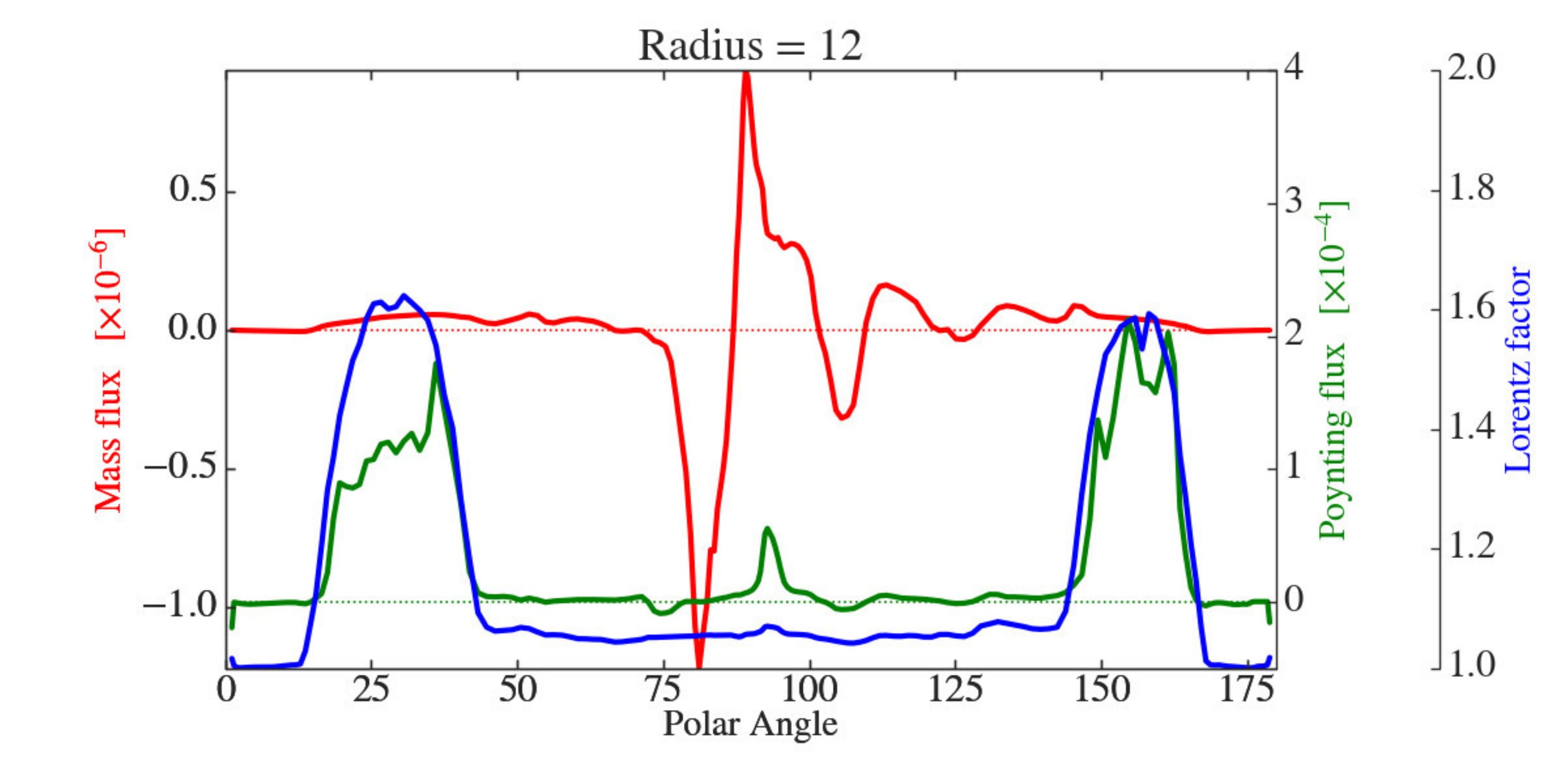} \\
    \includegraphics[width=0.95\columnwidth]{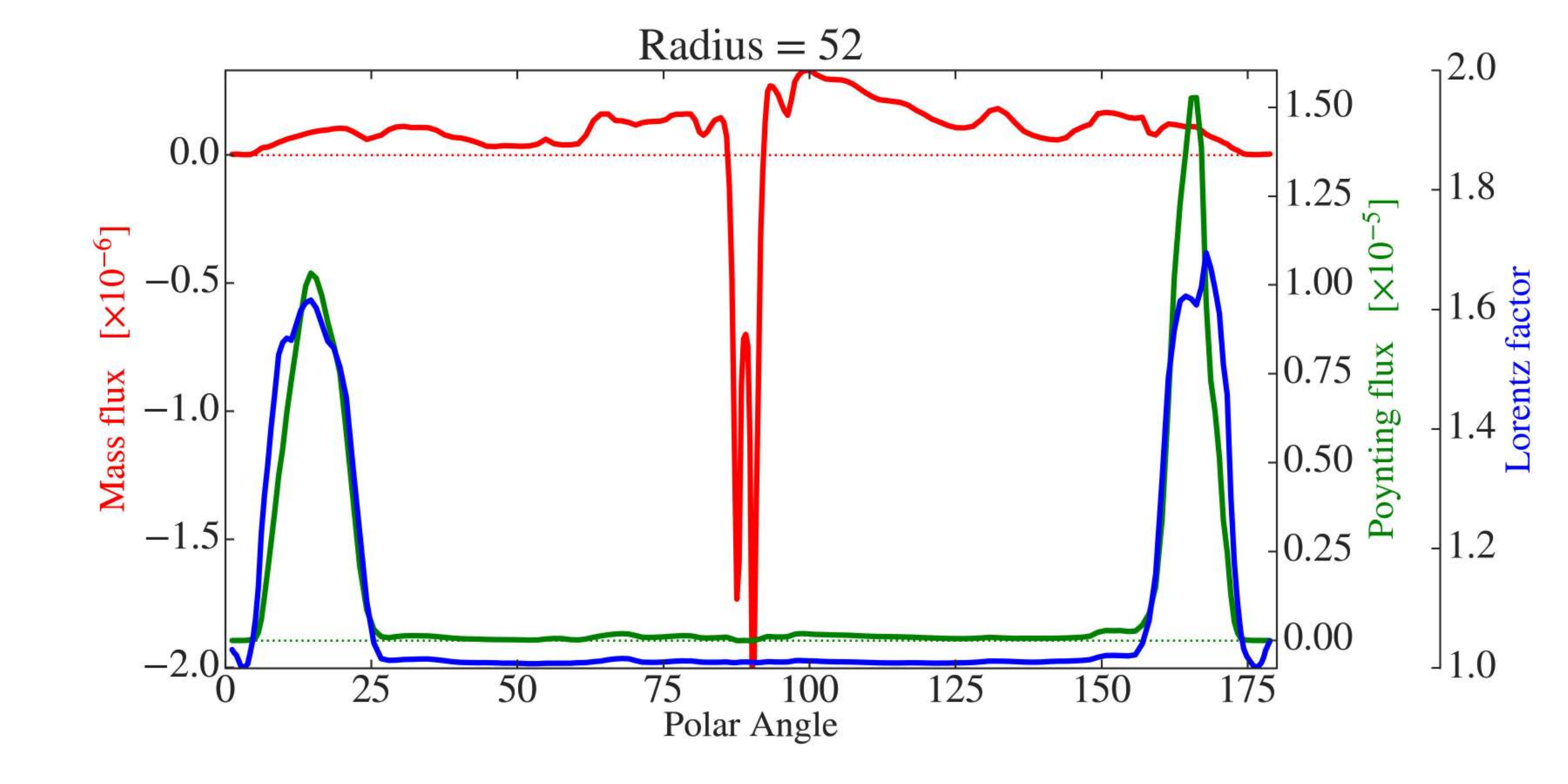}
    \includegraphics[width=0.95\columnwidth]{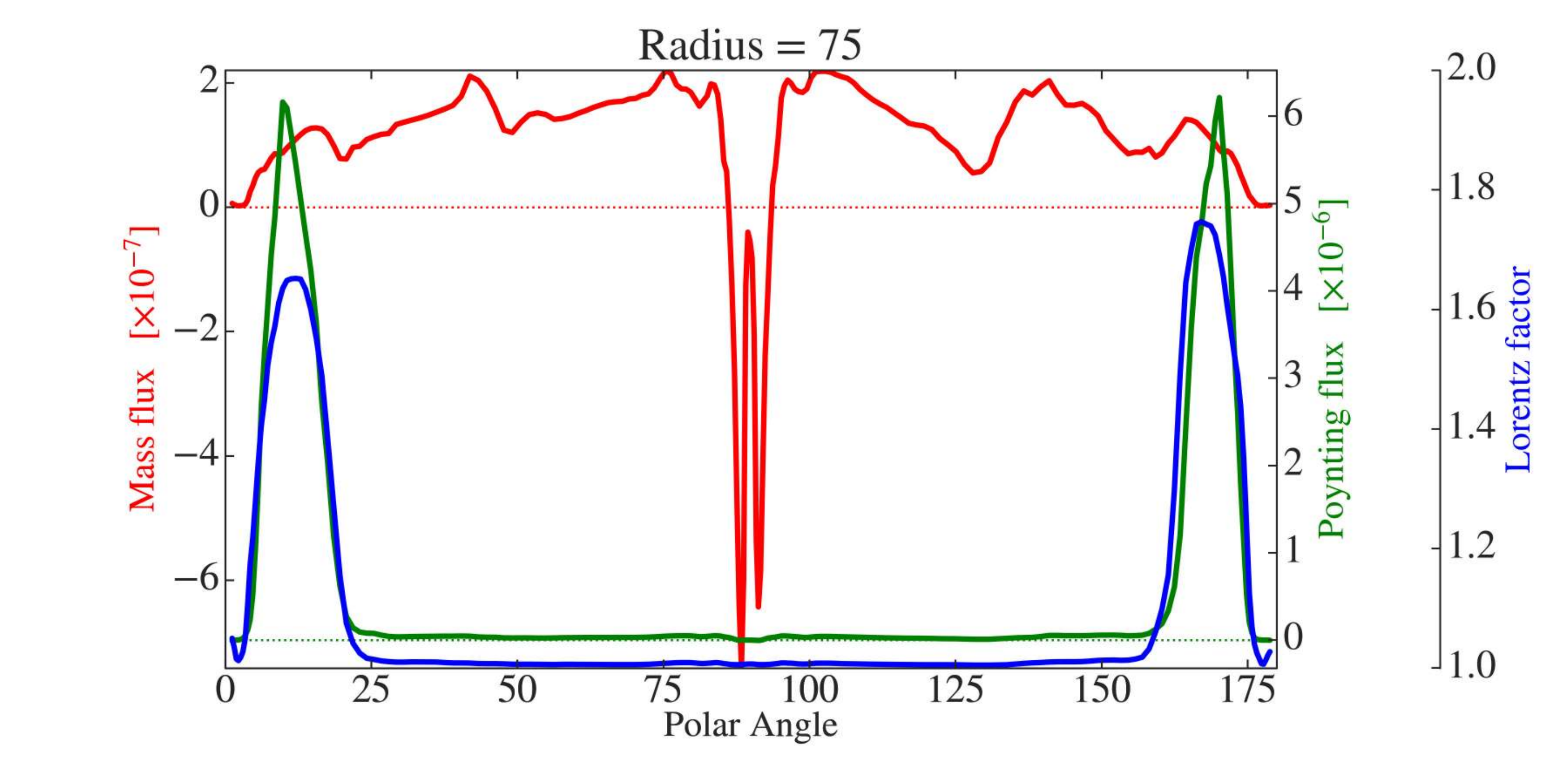}
    \caption{Comparison of the angular distribution of mass flux (red), Poynting flux per solid angle (green) 
    and Lorentz factor (blue) for the reference simulation at $t=4000$ at four radii, $r=4, 12, 52,75$. 
    Negative mass flux indicates accretion towards the black hole. 
    The BZ driven jet funnel is clearly distinguished by the peaks in Lorentz factor and electromagnetic energy flux.
    For increasing radii, the mass flux increases, demonstrating the matter-dominated disk wind.
    In low radii, between the rotational axis $\theta =0, 180$ and the jet funnel floor density material falls towards the 
    black hole.
    }
    \label{fig:h20.3_radii_4000}
\end{figure*}

The most prominent feature of our reference simulation (as visible in Figure \ref{fig:h20.3}) is the outflow that develops 
from the area around the black hole. 
It starts around $t = 1000$ with the advection of magnetic flux towards the black hole.
The field lines that enter the ergosphere are being twisted and turned along the toroidal direction creating eventually 
a jet toward the polar direction, according to the BZ mechanism \citep{BZ1977}.
Up to $t=3000$, this jet has been fully developed and it enters a quasi steady state until the end of the simulation 
($t = 9000$), even though its strength still depends on the advection of the magnetic flux, and through that, 
on the accretion rate of the disk.
The jet is identified by an parabolic-shaped funnel of high velocity fluid that originates from the area around 
the black hole and moves almost parallel (in the later stages) to the symmetry axis towards the outer parts of our domain.

The jet funnel can be seen clearly in Figure \ref{fig:h20.3_Vz_gamma_4000}, where we plot the z-component of the fluid frame
velocity and the Lorentz factor at time $t=4000$. 
The jet seems to consist of fast moving inner parts with $\gamma \approx 1.8$, a moderately fast moving envelope with 
$\gamma \approx 1.5$ and the outer part where the Lorentz factor values stay below $\gamma = 1.3$. 
The fast moving inner parts seem discontinuous and we can clearly distinguish 2-3 knots of high velocity in larger
radii ($r>50$) while closer to the black hole the high values of Lorentz factor seem to have a more continuous 
distribution (see Figure~\ref{fig:h20.3_4000_r15}).

We  select four radii, $r \approx 4, \; 12,\; 52, \; 75$, where high velocity knots appear.
In Figure \ref{fig:h20.3_radii_4000} we see how the radial velocity, the mass flux and the electromagnetic energy 
flux (Poynting flux) per solid angle are distributed along the polar angle in these radii.
In general, the Poynting flux distribution follows the high velocity areas  proving that the jet funnel has a strong 
electromagnetic component.
The mass flux in the funnel area does not show a significant increase in comparison with the disk wind area and the disk, 
where the mass density is considerably higher, since the accelerated material consists primarily of floor density values.

Figure~\ref{fig:h20.3_4000_r15} shows the Lorentz factor, the poloidal Alfv\'en Mach number, the plasma-$\beta$ and the magnetization over 
an area of 15 $\rg$ at time $t=4000$. 
The highly magnetized funnel coincides  with the high velocity area of the jet.
It starts as a sub-Alfv\'enic flow right outside of the ISCO however, even though in the area of the funnel is highly 
magnetized, ($\text{plasma-}\beta \approx 1, \; B^2/\rho \approx 1$), the flow is accelerated quickly to super-Alfv\'enic 
speed, indicating that it is dominated by kinetic energy.

\subsection{Evolution of the Poynting flux}
\begin{figure}[h]
    \centering
    \includegraphics[width=0.98\columnwidth]{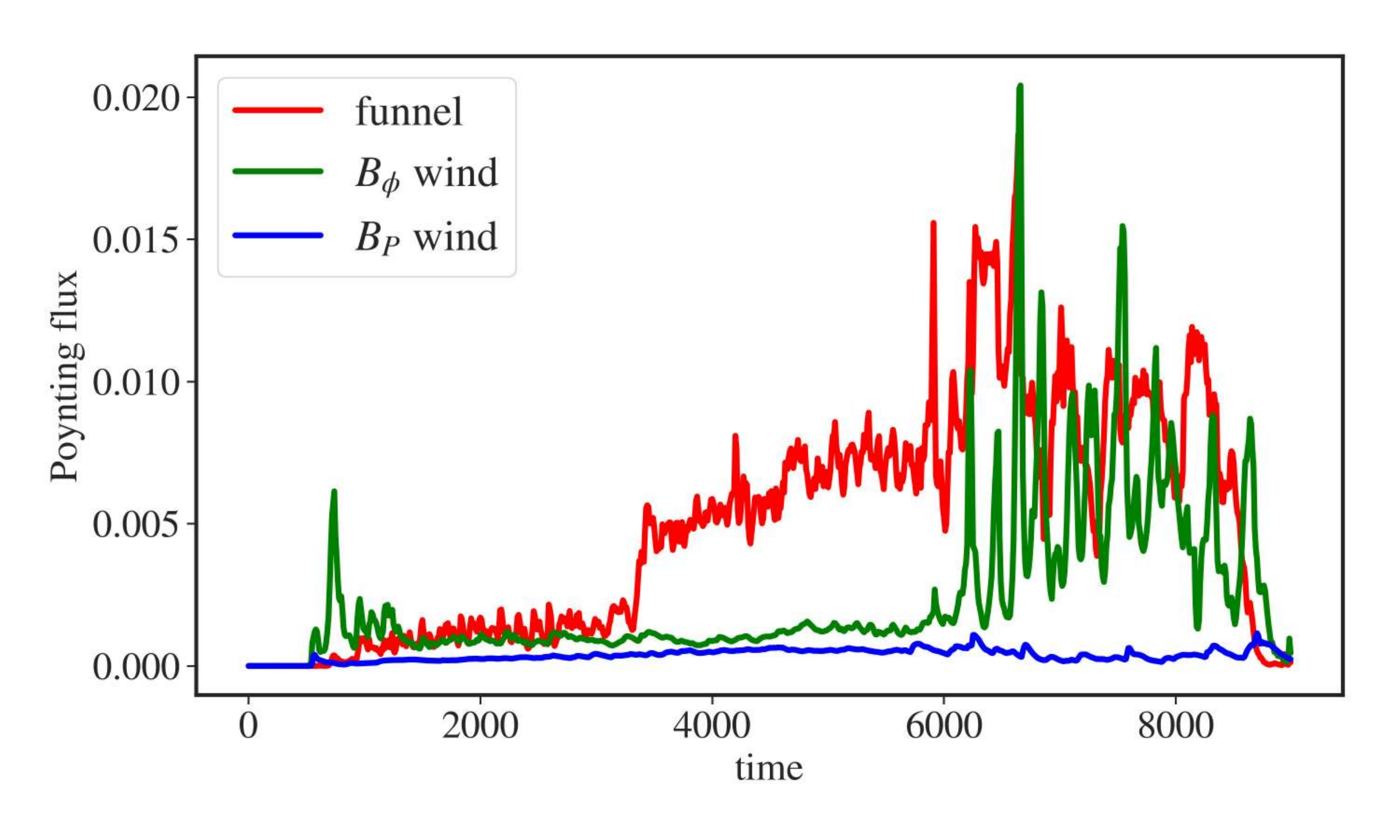}
     \caption{The evolution of the total Poynting flux for our reference simulation at radius $r=100$. 
     We split our domain in three regions.
     The first is between $0^{\circ} < \theta < 25^{\circ}$ expressing the Poynting flux from the relativistic jet funnel (red).
     The second is between $25^{\circ} < \theta < 65^{\circ}$ which includes the $B_{\phi}$-dominated disk wind (green).
     The third is between $65^{\circ} < \theta < 80^{\circ}$ which includes the $B_P$-dominated disk wind (blue).}
    \label{fig:h20.3_time_poyntingFlux}
\end{figure}

We now examine the electromagnetic energy fluxes (Poynting flux) of our reference simulation. 
In Figure \ref{fig:h20.3_time_poyntingFlux} we show the evolution of the integrated Poynting flux through a surface at radius $r=100$.
We further split our integration domain into the following three areas.
The first region is between $0^{\circ} < \theta < 25^{\circ}$ and mainly covers the funnel region hosting the relativistic 
jet from the black hole magnetosphere.
The disk wind area (covering larger polar angles) is split into two more regions (see also our Sect.~\ref{sec:diskwind}).
This is a region between $25^{\circ} < \theta < 65^{\circ}$ where the $B_{\phi}$-dominated disk wind evolves
and a region between $65^{\circ} < \theta < 80^{\circ}$ where the poloidal magnetic field dominates. \footnote{Section~\ref{sec:diskwind} discuss the different types of disk wind extensively.}
The chosen separation does not exactly follow the direction of the funnel as the geometry of the funnel flow changes 
with time.
However, it is a good approximation for the average location of the funnel especially in higher radii.
Note that even though the majority of the (bent) funnel jet is inside the opening angle we have just defined, at 
$r \approx 2-4$ it is rooted closer to the equatorial plane resulting in very low values of Poynting flux measured for 
the launching region and higher values for the disk wind regions (see also Figure~\ref{fig:h20.3_radii_4000}).

The three phases of the disk evolution can also be seen in the evolution of the Poynting flux of the jet funnel.
In the beginning, the flux remains almost constant, but after $t=3000$ drastically increases indicating the development of a 
strong jet.
This is mainly due to the advection of magnetic field and energy towards the black hole along with the mass accretion 
from the disk.
Beyond $t=6000$ -- due to the high disk mass loss -- the variability in the accretion rate triggers the Poynting flux,
leading to strong variations in the funnel and in most of the disk wind.

For the disk wind, at small radii the associated Poynting flux shows a steady increase with time. 
However, this is again an artifact due to integration area that cannot follow the bent geometry of the funnel flow.
Also, the base of the funnel flow is partly extending beyond the chosen integration domain (limited to $25^{\circ}$).
It is thus not accounted for the initial funnel Poynting flux, but contributing to the Poynting flux we measure for 
the ``wind''.
This is indicated clearly in Figure~\ref{fig:h20.3_radii_4000}.
At larger distances, the Poynting flux remains at low levels, now following the true geometry of the disk wind.
For the $B_P$-dominated disk wind the Poynting flux has very low but still positive values in the outer radii.

Table~\ref{tab:simulations} shows the time-averaged Poynting flux measured at radius $r=100$ for the three
previously mentioned angular regions.
As probably expected, the higher values of Poynting flux are detected in the jet funnel, about two times larger than the 
corresponding flux in the disk wind.
The $B_{\phi}$-dominated disk wind also drives a Poynting flux about six times larger than the flux in the 
$B_{\text{P}}$-dominated disk wind.
In total, the electromagnetic energy output of the disk is lead mainly by the Poynting-dominated jet from the black hole
where we also detect the highest velocities.

This seems to contradict earlier results \citep{QQ2} indicating a disk wind substantially contributing to the total electromagnetic 
flux.
We think that the reason for this difference is mainly the shorter live time of the simulation in \citet{QQ2}, in particular
for the simulation with high spin.
This is visible in Figure~\ref{fig:h20.3_time_poyntingFlux} where we see that for early times $t\simeq 500$ the Poynting flux
of the $B_{\phi}$-dominated wind (green curve) dominates the inner jet.

\subsection{The accretion disk wind}
\label{sec:diskwind}
The origin of accretion disk winds has been studied in the context of both AGNs and YSOs. 
Numerous works have investigated the launching mechanisms especially in the non-relativistic regime 
\citep{CK2002, Zanni2007, Somayeh2012, Stepanovs1}.
It has become clear since the seminal work of \citet{Ferreira1997} that the magnetic resistivity is a key parameter 
for the investigation of the disk wind since it allows the gas to penetrate the magnetic field lines and thus allows for both (i) advection towards the black hole and (ii) mass loading the disk wind.

In a strong disk magnetic field, magneto-centrifugally accelerated outflows can be driven once the material 
is lifted from the disk plane into the launching surface usually located around the magnetosonic surface.
\citet{QQ1} and \citet{QQ2} have extended the study of disk winds to the general relativistic regime.
However, they have found that - in contrary to non-relativistic disks - it is mainly the pressure gradient of the 
toroidal magnetic field that launches of disk winds, while the energy output by the disk wind can indeed
be comparable to the BZ outflow launched by the BH.
In addition (or rather a consequence) disk winds from relativistic disk are quite turbulent and do not
evolve in the smooth outflow structures that are known from non-relativistic cases.
In this section we continue the analysis of the disk outflows, extending their study to (physically) larger grids of
higher resolution.

\subsubsection{General overview}
\label{sec:DWoverview}
\begin{figure}
    \centering
    \includegraphics[width=0.45\columnwidth]{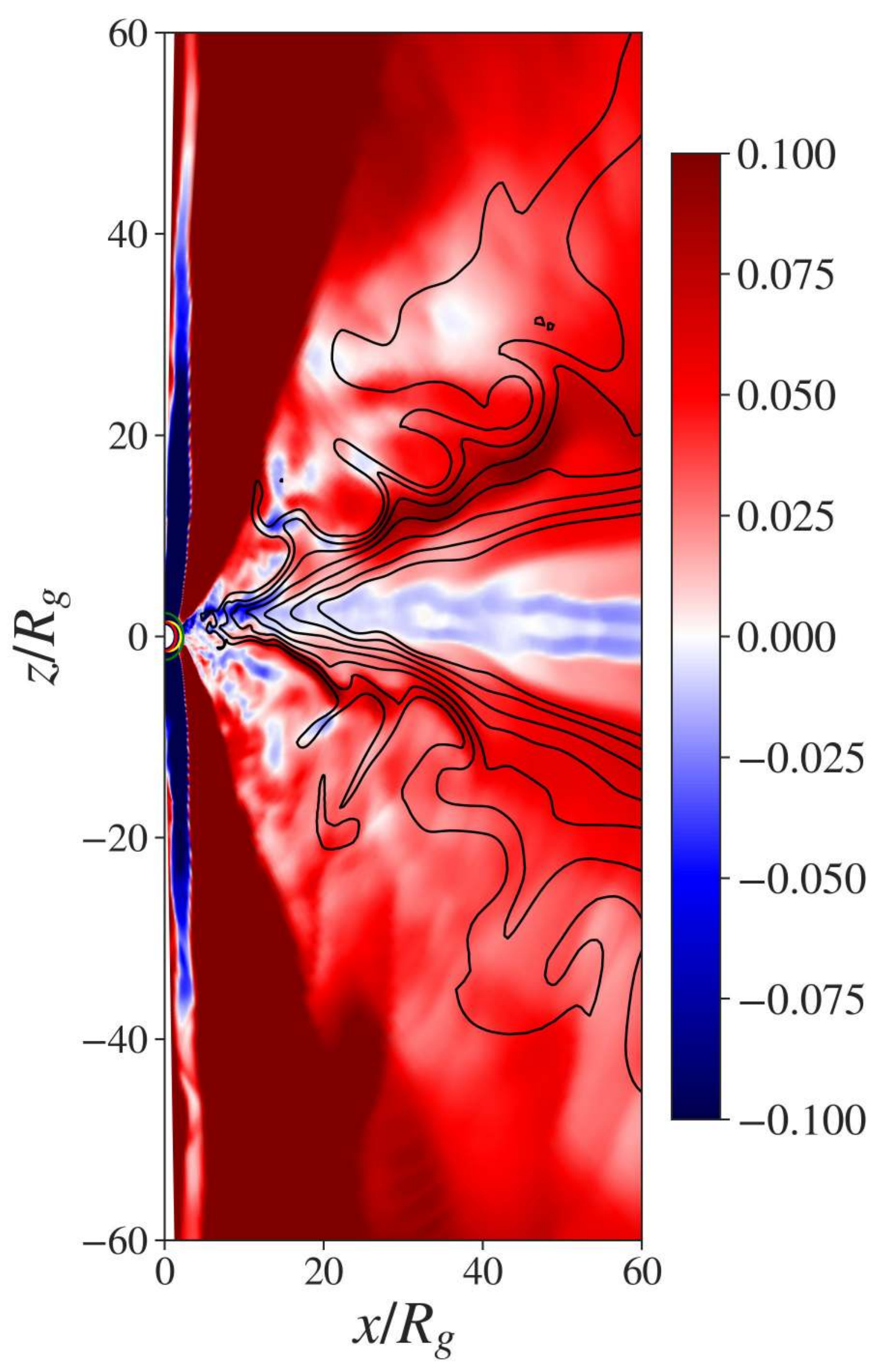}
    \includegraphics[width=0.45\columnwidth]{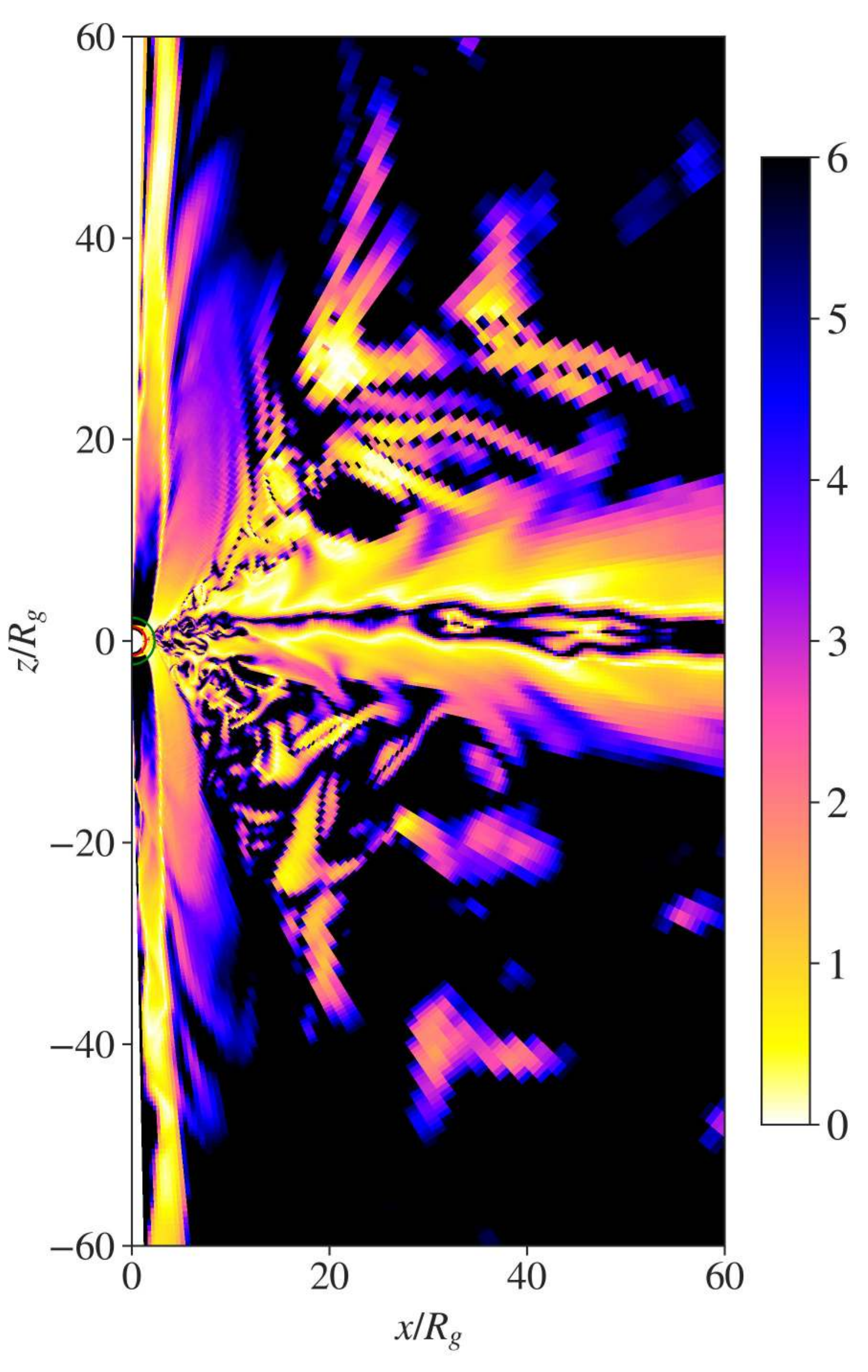}
    \caption{Reference simulation {\em sim0}.
    The radial velocity (left) with superimposed contours of the vector potential (black lines) and 
    the Alfv\'en Mach number (right), both at time $t=4000$.}
    \label{fig:wind1}
\end{figure}

\begin{figure*}
    \centering
    \includegraphics[width=0.98\columnwidth]{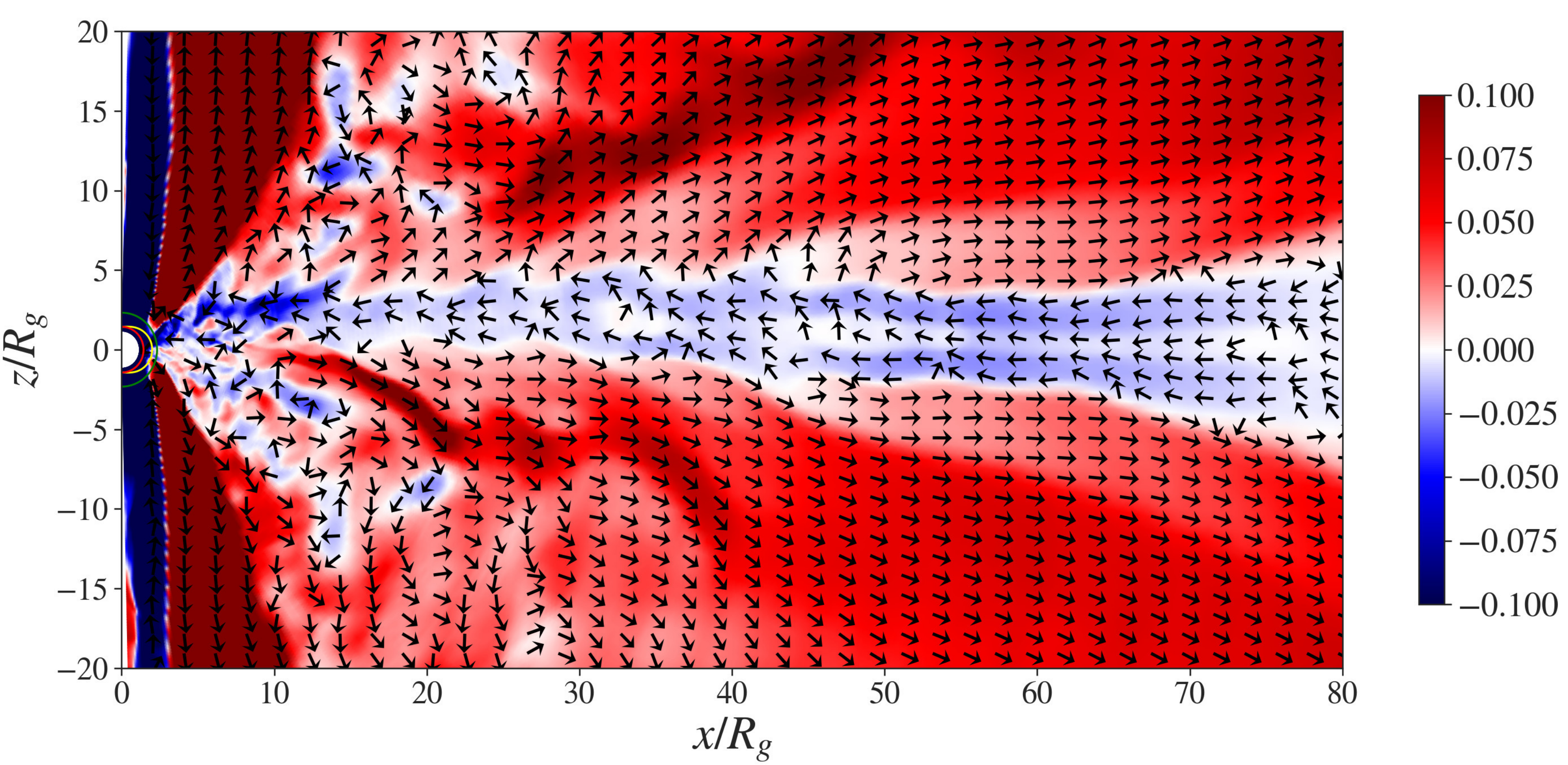}
    \includegraphics[width=0.98\columnwidth]{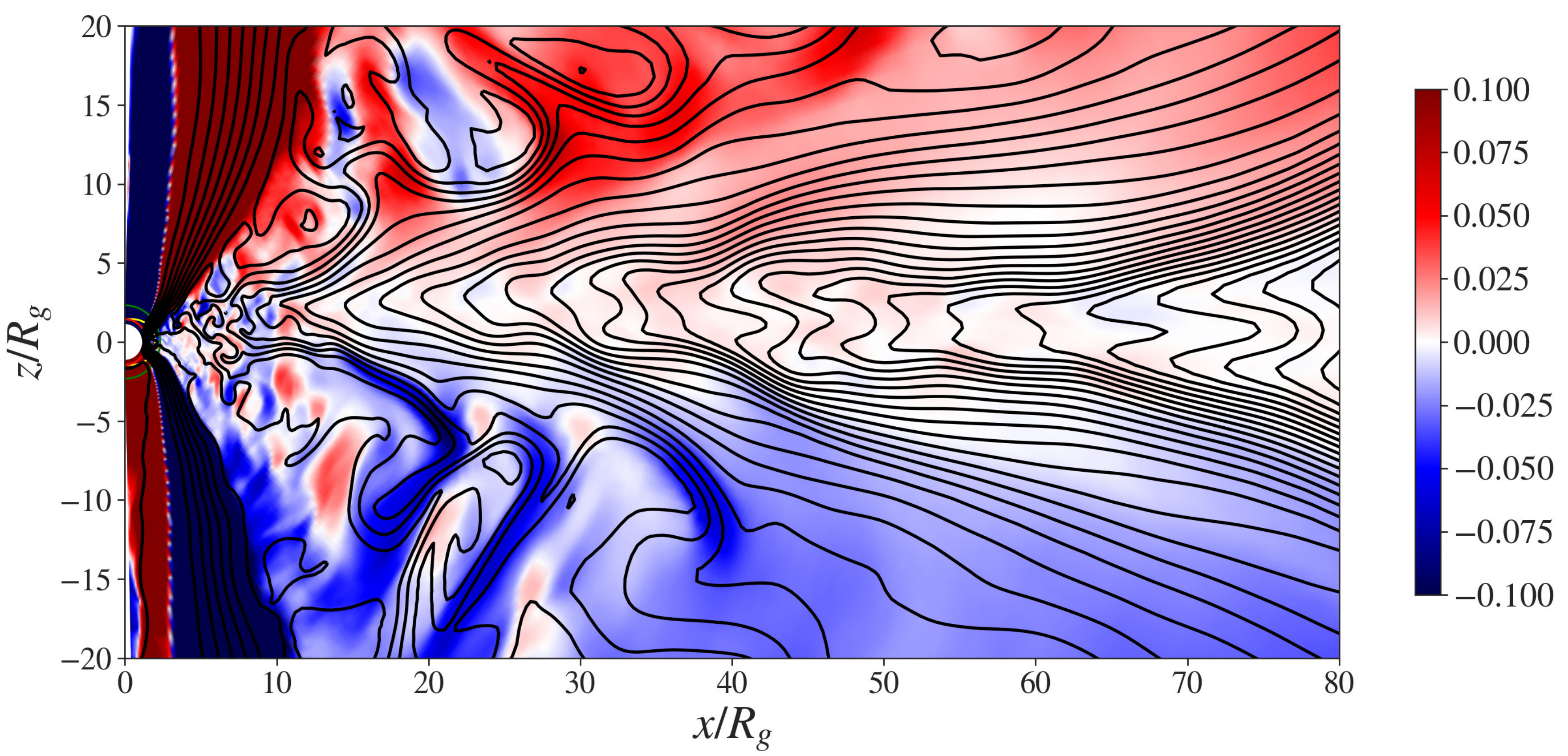}
    \includegraphics[width=0.98\columnwidth]{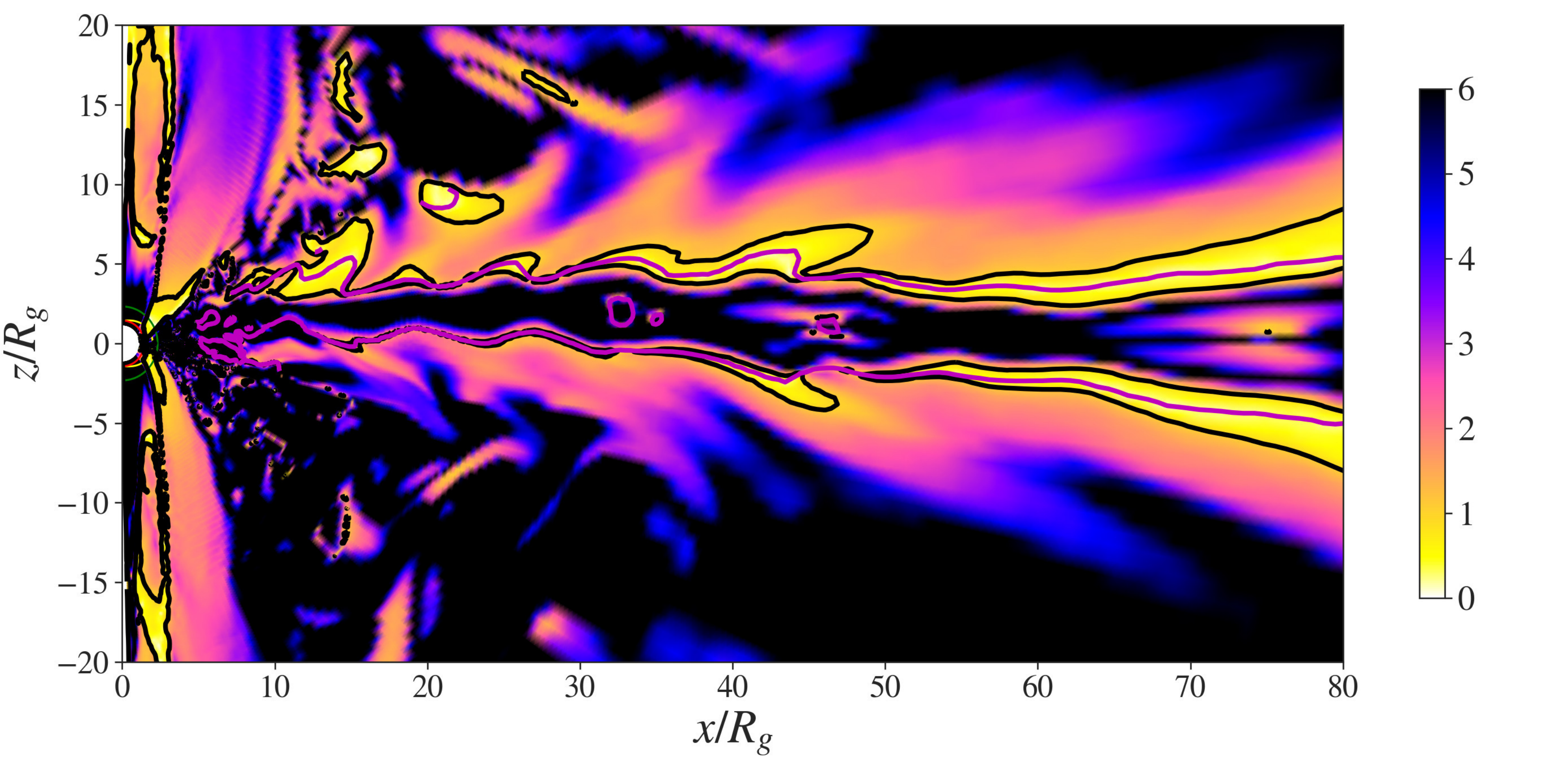}
    \includegraphics[width=0.98\columnwidth]{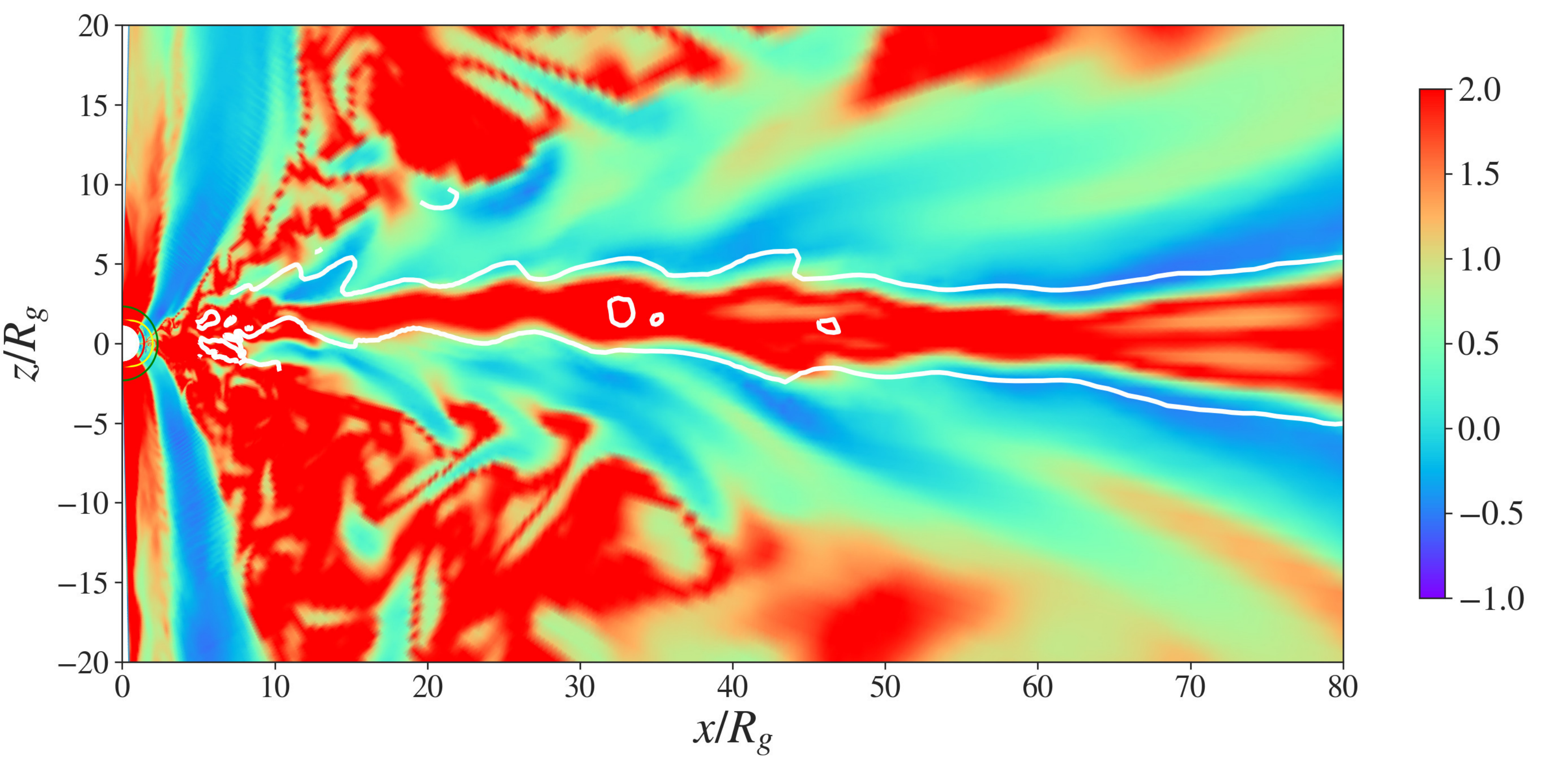}
    \includegraphics[width=0.98\columnwidth]{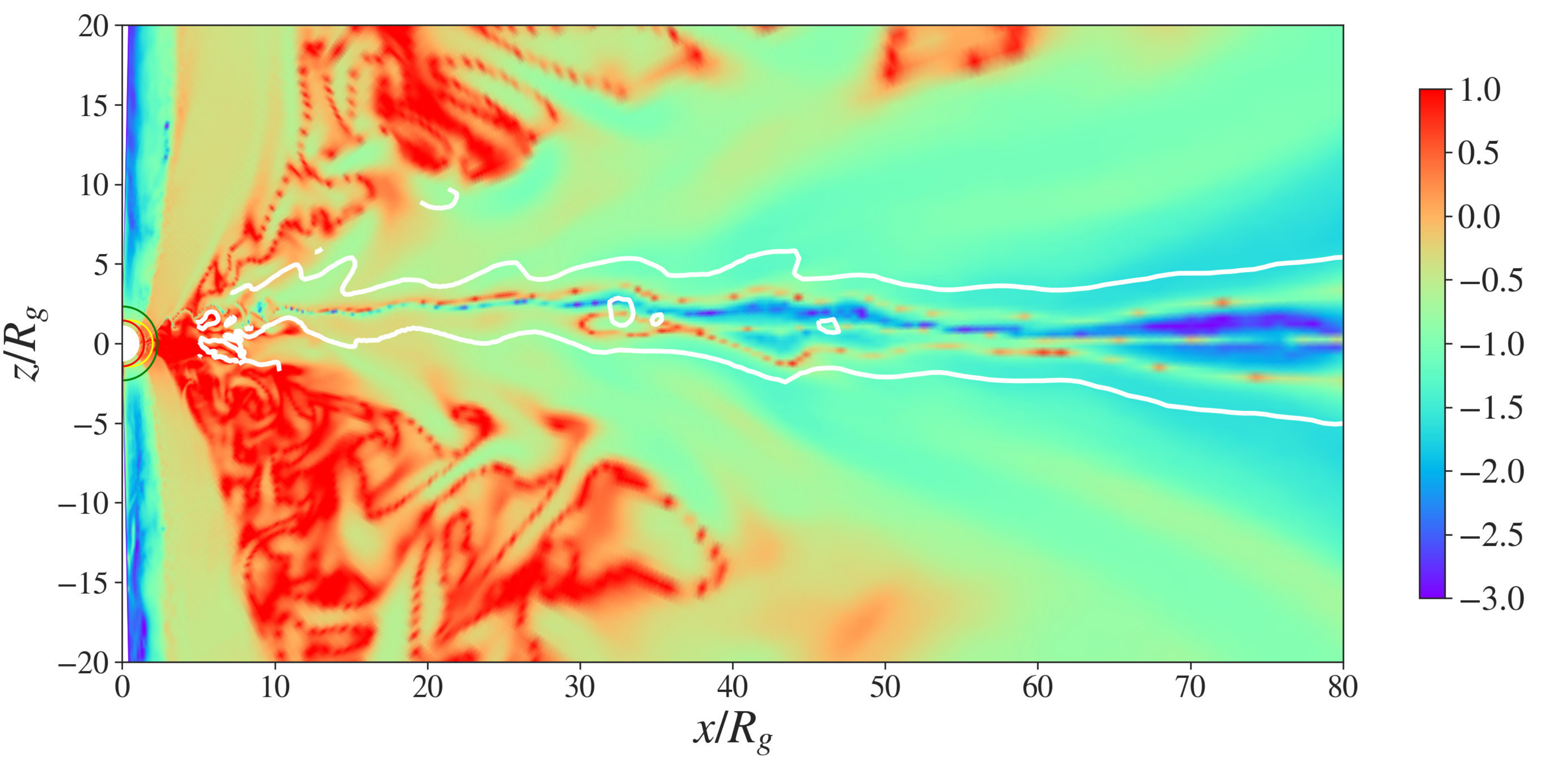}
    \includegraphics[width=0.98\columnwidth]{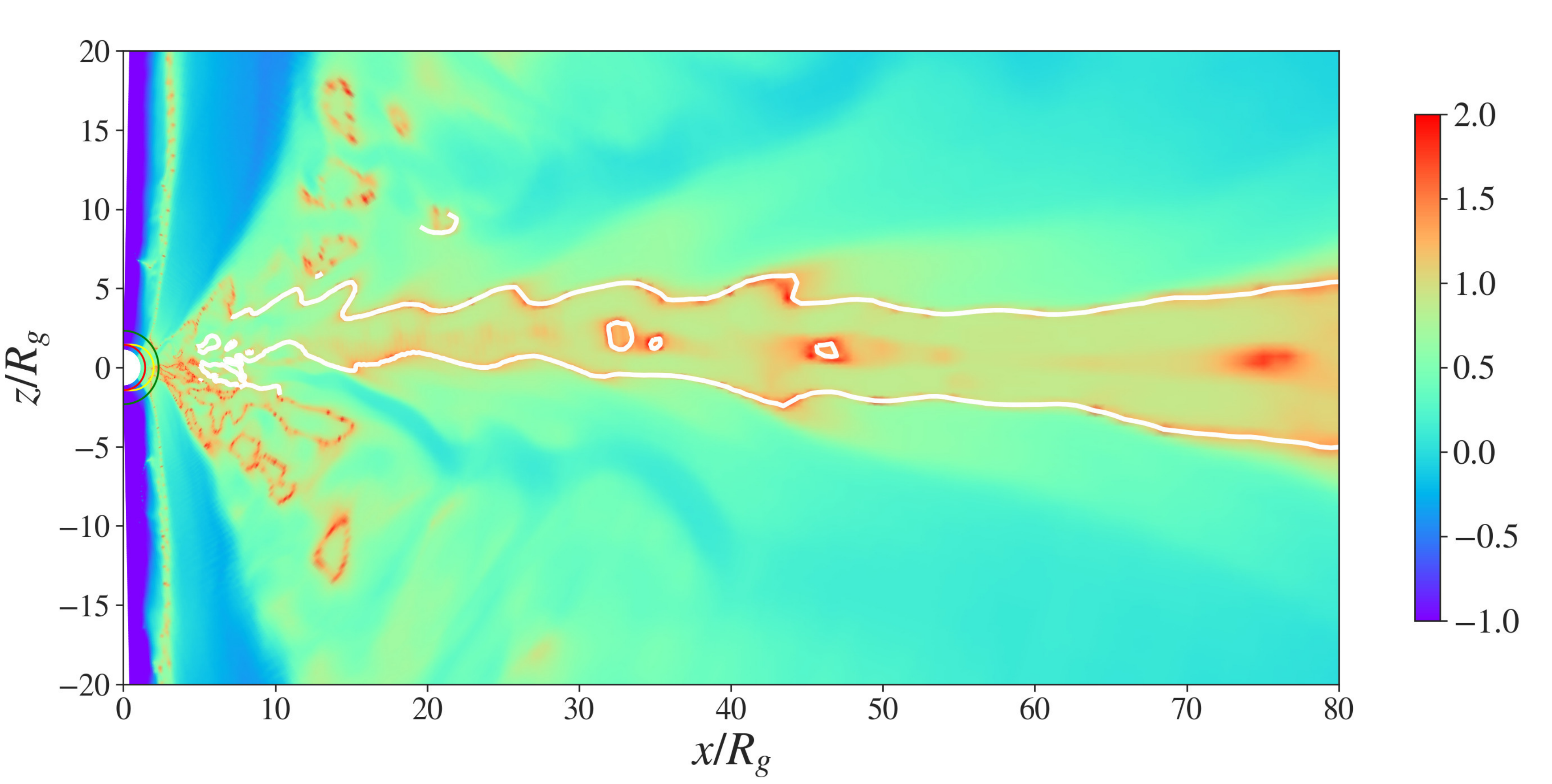}
    \caption{Snapshots of the disk area for different physical variables for simulation {\em sim0} at time $t=4000$. 
    From top left to bottom right we show 
    the radial velocity (colours) with poloidal velocity black arrows;
    the vertical velocity (colors) with magnetic field lines (black lines);
    the poloidal Alfv\'en Mach number superimposed with the Alfv\'en surface (black lines);
    the plasma-$\beta$; 
    the ratio between the toroidal and poloidal magnetic field components $| B_{\phi} / B_{\text{P}} |$ (logscale) and
    the ratio between toroidal and poloidal velocity $| u_{\phi} / u_{\text{P}} |$ (logscale). 
    The white line defines the area where $u^r$ changes sign, $u^r = 0$.}
    \label{fig:wind2}
\end{figure*}

In Figures ~\ref{fig:wind1} and \ref{fig:wind2} we present the velocity structure, the Alfv\'en Mach number and the plasma-$\beta$ 
for different areas of the disk wind.
In order to emphasize the dynamic range of the disk wind, we restrict the velocity plots to $v_{\text{p}} < 0.1c$.

The plots of radial velocity (Figure~\ref{fig:wind1} left, Figure~\ref{fig:wind2} top) nicely demonstrate the wind launching surface
where the radial velocity changes sign, thus indicating the transition from accretion to ejection.
The total poloidal velocity vectors start from inside the disk, where accretion dominates, then continue across zero-velocity surface
into the disk wind. 
The radial disk wind velocity increases as the wind leaves the disk surface, reaching up to $u^r = 0.1c$ and following the 
magnetic field lines.
Our vectors clearly demonstrate the connection between disk accretion and wind ejection.

In Figure~\ref{fig:wind2} we show the poloidal Alfv\'en Mach number $M_{\rm A,p}$.
The Alfv\'en surface is located slightly above the disk surface (which we defined by $u^r=0$), implying that the fluid leaves the 
disk surface with sub-Alfv\'enic speed, $M_{\rm A,p} < 1$.
However, it quickly accelerates to super-Alfv\'enic velocity.
This is a major difference to the non-relativistic launching simulations we have cited above, where the extension of the 
sub-Alfv\'enic regime is more comparable to the self-similar solution described by 
\citet{BP1982}, in which the flow in the area close to the disk is magnetically dominant, with matter accelerated along the 
field lines by the magnetic stress (or so-called {\em magneto-centrifugally}).
The flow then consecutively passes the Alfv\'en and the fast-magnetosonic surface, before it becomes collimated
by magnetic tension.

That mechanism may work as well for relativistic jets has been suggested by numerical simulations by \citet{Porth2010}, however
without considering the launching process out of the accretion disk.
In our reference simulation, the picture is quite different with an Alfv\'en surface much closer to the disk surface.
The flow reaches super-Alfv\'enic speed of $M_{\rm A,p} > 5$ already in the altitude of $z<10$ from the disk midplane.
Thus, we conclude that we do not find evidence for a large BP-driven region of the disk wind, and the outflow is most probably 
driven by the magnetic pressure gradient of the toroidal field, thus as a so-called magnetic tower \citep{LBell1996}.

We also need to compare the magnetic pressure to the gas pressure.
This is done in Figure~\ref{fig:wind2} where we present the distribution of plasma-$\beta$ in the area of the disk.
Inside the disk, we find plasma-$\beta > 100$ (as prescribed by our initial condition), but as we move away from the disk surface,
the plasma-$\beta$ quickly starts decreasing to values between 10 and 1 or even lower.
This finding supports the idea of a magnetic pressure-driven disk wind.

Interestingly, we find that the disk wind separates into two components considering the plasma-$\beta$.
There is an inner component of the disk wind which develops from the innermost part of the accretion disk ($r \lesssim 10$).
This wind component has a rather high gas density and pressure resulting in high poloidal plasma-$\beta$ and low magnetization,
$B^2 / \rho \sim 0.0001 $.
The second wind component originates from larger radii and it is dominated by the poloidal magnetic field.
We will first describe the inner wind component.

\subsubsection{$B_{\phi}$ dominated disk wind}
\label{sec:BphiDW}
Considering the strength of the magnetic field components, we see that the toroidal field dominates the poloidal magnetic field.
This is shown in Figure~\ref{fig:wind2} where we plot the ratio $| B_{\phi} / B_{\text{P}} |$.
In particular, the wind from the inner disk carries a toroidal field ten times larger than the poloidal component.
We believe that this results from the fact that at this time the innermost part of the disk has completed a
larger number of orbits:
at time $t=4000$ and at $r=5$ we have almost 50 orbits compared to about 18 at $r=10$ and only ten at $r=15$.
So, simply the twist of the originally poloidal magnetic field may induce such a strong toroidal field component.
If the simulation would evolve further, we expect this area of a toroidaly dominated magnetic field to grow along the disk.

We find that the radial velocity of the disk wind is not homogeneously distributed, but contains patches of negative speed.
These patches coincide with areas of strong toroidal velocity which usually accompanies the toroidal magnetic field 
in the super-Alfv\'enic flow regime (Figure~\ref{fig:wind2}, last panel).

The turbulent nature of the wind seems to damp down as the wind moves further away from its source.
Unsteady, super-Alfv\'enic outflows are well known from non-relativistic simulations.
For example, \citet{Somayeh2012} observe a similar structure for the overall disk wind in high plasma-$\beta$ simulations.
These outflows are dominated by the toroidal magnetic field component also known as tower jets (see below),
and are accelerated by the vertical toroidal magnetic field pressure gradient.
However, in our simulations we notice that this turbulent outflow layer has a certain, rather narrow opening angle.
If we assume that the extend of this layer defines a characteristic length, we may also assume that the extension
of this structure in $\phi$-direction may be similar, possibly hinting to a series of outflow tubes around the disk.
Interestingly, \citet{Britzen2017} have recently suggested that such turbulent loading of jet channels
may happen in M87, leading to large-scale episodic wiggling of the overall jet-structure.

In Figure~\ref{fig:wind2}(top right) we show the z-component of the velocity where we can distinguish a 
number of ''branches" with values higher than in the adjacent area.
These branches are actually part of the $B_{\phi}$-dominated disk wind. 
They seem to stay connected to the surface of the disk from where they are originally launched and then continue 
through the $B_{\phi}$-dominated wind following the poloidal magnetic field lines.
The footpoint of the branches coincides with highly magnetized disk areas.
This might explain the acceleration within the branches - on the other hand, when this material enters the 
$B_{\phi}$-dominated wind, the plasma-$\beta$ {\em increases} without weakening the acceleration.
We note that the strong $V_z$ component pushes the disk wind material towards the boundaries of the funnel outflow.
As for an alternative scenario we may think of a magnetic pressure-driven radial outflow which drags the poloidal field 
with it, thus stretching it into a radially aligned poloidal field distribution.

\subsubsection{$B_{\text{p}}$ dominated disk wind}
\label{sec:BpDW}
We now discuss the second wind component that originates in the outer, main body of the disk.
Here, for radii $r \gtrsim 10$, the $| B_{\phi} / B_{\text{P}} |$ ratio decreases with radius and the poloidal 
field starts to dominate. 
This outer wind becomes launched almost parallel to the magnetic field lines (see velocity streamlines and poloidal field lines in Figure~\ref{fig:wind2}) and it retains that direction as well for larger distances.
The vertical velocity component is substantially lower compared to the inner disk wind, implying a weaker acceleration 
despite the higher magnetization.
When comparing the local escape speed with the local poloidal velocity of the disk wind,
we find that the disk wind is launched with sub-escape velocity. 
However, the wind becomes further accelerated to $u_{\text{P}} > u_{\text{esc}}$ and becomes eventually fast enough to escape 
the gravity of the black hole.

In Figure~\ref{fig:wind2} (first panel), we notice that in the area where the disk wind develops, the wind tends to follow 
the radial direction in general. 
However, in Section~\ref{sec:evo} above we quantified the launching of the disk wind as mass flux escaping the disk surface
in polar direction ($\theta$-component of the velocity). 
Thus, after being launched vertically from the disk surface the wind further develops into a kind of radial outflow.
This overall picture connecting between the launching in polar direction and the radial outflow can be verified
by calculating the mass fluxes through the respective boundaries.

\subsubsection{Connecting the vertical and radial disk wind}
\label{sec:connectingDWs}
\begin{figure}
    \centering
    \includegraphics[width=0.95\columnwidth]{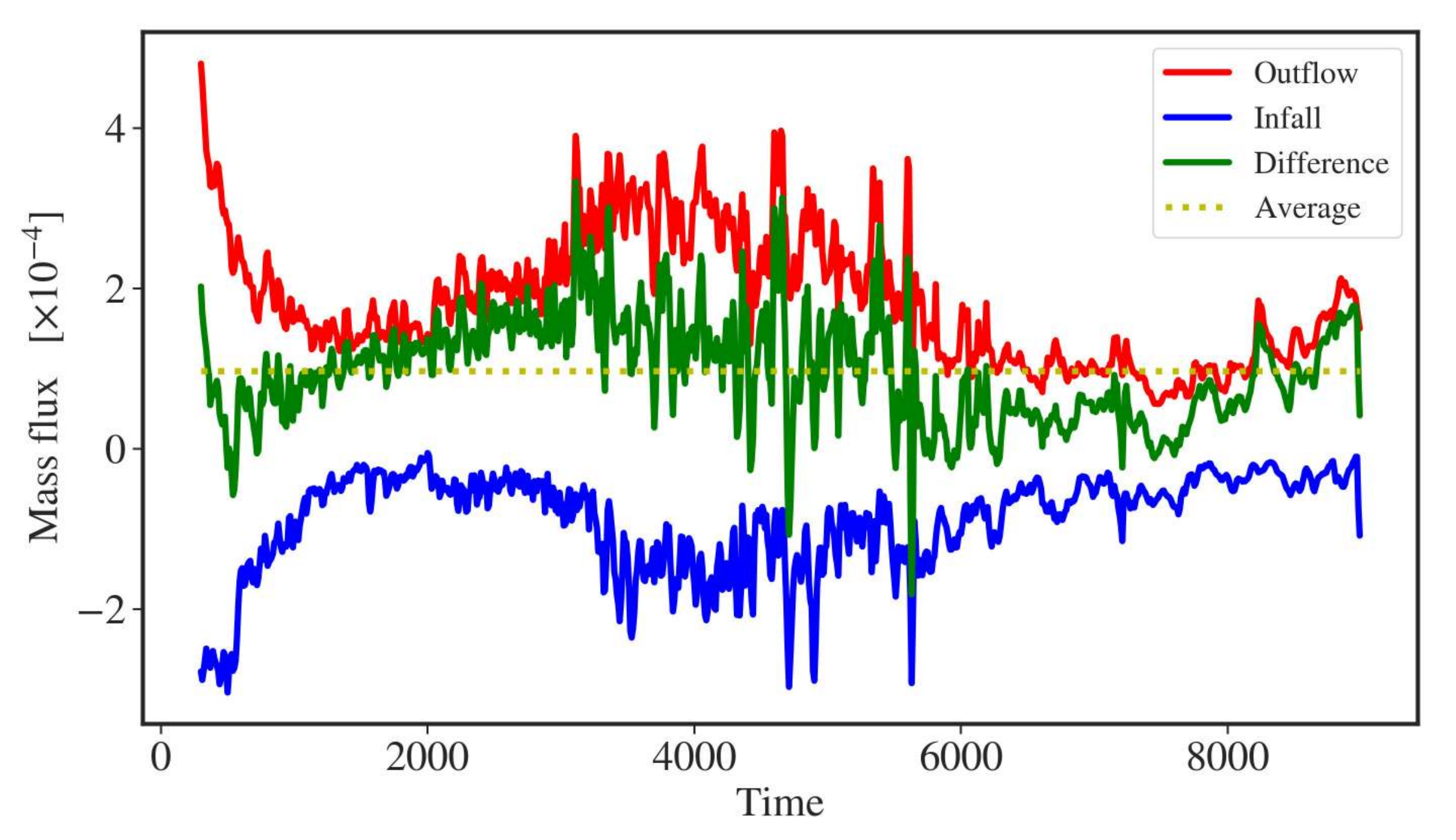}
     \caption{Mass fluxes in the reference simulation {\em sim0}.
     Shown are the outflow (red) and the inflow (blue) mass fluxes integrated along a constant opening angle 
     (considered as the initial "disk surface" ) as a function of time.
     Mass flux away from (towards) the equatorial plane is counted as positive (negative).
     The difference of the two contributions is shown in green color, while the average is shown by 
     the yellow line.}
    \label{fig:h20.3_disk_time_polarmassflux}
\end{figure}

Following the considerations of the disk wind towards the end of Section~\ref{sec:evo},
we show in Figure~\ref{fig:h20.3_disk_time_polarmassflux} the disk wind mass fluxes, 
only in this case we are interested in the situation at smaller radii where the disk outflow is 
stronger.
We calculate the mass fluxes vertical to surfaces of constant opening angle of $\theta= 80^\circ, 100^\circ$ that 
approximate the opening angle of the initial disk density distribution.
We integrate the mass fluxes in the range $r=\left[4,\;100 \right]$ where we also separate between infall (motion towards 
the disk surface) and outflow (motion away from the disk surface), thus providing the net vertical fluxes.
We find that the disk wind seems to increase for $0<t<3000$ while it decreases for $3000<t<6000$, over all we 
measure an average mass flux of $\langle\dot{M}\rangle = 9.72 \times 10^{-5} M_{0,\text{disk}}$.
The variations in the mass flux during the second phase are much stronger, this is consistent with a similar behaviour 
in the accretion rate (see \ref{sec:evo} and Figure~\ref{fig:h20.3_disk_time_totalaccretion}).

A similar behaviour is observed in the radial mass fluxes. 
These increase or decrease with radius, depending on the phase during the simulation.
Simultaneously, this is visible as an increase or decrease, respectively, in the mass load of the disk wind.
Note that the latter, we can also observe by measuring the difference in the two mass fluxes.
The overall time averaged mass flux is $\langle\dot{M}\rangle = 4.56 \times 10^{-5} M_{0,\text{disk}}$ 
across a spherical surface at $r=100$.
The difference in the two mass fluxes is deposed as mass in the area of the disk wind increasing its density.
Taking into account this mass sink as well as all mass fluxes through the surfaces of the integration area,
we find a good agreement between the radial and the disk wind fluxes
for small time intervals.
The remaining difference is due to the jet funnel that is constantly loaded by the floor model for the density
and which naturally contributes to the radial mass fluxes and also increases the mass load in the radial wind.

Our detection of a $B_{\phi}$-dominated disk wind confirms the results of \citet{QQ2},
who interpreted their results in terms of a tower jet \citep{LBell1996, Ustyugova1995}.
However, the whole disk wind in \citet{QQ2} is entirely dominated by the $B_{\phi}$, 
while in our simulation it is restricted to the disk wind from the inner disk only.
As our new simulations have a higher resolution, \citet{QQ2} may have not been able to resolve the inner part of the disk 
wind properly.

\subsubsection{Magnetic reconnection and ohmic heating}
Since the disk evolves in a resistive environment we expect the generation of ohmic heating which will affect the internal
and magnetic energy in the disk.
As we do not use radiative transfer, we cannot directly compare the energetics of ohmic heating with the emitted radiation.

However, we can attempt an estimation of the generated heating.
For the reference simulation, we calculated an approximation of ohmic heating as $\eta \vec{J}^2$ and compared it with 
the internal and magnetic energy of the fluid.
We separated the area into two parts -- 
the first one is from $r=5$ to $r=20$ and the second one from $r=20$ to $r=50$. 
Since the resistivity is concentrated to the accretion disk (and thus, the ohmic heating), we also constrain the area
between $5^{\circ}$ above and below the equatorial plane.
The ohmic heating is mostly generated from the inner part of the disk, as the magnetic field gradients 
($j \propto \rot B$) are largest over there.

We find that up to time $t=5000$ ohmic heating generates a total energy of $1.5 \times 10^{-4}$ (in code units).
This is somewhat higher than the total magnetic energy in this disk area, but substantially lower than the internal 
energy of the disk.
At larger radii, from $20 < r < 50$ the ohmic heating rate is even lower making it overall negligible in comparison with the 
magnetic and internal energy.

Another physical mechanism that contributes to the heating of our fluid is magnetic re-connection.
It has been shown \citep{DeGouveia2005, DeGouveia2010} that in AGNs, the magnetic re-connection episodes that occur 
mostly in the inner disk and the black hole magnetosphere can heat up the disk material and at the same time accelerate 
the ejected disk wind.

\begin{table*}[t]
    \caption{Mass and energy fluxes for simulations applying different black hole spin $a$ and diffusivity $\eta_0$. 
    The average mass fluxes in units of $10^{-5}$ are measured over the whole simulation period and are 
    normalized by the initial disk  mass.
    The average Poynting fluxes are in code units of $10^{-3}$.
    The average vertical wind mass flux is integrated along the radius vector along $80^{\circ}$ or $100^{\circ}$ up to $r=100$.
    The average radial wind flux is integrated along a spherical surface at $r=100$. 
    Note that simulations {\em sim3, sim5, sim6} end before $t=6000$.
    The columns show from left to right 
    the simulation run ID; the spin parameter $a$; the maximum diffusivity $\eta_0$;
    the average accretion rate at $r=2$, $\langle \dot{M}_{\text{acc}} \rangle $;
    the average vertical mass flux $\langle \dot{M}_{\theta} \rangle $;
    the average total radial mass flux $\langle \dot{M}_{r} \rangle $ ($0_{\circ}<\theta<80^{\circ}$);
    the average mass flux in the jet funnel $\langle \dot{M}_{r} \rangle_{\rm fun} $ ($0^{\circ}<\theta<25^{\circ}$);
    the average mass flux in the $B_{\phi}$-dominated disk wind $\langle \dot{M}_{r} \rangle_{B_{\phi}}$ ($25^{\circ}<\theta<65^{\circ}$);
    the average mass flux in the $B_{p}$-dominated disk wind $\langle \dot{M}_{r} \rangle_{B_{\text{P}}}$ ($65^{\circ}<\theta<80^{\circ}$);
    the electromagnetic energy flux in the funnel $\langle \dot{E}_{\text{EM}} \rangle_{\rm fun}$ ($0^{\circ}<\theta<25^{\circ}$);
    the electromagnetic energy flux in the $B_{\phi}$-dominated disk wind $\langle \dot{E}_{\text{EM}} \rangle_{B_{\phi}}$ ($25^{\circ}<\theta<65^{\circ}$);
    the electromagnetic energy flux in the $B_{p}$-dominated disk wind $\langle \dot{E}_{\text{EM}} \rangle_{B_{\text{P}}}$ ($65^{\circ}<\theta<80^{\circ}$);
    Values in parentheses show the percentage of each individual radial mass flux over the total radial mass flux.
    }
    \centering
    \begin{tabular}{cccccccccccc}
    \hline
    \noalign{\smallskip}
        run   &  $a$  & $\eta_0$ & $\langle \dot{M}_{\text{acc}} \rangle $ & $\langle \dot{M}_{\theta} \rangle $ & $\langle \dot{M}_{r} \rangle $ & $\langle \dot{M}_{r} \rangle_{\rm fun} $ & $\langle \dot{M}_{r} \rangle_{B_{\phi}}$  & $\langle \dot{M}_{r} \rangle_{B_{\text{P}}} $ & $\langle \dot{E}_{\text{EM}} \rangle_{\rm fun}$ & $\langle \dot{E}_{\text{EM}} \rangle_{B_{\phi}}$ & $\langle \dot{E}_{\text{EM}} \rangle_{B_\text{P}}$  \\ 
    \noalign{\smallskip}     
    \hline 
    \noalign{\smallskip}
        sim0  & 0.9 & 0.001  & -0.75 & 9.72 & 4.15 & 1.02 (25) & 2.40 (58) & 0.73 (16) & 4.89 & 2.38 & 0.38 \\ 
        sim1  & 0   & 0.001  & -1.59 & 6.20 & 1.83 & 0.26 (14) & 1.02 (56) & 0.55 (30) & 0.44 & 0.55 & 0.23 \\ 
        sim2  & 0.5 & 0.001  & -1.57 & 7.51 & 2.88 & 0.67 (23) & 1.57 (54) & 0.64 (22) & 2.87 & 1.29 & 0.32 \\ 
        sim3  &-0.9 & 0.001  & -1.27 & 5.77 & 4.88 & 0.96 (20) & 3.48 (71) & 0.44 (9) & 3.00 & 4.52 & 0.21 \\ 
        sim4  & 0.9 & 0.01   & -0.53 & 5.61 & 3.17 & 0.79 (25) & 1.94 (61) & 0.44 (14) & 1.82 & 1.93 & 0.19 \\ 
        sim5  & 0.9 & 0.0001 & -1.24 & 12.8 & 3.70 & 1.43 (39) & 1.81 (49) & 0.46 (12) & 4.11 & 1.93 & 0.24 \\ 
        sim6  & 0.9 &$10^{-10}$& -1.24 & 11.6 & 3.04 & 1.37 (45) & 1.33 (44) & 0.33 (10) & 4.17 & 1.72 & 0.25  \\ 
         
    \noalign{\smallskip}
    \hline
    \end{tabular}
    \label{tab:simulations}
\end{table*}

\section{Comparison study} 
\label{sec:comparison}
We will now compare our reference run {\em sim0} with a number of simulations that apply different 
physical parameters such as black hole spin, magnetic field strength, or magnetic diffusivity (see Table~\ref{tab:simulations}).

\begin{figure}
    \centering
    \includegraphics[width=0.95\columnwidth]{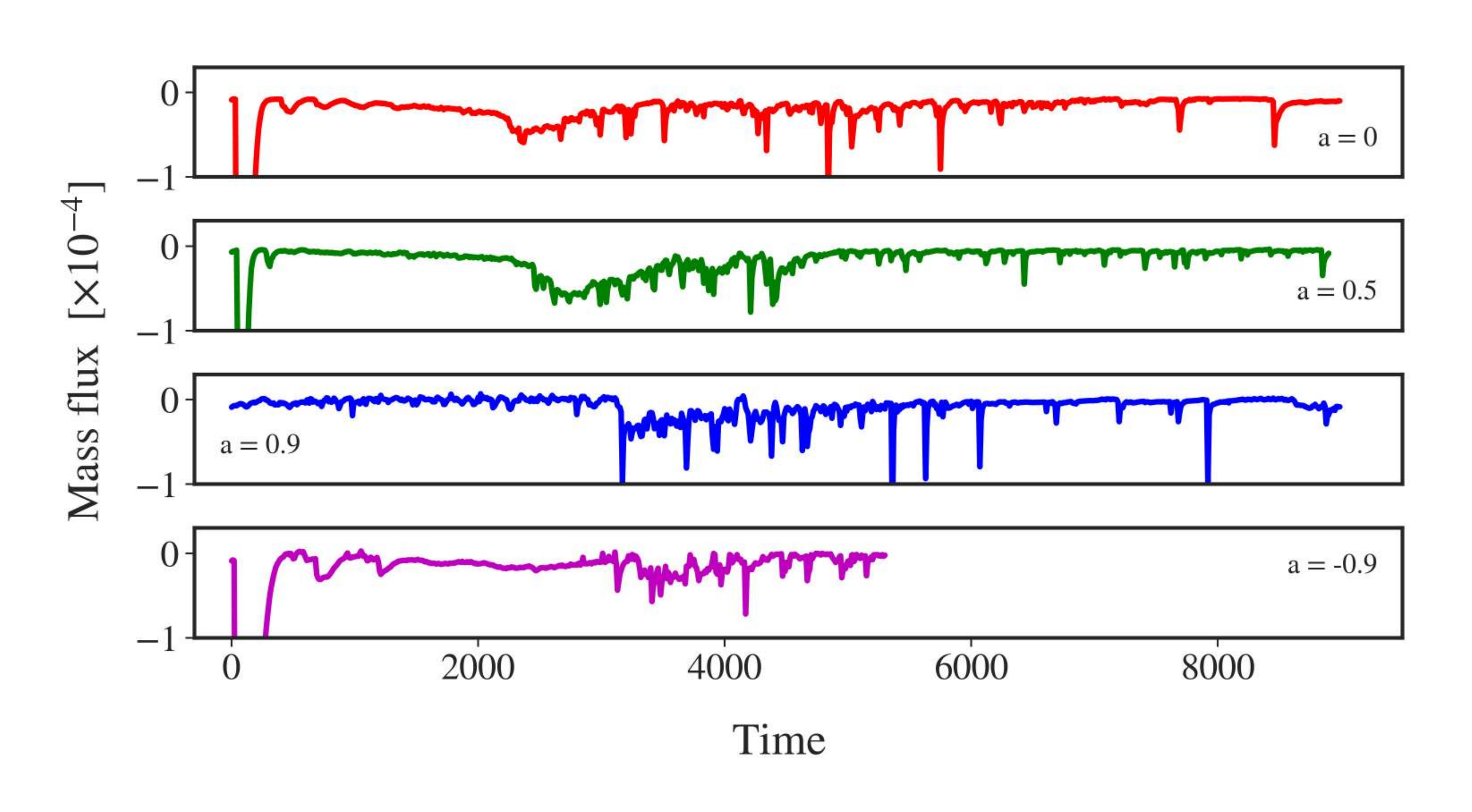}
     \caption{Accretion rate and black hole spin. 
     Comparison of the accretion rates measured at $r \approx 2$ for simulation runs applying a Kerr parameter 
     $a=0$ ({\em sim1}, red), 
     $a=0.5$ ({\em sim2}, green), 
     $a=0.9$ ({\em sim0}, blue), and 
     $a=-0.9$ ({\em sim3}, magenta).}
    \label{fig:BHa_compare_accretion}
\end{figure}

\begin{figure}
    \centering
    \includegraphics[width=0.95\columnwidth]{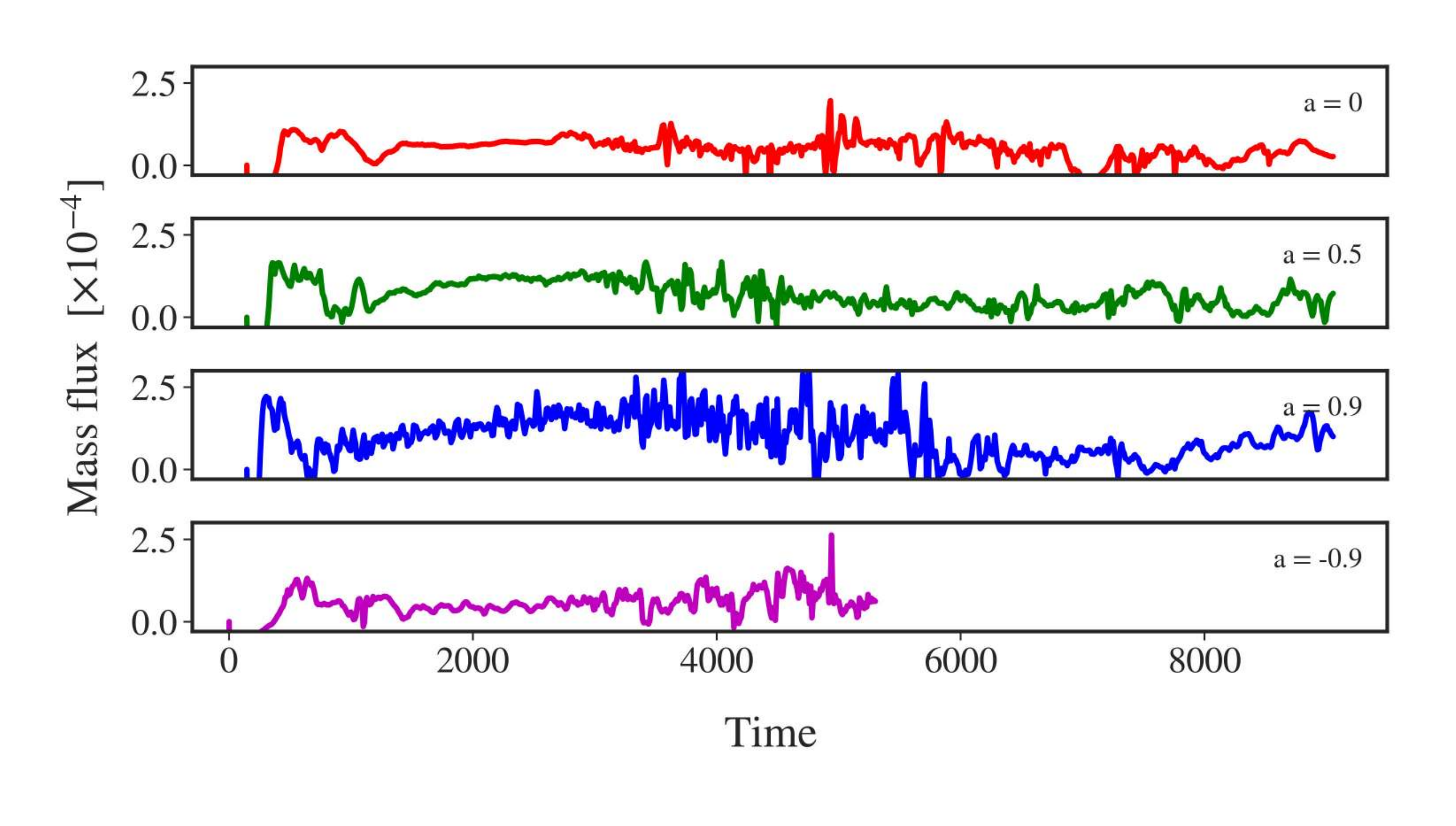}
     \caption{Ejection rate and black hole spin. 
     Comparison of the vertical mass fluxes associated with the disk wind for simulation runs applying a Kerr parameter
     $a=0$ (red), $a=0.5$ (green) and $a=0.9$ (blue) and $a=-0.9$ (magenta), 
     integrated along the surfaces of constant angle at $80^{\circ}$ and $100^{\circ}$.}
    \label{fig:BHa_compare_diskwind}
\end{figure}

\subsection{Accretion-ejection and black hole rotation}
We now discuss how the dynamical evolution of accretion-ejection interrelates with the black hole rotation,
i.e. the Kerr parameter $a$.

We first concentrate on the disk accretion.
Figure~\ref{fig:BHa_compare_accretion} shows the disk accretion rates at $r \approx 2$ for the simulation runs
{\em sim1}, {\em sim2} and {\em sim0}, each normalized with the mass of the respective initial disks. 

While for {\em sim0} the accretion rate in the first stages of the evolution ($t \in [0,3000]$) is constant and very low, 
for slower rotating black holes the accretion rate shows a noticeable increase.
Also, this first stage that looks different from the later evolution last longer in the case of $a=0.9$.
We think that this is due to the fact that the horizon ($r = 2$), and the ISCO ($r = 6$) are  located closer to the initial disk radius.
Therefore, it take less time to bring disk material to the ISCO from which it falls to the horizon.
At later stages, all simulations show a similar behaviour, with only the accretion spikes in {\em sim0} being slightly stronger.

On average, for the duration of the simulation, the normalized accretion rate at $r=2$ for the Schwarzschild black hole is slightly higher, 
$\langle\dot{M}\rangle = -1.59 \times 10^{-5} M_{0,\text{disk}}$,
while for the case of $a=0.9$ we find $
\langle \dot{M} \rangle = -7.49 \times 10^{-6} M_{0,\text{disk}}$.
Specifically, the three systems accrete $18.4 \%$,  $17.3 \%$ and $11 \%$
of their initial disk mass into the black hole for the duration of the simulations.

In Figure~\ref{fig:BHa_compare_diskwind} we compare the disk wind that is launched from the disk surface.
The disk mass flux is in general positive with few exceptions\footnote{Most of the negative flux occurrences appear in the late stages of {\em sim1}},
meaning that there is a substantial mass injection from the disk into the outflow. 

Following the same method as described towards the end of  Section~\ref{sec:connectingDWs} we measure a normalized mass flux for the disk wind 
of $\langle \dot{M}\rangle = 6.2 \times 10^{-5} M_{0,\text{Disk}}$ for the case of $a=0$, 
a flux of $\langle\dot{M}\rangle = 7.41 \times 10^{-5} M_{0,\text{Disk}}$ for the case of $a=0.5$,
and a flux of $\langle \dot{M}\rangle = 9.72 \times 10^{-5} M_{0,\text{Disk}}$ for the case of $a=0.9$.
This implies that the three accretion-ejection systems accumulate a mass loss of 
$49 \%$, $62.3 \%$ and $80.7 \%$ of their initial disk mass by the disk wind.
The cases $a=0$ and $a=0.9$ differ by almost $30\%$ in the disk wind mass flux.
For the radial fluxes there is a similar increase by $178 \%$ between the simulations applying $a=0$ and the $a=0.9$
(see Table~\ref{tab:simulations}).
Thus, as an overall trend we find that the disk wind mass flux increases for higher black hole spin.

We understand that this is due to the ejection of mass that is launched from the innermost radii of disk accretion for high $a$ (see Figure~\ref{fig:h20.3_disk_time_totalaccretion}, middle panel).
These ejections, thus positive radial mass fluxes inside the disk, do not appear for the cases of low spin $a=0 \text{ and } a=0.5$,
for which accretion dominates, and  which result in an overall lower disk wind ejection rate
(see in Figure~\ref{fig:BHa_compare_diskwind}).
There is also the interplay between the evolution of the disk structure in respect to the distribution of magnetic diffusivity.
As the ISCO radius is affected by the Kerr parameter, the disk is located completely inside the high diffusivity area for $a=0$, 
while part of the inner radii has lower diffusivity for the case of $a=0.9$.

Note that the radius $r=3$ is just outside the ISCO for simulation {\em sim0}, but inside the ISCO for {\em sim1} and
{\em sim2}, which we think explains why no ejection is visible in the case of the latter two simulations. 
In order to check this hypothesis, we also measured the mass flux at one and two $\rg$ outside of the ISCO
for each of our simulations.
Only in simulation {\em sim0} there appears a positive mass flux from this radius, subsequently contributing to the 
increased mass flux we measure in the disk corona.

We further investigate the radial mass fluxes through a surface of radius $r=100$.
We find that the increase in the mass flux is much higher than in the vertical fluxes.
We have also analyzed the radial mass flux of the disk wind by comparing the fluxes in three domains of the outflow
(see Table~\ref{tab:simulations} for numerical values).
The innermost flow area is from $0^{\circ}$ to $25^{\circ}$ and it indicates the mass flux in the Poynting-dominated jet.
The adjoined area  from $25^{\circ}$ to $65^{\circ}$ covers the $B_{\phi}$-dominated wind launched in the innermost disk.
The third domain from $65^{\circ}$ to $80^{\circ}$ contains the mass flux from the $B_{\text{P}}$-dominated disk wind.
Obviously, we also include the fluxes from the lower hemisphere.

We recognize that our choice for the limits in the polar angle will not always coincide perfectly with the physical part 
of the flow we want to study.
This holds especially in the earlier and later times of the simulations when both the jet and the disk wind are strongly 
evolving, either further being developed (early) or are dying off because of the disk mass loss (late). 
For the Poynting-dominated jet, the floor density model that dominates this area obviously determines most 
of the mass flux .

Comparing the simulations, we find
that the relative contribution of the $B_{\phi}$-dominated disk wind to the overall mass flux is similar 
for  simulations {\em sim1}, {\em sim2} and {\em sim0} - even though in absolute values the wind mass flux increases with 
black hole spin.
The relative contribution of the $B_{\phi}$ and the $B_{\text{P}}$ dominated disk winds, however, depends on on the black hole spin.
In the case of a Schwarzschild black hole the $B_{\phi}$ dominated disk wind contributes $65\%$ to the total disk wind 
mass flux while for the case of $a=0.9$ the contribution is at $77\%$.
For the counter-rotating black hole the contribution increases to $89\%$ while it shows the strongest wind also in absolute values.
We conclude that the black hole rotation increases not only the disk outflow mass flux in general, but also contributes 
substantially in the $B_{\phi}$ dominated disk wind as it is generated from the inner part of the disk.

\begin{figure}
    \centering
    \includegraphics[width=0.95\columnwidth]{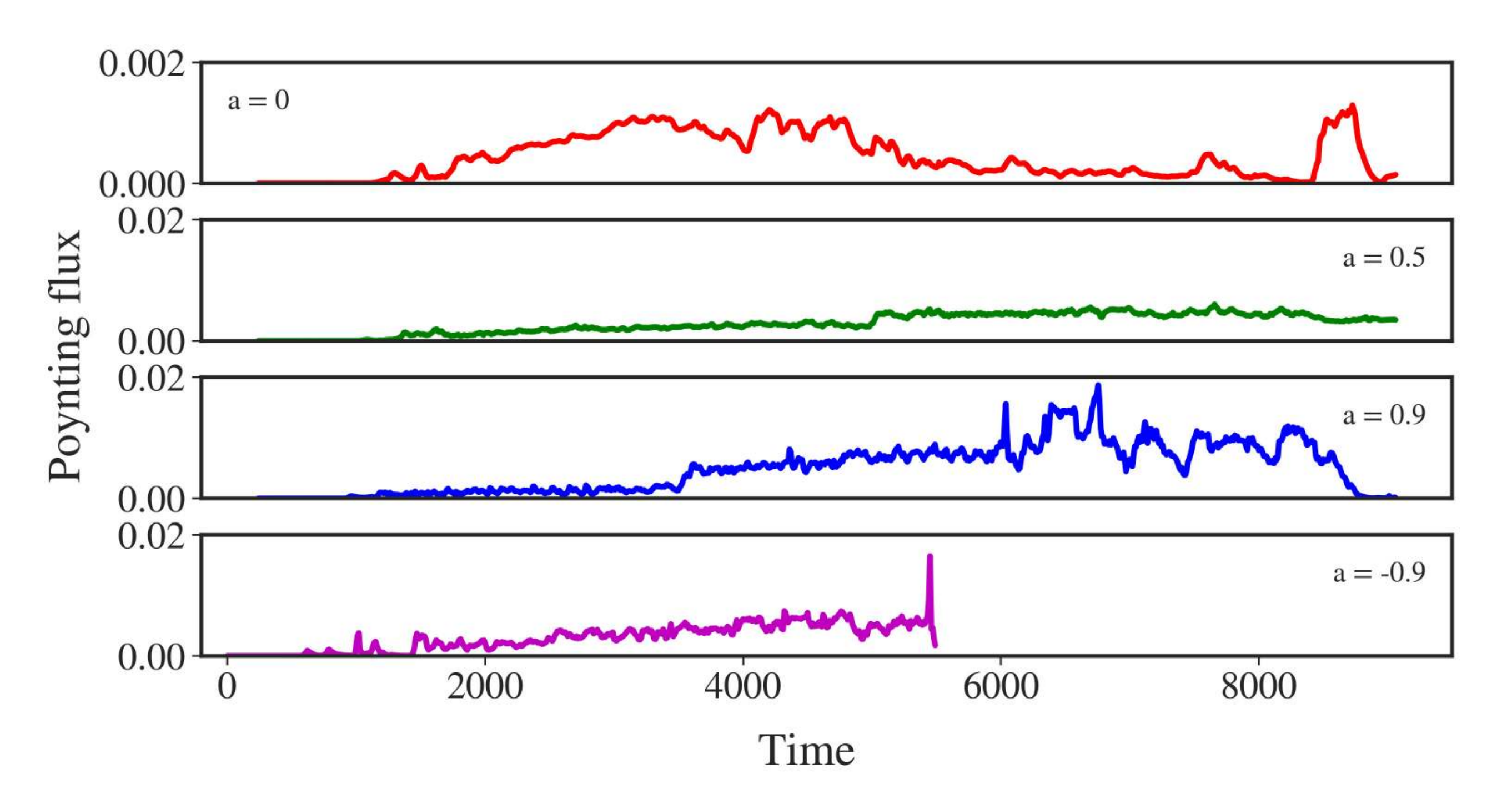}
     \caption{Poynting flux and black hole spin. 
     Comparison of the radial Poynting fluxes at $r=100$ for simulation runs applying a Kerr parameter 
     $a=0$ ({\em sim1}, red), 
     $a=0.5$ ({\em sim2}, green), 
     $a=0.9$ ({\em sim0}, blue), and 
     $a=-0.9$ ({\em sim3}, magenta).
     Note that simulation {\em sim1} is scaled 10 times lower than the others.
     }
    \label{fig:BHa_compare_poynting}
\end{figure}

Finally, we compare the Poynting flux in our simulations.
Figure~\ref{fig:BHa_compare_poynting} shows the time evolution of the Poynting flux through a surface at $r=100$ in the 
area of the funnel flow for the four different cases of black hole spin.
There is a clear trend that the Poynting flux from the jet funnel increases with spin parameter.
The highest Poynting flux appears in the reference simulation with $a=0.9$.
For simulation {\em sim1} the flux is substantially (factor 10) lower than for the simulation with a rotating black 
hole.
Also, in {\em sim1} the absence of black hole rotation results in a relatively higher flux from the disk wind.
A question arises on what drives the Poynting flux from a non-spinning black hole.
We believe that this Poynting flux is driven by the rapidly-rotating (infalling) material that is just outside the horizon in a 
fashion similar to the BZ mechanism.
The magnetic field lines are twisted by the rotating disk creating a jet with smaller  electromagnetic energy flux.

\subsubsection{A counter-rotating black hole}
We now investigate how a counter-rotating black hole affects the overall jet launching.
It has been suggested that the efficiency of the BZ process in prograde systems is slightly higher compared to 
retrograde black hole-torus systems \citep{TchekhMcKinney2012}.
Here we extend this analysis for resistive GR-MHD and for thin accretion disks.
We have setup simulation run {\em sim3} with a negative Kerr parameter $a=-0.9$, but otherwise identical to our 
reference simulation.

A first comparison shows the accretion rate at radius $r=2$ (see Figure~\ref{fig:BHa_compare_accretion})
and the disk wind mass flux (see Figure~\ref{fig:BHa_compare_diskwind}) for both simulations.
For $a=-0.9$ the ISCO is located at $r \approx 8.7 $.
As a result, since the inner radius of the initial disk is located further in at $r=7$, accretion towards the black hole
starts immediately with a sudden infall of the disk area inside ISCO.
Furthermore, the disk immediately looses a substantial fraction of mass, about $30\%$ until $t=300 \tg$.
Afterwards, the disk structure adjusts such that its inner radius remains outside the ISCO and the normal -- slow -- accretion begins
as soon as angular momentum is removed from the disk material.

\begin{figure}
    \centering
    \includegraphics[width=4.25cm]{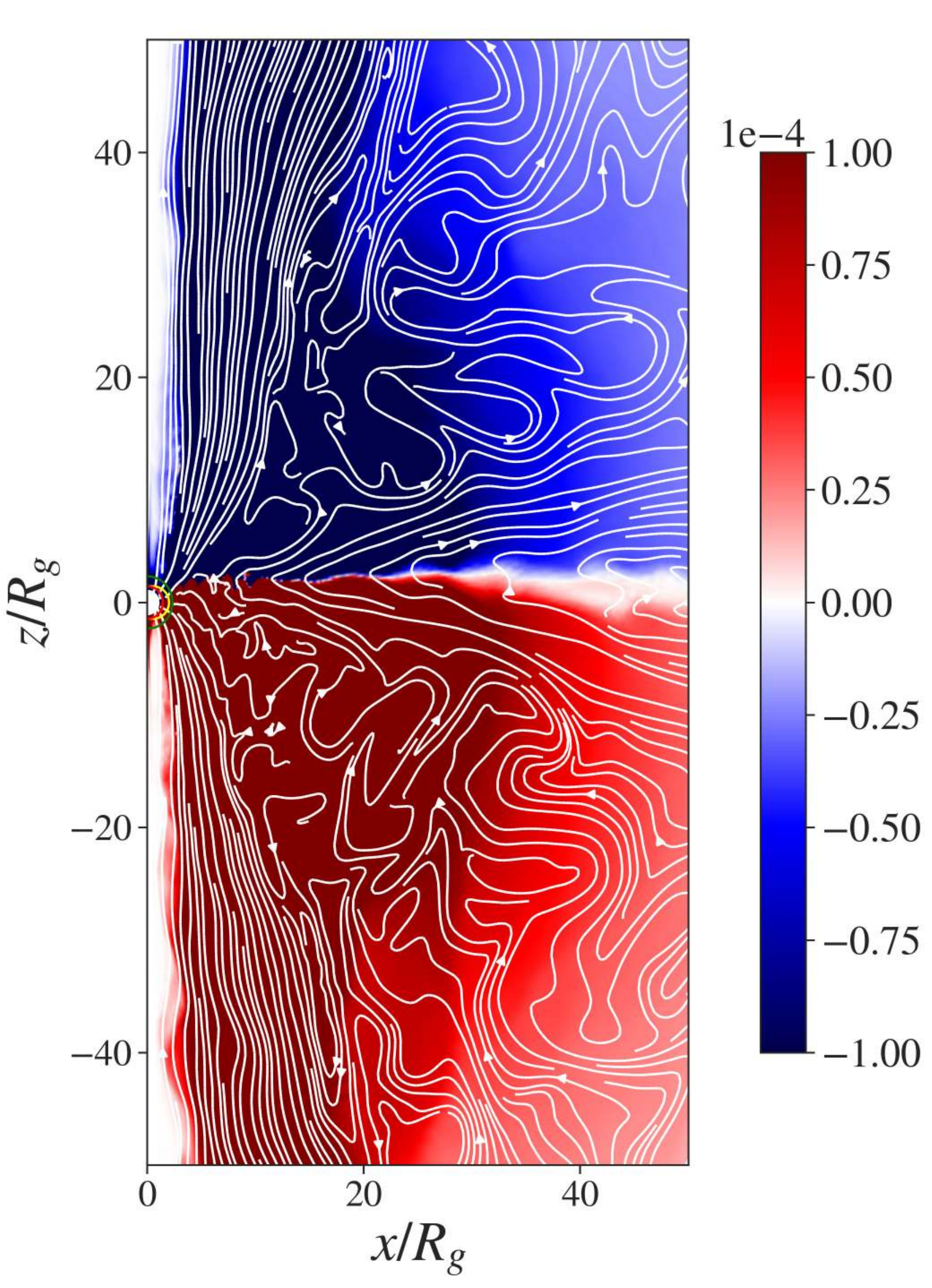}
    \includegraphics[width=4.25cm]{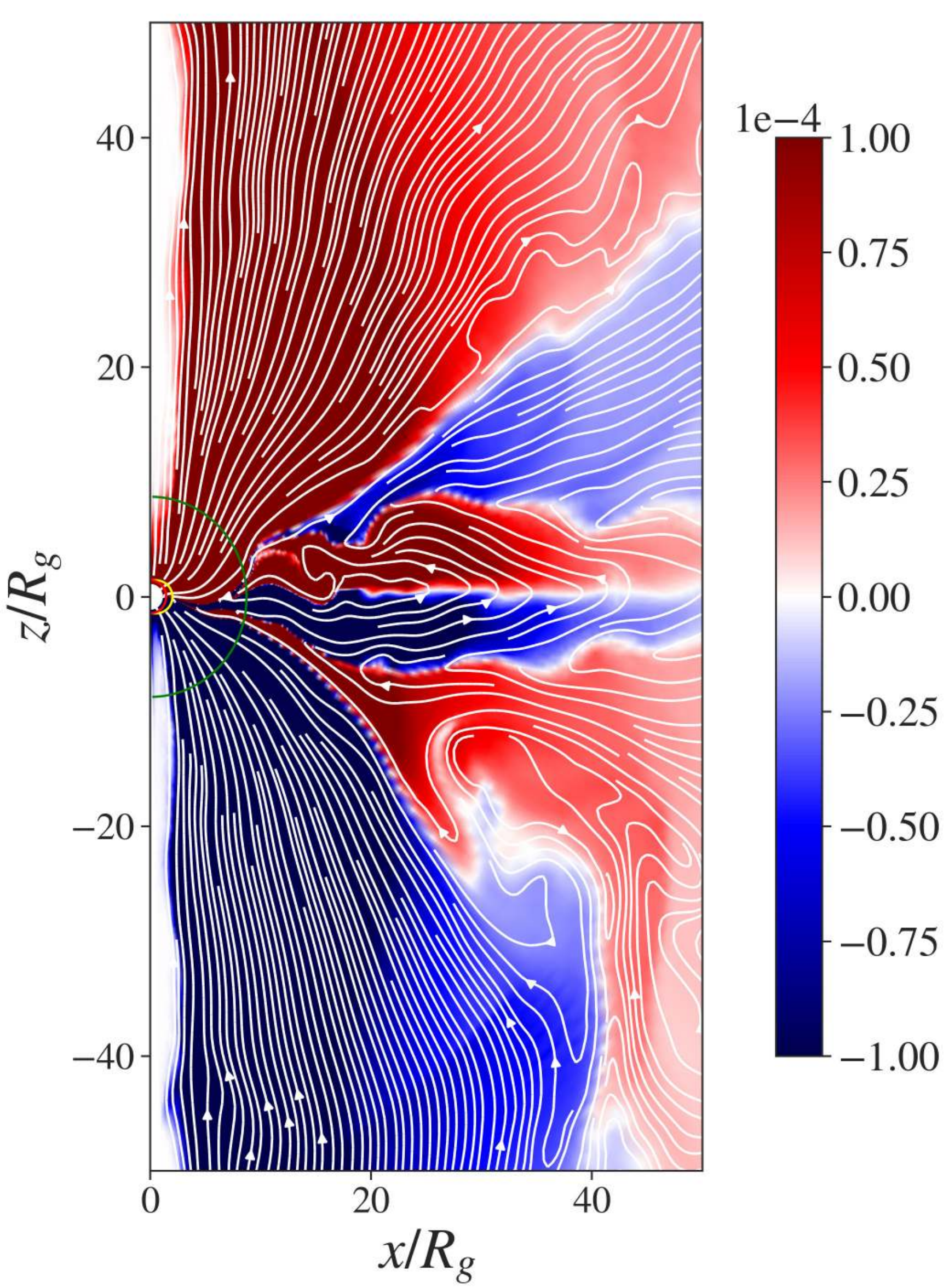}
     \caption{A counter-rotating black hole.
     Comparison of the toroidal magnetic field component for simulation 
     {\em sim0} with $a=0.9$ (left) and 
     {\em sim3} with $a=-0.9$ (right) at $t=4000$. 
     The white lines show poloidal magnetic field lines.}
    \label{fig:compare_sim0-sim3_a}
\end{figure}

\begin{figure}
    \centering
    \includegraphics[width=4.2cm]{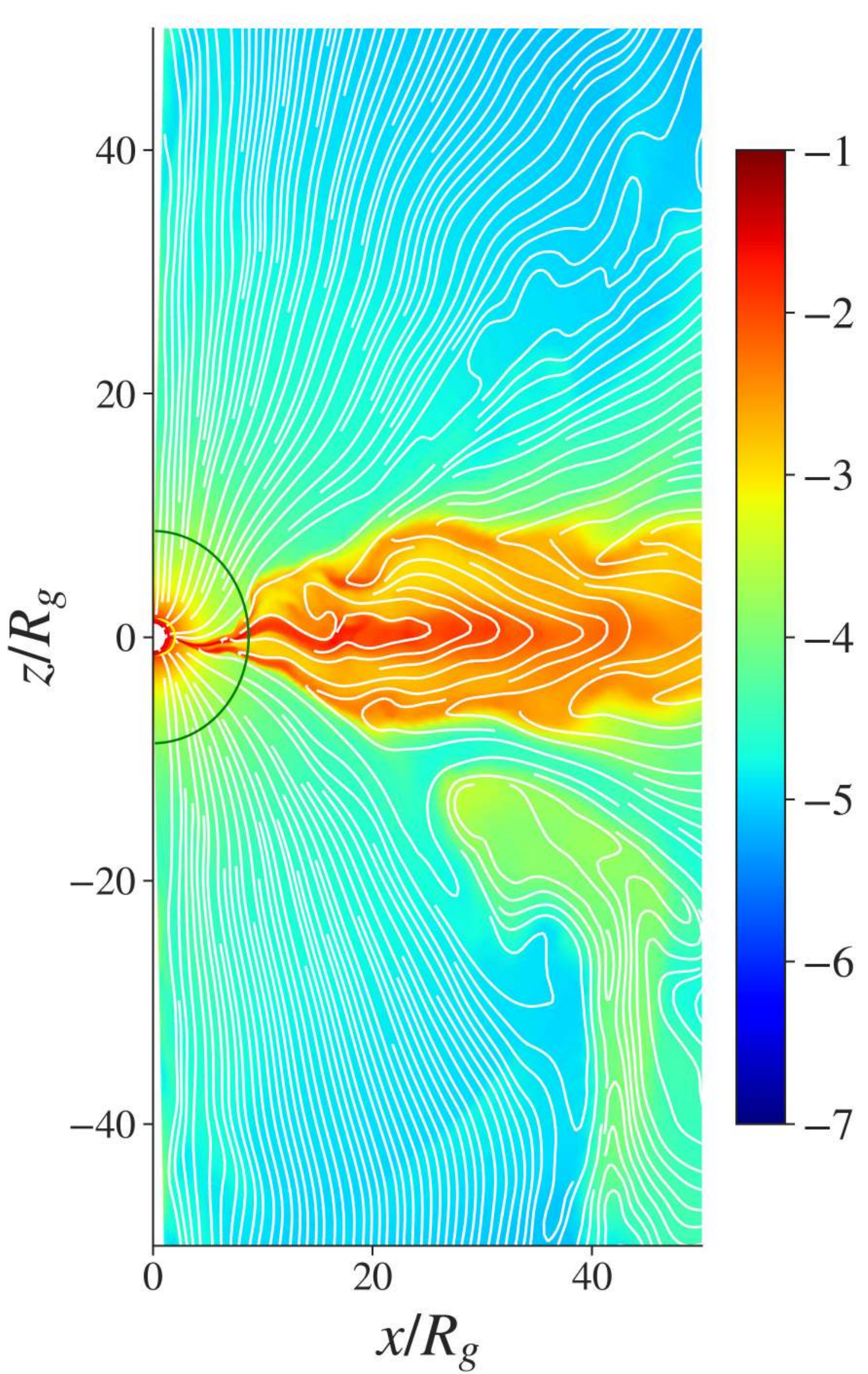}
    \includegraphics[width=4.3cm]{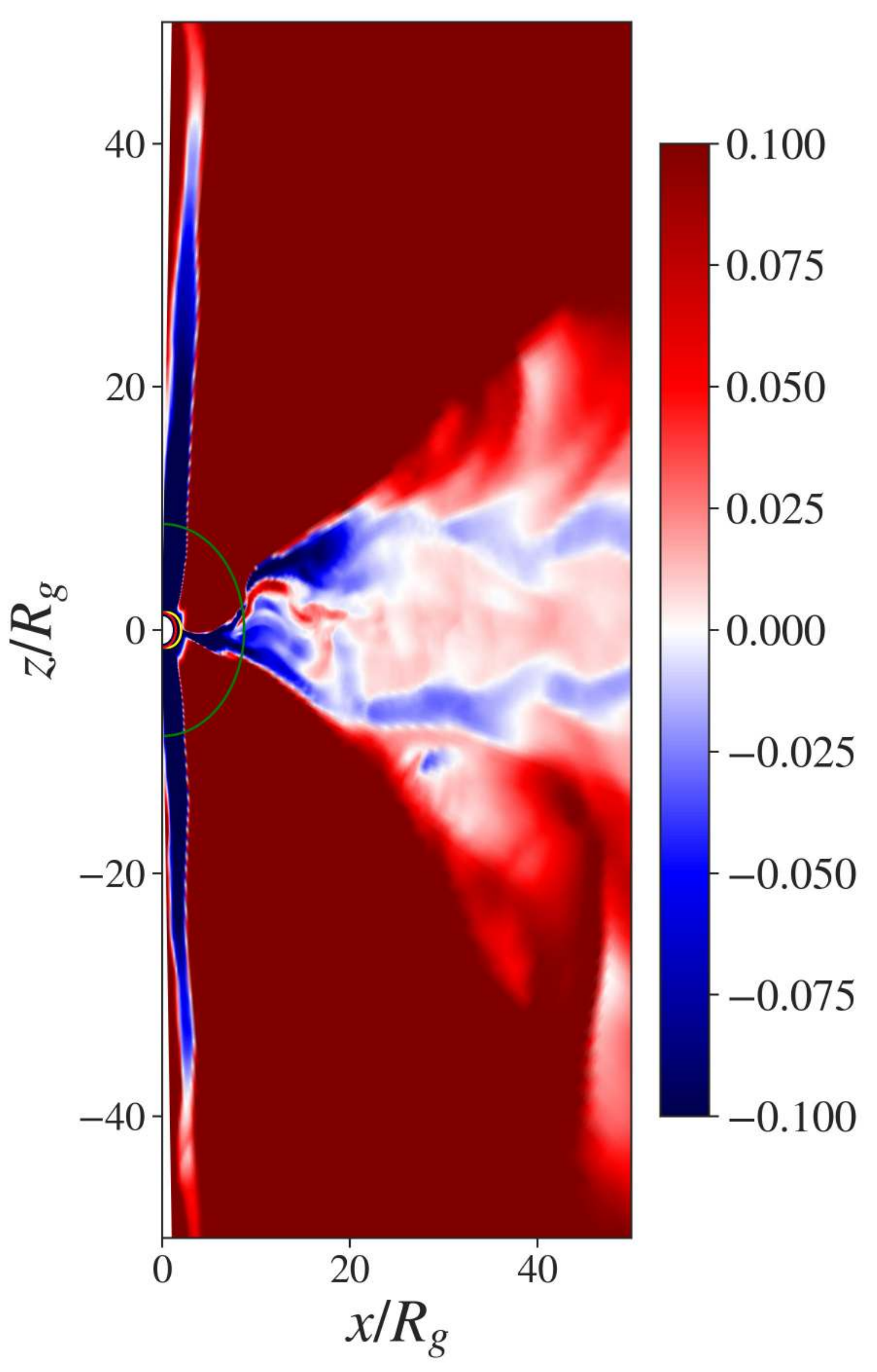}
     \caption{A counter-rotating black hole.
     Density (log scale, left) and radial velocity (right) for simulation
     {\em sim3} with $a=-0.9$ at $t=4000$.}
    \label{fig:compare_sim0-sim3_b}
\end{figure}

All simulations start with an initial setup with $B_{\phi}=0$.
However, by the rotation of the footpoints of the field lines (accretion disk or space time) a toroidal field is induced.
In the prograde simulations, the $B_{\phi}$ in the disk wind and the black hole magnetosphere have the same sign 
since both the disk and the black hole have rotate in the same direction.
At the equatorial plane $B_{\phi}$ changes sign (see Figure~\ref{fig:compare_sim0-sim3_a}, left), since the magnetic 
field lines are anchored at infinity.

In contrast, for the case of retrograde black hole rotation, simulation {\em sim3}, the $B_{\phi}$ in the black hole magnetosphere 
and in the outflow launched from there, is induced with the opposite sign compared to the disk wind 
(see Figure~\ref{fig:compare_sim0-sim3_a}, right),
resulting in another boundary layer with $B_{\phi} = 0$ to appear between the jet funnel and the disk wind.

While we expect (and find) the black hole driven outflow to have a different sign for negative Kerr parameter,
we would expect the disk wind to have $B_{\phi}$ with the same sign for positive and negative Kerr parameter,
again with $B_{\phi} = 0$ and a change of sign at the equatorial plane.
However, to our surprise, we find that in the disk area close to the inner disk radius, the $B_{\phi}$ changes sign 
three times (instead of only once,  see Figure~\ref{fig:compare_sim0-sim3_a}).
In fact, the $B_{\phi}$ in the wind above the disk surface is directed {\em opposite} to the $B_{\phi}$ below the 
disk surface\footnote{Of course similar for the upper and lower hemisphere respectively}. 
Along the disk surface $B_{\phi}=0$.

This also affects the poloidal component of the magnetic field (mainly the radial component) as it is visible purely 
from the shape of the field lines.
The change of sign in $B_{\phi}$ close to the equatorial plane is intrinsically connected to the type of accretion:
Figure~\ref{fig:compare_sim0-sim3_b} shows the radial velocity for simulation {\em sim3} and clearly indicates that inside 
the disk some material is moving {\em outwards}, while accretion happens along the surface layers of the disk.
For the case of prograde rotation, accretion is mainly along the equatorial plane.
This unexpected behaviour, however, does not affect the overall accretion rate.

For simulation {\em sim3} with $a=-0.9$ we find -- similar to the prograde case -- an outgoing Poynting flux, 
which is indicative of Blandford-Znajek launching.
The Poynting flux in the funnel area increases with time, with a time average value of 
$\langle \dot{E}_{\text{EM}} \rangle = 3 \times 10^{-3} $ at radius $r=100$.
For comparison, the Poynting flux at $r=100$ for the prograde simulation {\em sim0} is 
$\langle \dot{E}_{\text{EM}} \rangle = 4.89 \times 10^{-3} $.
Furthermore, the Poynting flux from the disk wind appears to be stronger than the one from the funnel having a time average of $\langle \dot{E}_{\text{EM}} \rangle = 4.52 \times 10^{-3} $ at $r=100$.
We do not find significant differences in the electromagnetic energy emitted within the funnel flow
between the prograde and retrograde simulations

It would have been interesting to follow the retrograde setup for longer time, but the simulation stopped 
at $t \approx 5500$,
most probably due the high mass loss and also the complex magnetic field and velocity structure.

Although we find for the retrograde black hole rotation a few remarkable and also unexpected features that can be 
astrophysically interesting, we do not want to over-interpret, as we think that the retrograde case is not likely 
realized in nature.
Retrograde black hole rotation may be realized by galaxy mergers with accompanied binary black hole mergers, but not from
pure disk accretion. Similarly, counter-rotating black hole-disk systems may be expected from specific initial conditions for neutron 
star mergers and thus may affect the subsequent gamma ray burst activity.

\subsection{Impact of magnetic diffusivity}
The magnetorotational instability is thought to be the main driver of turbulence in accretion disks \citep{BH1991, BH1998}.
The feasibility of the MRI has been demonstrated also in GR-MHD simulations \citep{Penna2010, McKinneyTchekh2012}.
Overall, turbulence results in a dissipative effect for the magnetic field which we express through a mean magnetic diffusivity,
in analogy to the $\alpha$-effect for turbulent viscosity \citep{SS1973}.

In contrast with ideal MHD, the disk material is now able to move {\em across} the magnetic field (lines) while
accreting towards the black hole.
The advection of magnetic flux is reduced due to the weaker coupling between magnetic field and mass.
It is thus worth investigating the effect of diffusivity on the accretion-ejection mechanism and the launching of outflows
and jets.
As described above, we have implemented a fixed in time and space background diffusivity that mainly follows the
disk structure (see Sect.~\ref{sec:diffusivity}).

In the following we focus on varying the strength of the disk magnetic diffusivity.
Further studies considering the scale height or the radial profile need to be done, as it has been worked out
for non-relativistic studies of jet-launching simulations (see  e.g. \citet{Somayeh2012, Stepanovs1}). 

We have run three further simulations, 
that are identical to our reference simulation but consider $\eta_0 = 10^{-2}$ ({\em sim4}),  $10^{-4}$ ({\em sim5}), 
and $10^{-10}$ ({\em sim6}), respectively (see Table~\ref{tab:simulations}).
We observed that a higher magnetic diffusivity stabilizes the simulation run, simulations {\em sim4} runs until $t=15000$.
Simulations with lower diffusivity levels were terminating earlier, however still providing enough information for a
comparison.

\begin{figure}
    \centering
    \includegraphics[width=0.98\columnwidth]{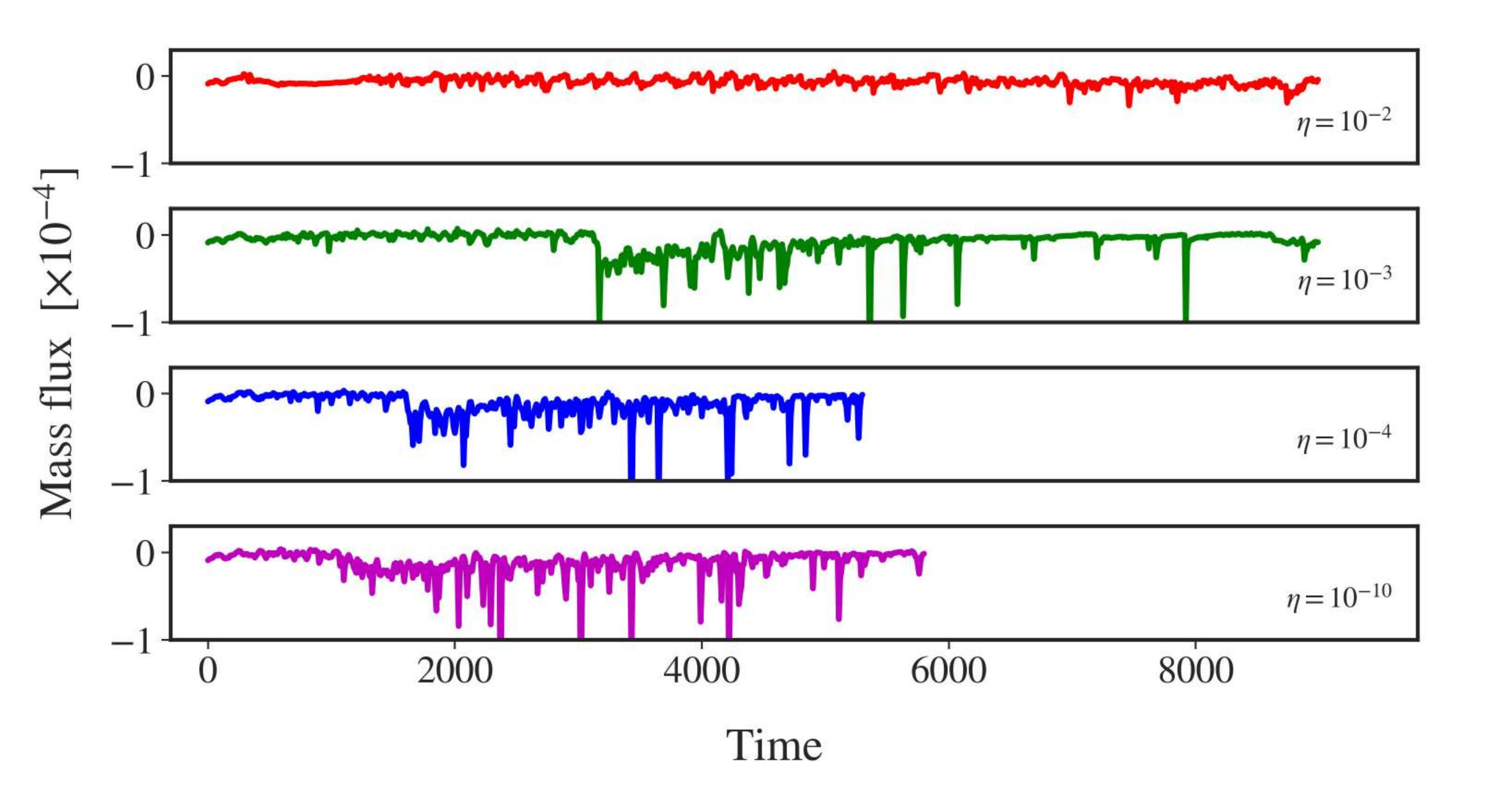}
     \caption{Accretion rate and resistivity.
     Comparison of the accretion rates measured at $r \approx 2$ for the simulation runs with 
     $\eta_0=10^{-2}$ ({\em sim4}, red), 
     $\eta_0=10^{-3}$ ({\em sim0}, green), 
     $\eta_0=10^{-4}$ ({\em sim5}, blue) and 
     $\eta_0=10^{-10}$ ({\em sim6}, magenta).}
    \label{fig:eta_compare_accretion}
\end{figure}

\begin{figure}
    \centering
    \includegraphics[width=0.98\columnwidth]{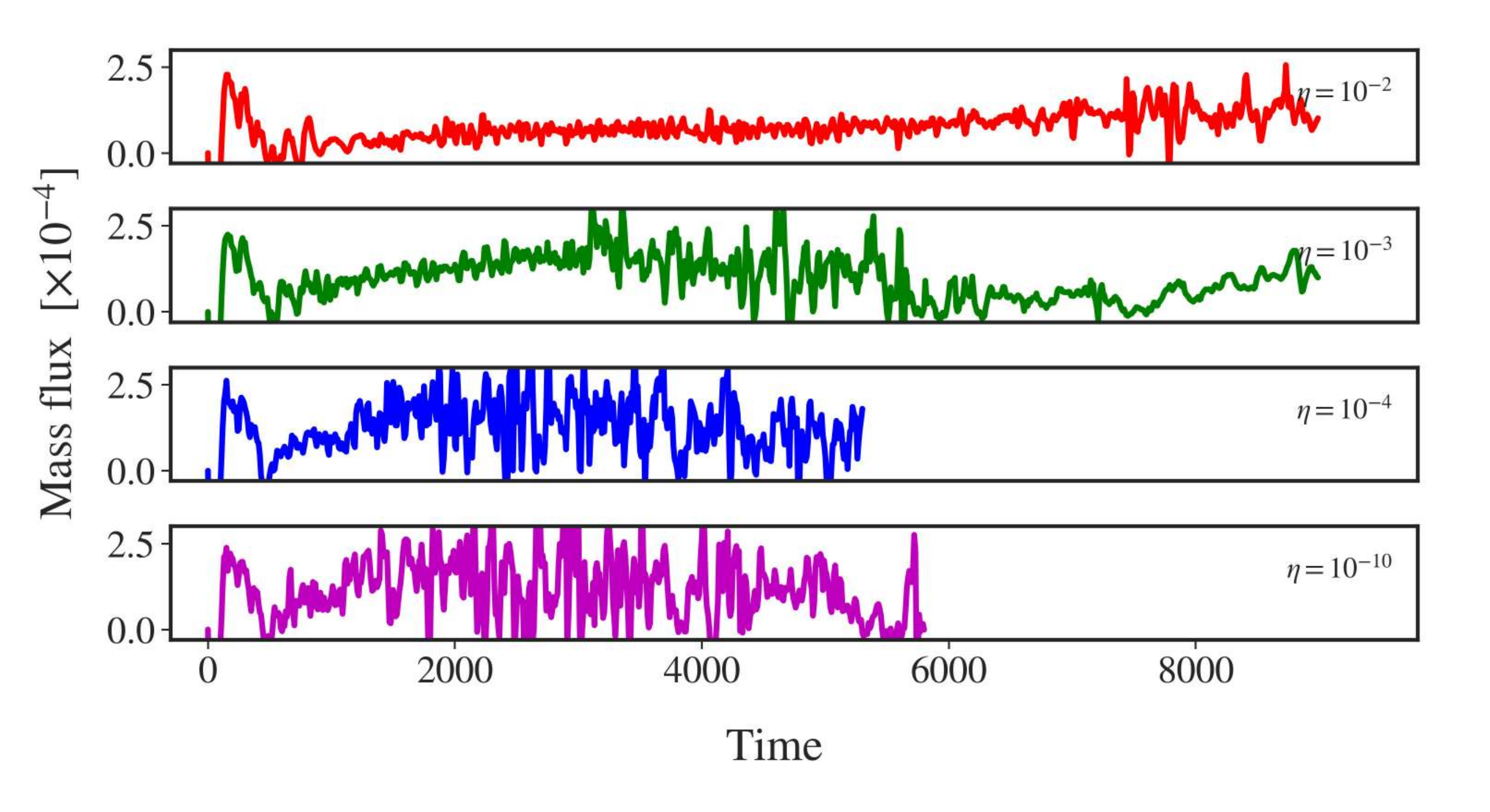}
     \caption{Ejection rate and resistivity.
     Comparison of the mass flux associated with the disk wind for the simulation runs with 
     $\eta_0=10^{-2}$ ({\em sim4}, red), 
     $\eta_0=10^{-3}$ ({\em sim0}, green), 
     $\eta_0=10^{-4}$ ({\em sim5}, blue) and 
     $\eta_0=10^{-10}$ ({\em sim6}, magenta), integrated along the surfaces of constant angle at $80^{\circ}$ and $100^{\circ}$.}
    \label{fig:eta_compare_diskwind}
\end{figure}
 
In Figure~\ref{fig:eta_compare_accretion} we compare the accretion rate at radius $r=2$ for different levels of magnetic 
diffusivity.
For simulation {\em sim4} with the highest level of diffusivity we notice an almost constant (in comparison with the 
other simulations) accretion rate without any spikes.
Still some spikes start appearing after $t=9000$ when we plot the long term accretion evolution of {\em sim4} even 
though the background accretion does not change much.
Overall, for this simulation we cannot identify the three phases of accretion rate we found in the reference simulation,
even with the longer simulation time.

For lower levels of diffusivity the evolution of the accretion rate has more similarities to simulation
{\em sim0}.
We identify similar phase changes as we detected in our reference simulation, however, unfortunately the simulations
stop before they reach a time scale that is comparable to that of the reference simulation.
Even in this case though, for {\em sim5} the second phase starts at $t \approx 1600$ while for {\em sim6} it 
starts at $t \approx 1100$, however it is not as clear as in the reference simulation.

For the vertical flux of the disk wind we observe a similar behaviour -- a larger disk wind mass flux resulting for lower levels
of diffusivity (see Figure~\ref{fig:eta_compare_diskwind}).
It therefore seems that high diffusivity reduces the efficiency for the magnetic field to a launch disk wind.
This is straightforward to understand and has been observed in non-relativistic simulations \citep{Somayeh2012}: For a magnetic driving
of outflows (Blandford-Payne or magnetic pressure-driven) a strong coupling between magnetic field and matter is essential.

For the radial mass flux we detect a different behaviour.
A high radial mass flux appears for the reference simulation with $\eta_0 = 0.001$, while for both higher and lower diffusivity levels
the mass flux decreases to approximately similar levels.
The area where we find the $B_{\phi}$ dominated wind has a lower diffusivity level than the equatorial plane, 
but for simulation {\em sim4} it is still significant enough to weaken the wind.
The area of the $B_{\phi}$-dominated wind increases with the increase of diffusivity. 

\begin{figure}
    \centering
    \includegraphics[width=0.98\columnwidth]{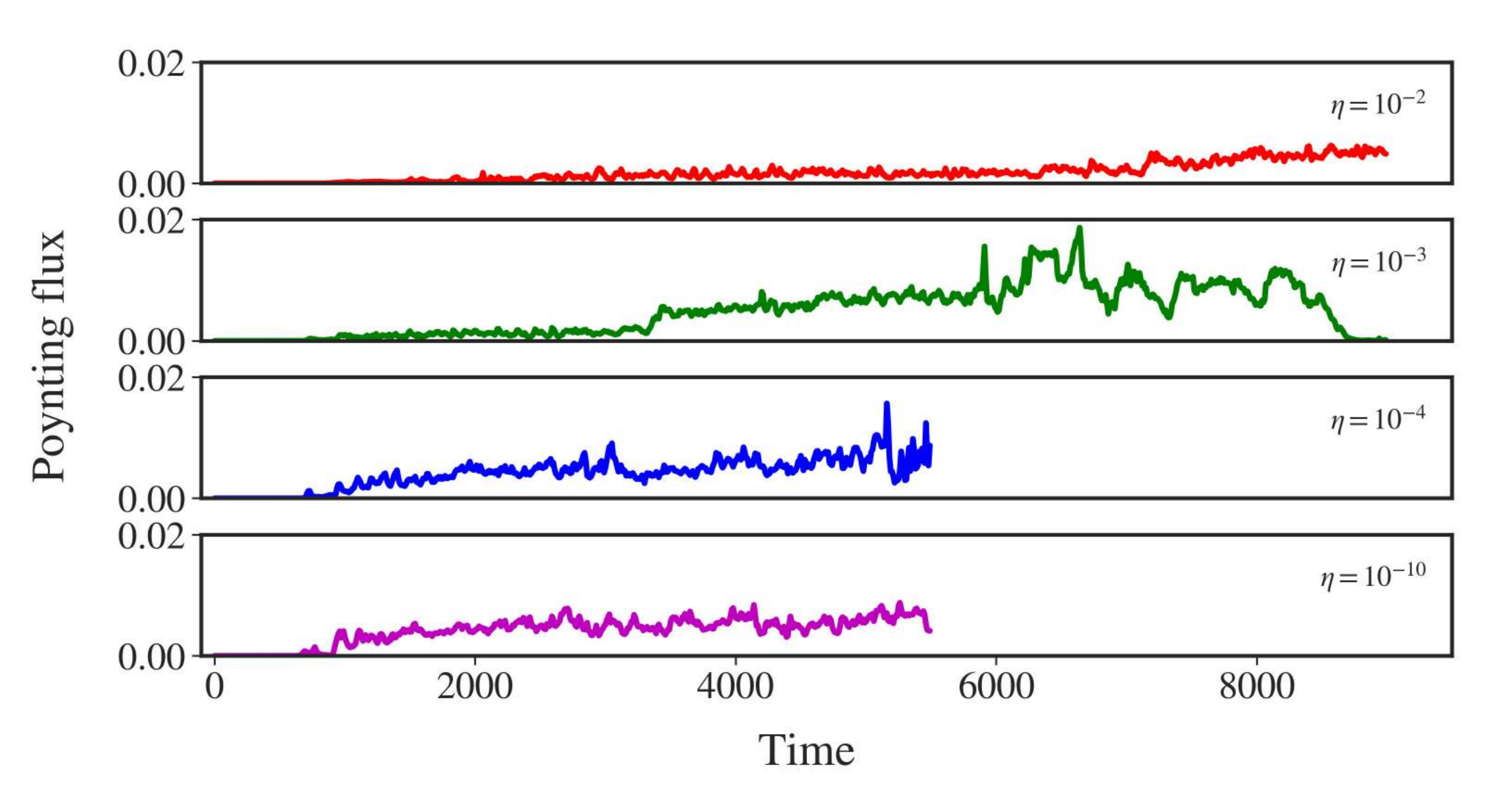}
     \caption{Poynting flux and resistivity. Comparison of the radial Poynting fluxes at $r=100$ in the jet funnel for simulation runs applying diffusivity 
     $\eta_0=10^{-2}$ ({\em sim4}, red), 
     $\eta_0=10^{-3}$ ({\em sim0}, green), 
     $\eta_0=10^{-4}$ ({\em sim5}, blue) and 
     $\eta_0=10^{-10}$ ({\em sim6}, magenta).}
    \label{fig:eta_compare_poynting}
\end{figure}

Finally, we investigate the Poynting fluxes for the different levels of diffusivity.
Figure~\ref{fig:eta_compare_poynting} shows the Poynting flux through the jet funnel at radius $r=100$ for various $\eta$.
The flux increases in time for all cases, however, comparing simulation {\em sim4} (largest $\eta$) with the reference simulation
the increase is much slower.
Simulations {\em sim5} and {\em sim6} show again very similar behaviour following the trend we observed in the accreting and vertical 
mass fluxes.
Also, in the case of {\em sim4} the flux from the disk wind is slightly stronger than the flux from the jet funnel.

The previous findings hint on preferred levels of diffusivity (or a preferred level of turbulence) that supports the launching of 
a disk wind.
For higher diffusivity, the coupling between matter and field may not be efficient enough for launching, while for lower levels of diffusivity
the mass loading becomes inefficient.

What is the mechanism behind these findings of a threshold value for the magnetic diffusivity of $\eta = 10^{-3} ... 10^{-2}$ where
the flow becomes smooth and never MAD-like?
We believe that is is the interplay between magnetic re-connection, magnetic diffusion and ohmic heating that governs the disk 
structure at these scales.
Magnetic re-connection destroys magnetic flux that is needed to launch strong outflows.
It also generates turbulence to the flow.
We would thus expect a high resistivity to weaken the outflow launching.
On the other hand a higher resistivity enables a more efficient mass loading of the outflow.
Thus a smaller resistivity would decrease the mass load of the outflow, but potentially may produce outflows with higher speed
(for the same magnetic flux available).
Ohmic heating of the launching area would in contrary increase the mass loading (in classic MHD steady-state theory 
the mass load is determined by the sound speed at the launching radius).

Overall, our simulations seem to follow these trends. 
For low resistivity, resistive mass loading becomes less efficient, assisted by low ohmic heating.
For high resistivity, re-connections weakens the outflow.
For a critical resistivity in-between, outflow launching becomes most efficient.

\section{A black hole shadow?}
\label{sec:EHT}
Motivated by the recent detection of a black hole shadow in the jet launching core of M87, we here discuss a 
few features of our simulation that can possibly interrelated with these new findings.
{ As we do not consider radiation in our simulations, we cannot provide emission maps of direct or lensed 
radiation.
However, we can estimate the opacities in our disk-outflow system and thus the visibility of the innermost central 
region around the black hole.
Obviously, if the black hole-surrounding medium is opaque, a black hole shadow cannot be seen from a distant observer.
This is interesting because the accretion structure and the metric depend on quantities that are not really known,
that is the black hole spin, the black hole mass, and the accretion rate.
Our question here is whether we could derive some general features that allow (or not) to observe a signal lensed into
the photon sphere as claimd for M87.
}

Considerably one of the most critical points in detecting a black hole shadow is the structure of the accretion 
flow close to the black hole.
Here, we expect to find differences when considering a (turbulent) magnetic diffusivity for accretion or ideal MHD.
One reason is that the coupling between matter and magnetic field is different.
The other reason is that physical resistivity allows for re-connection of the magnetic field.
This is a particular strong effect along the equatorial plane close to the horizon - the reason being that 
advection of large-scale magnetic flux from the disk that connects to the horizon, leading to a field reversal 
across the mid-plane.

\begin{figure*}
    \centering
    \includegraphics[width=0.66\columnwidth]{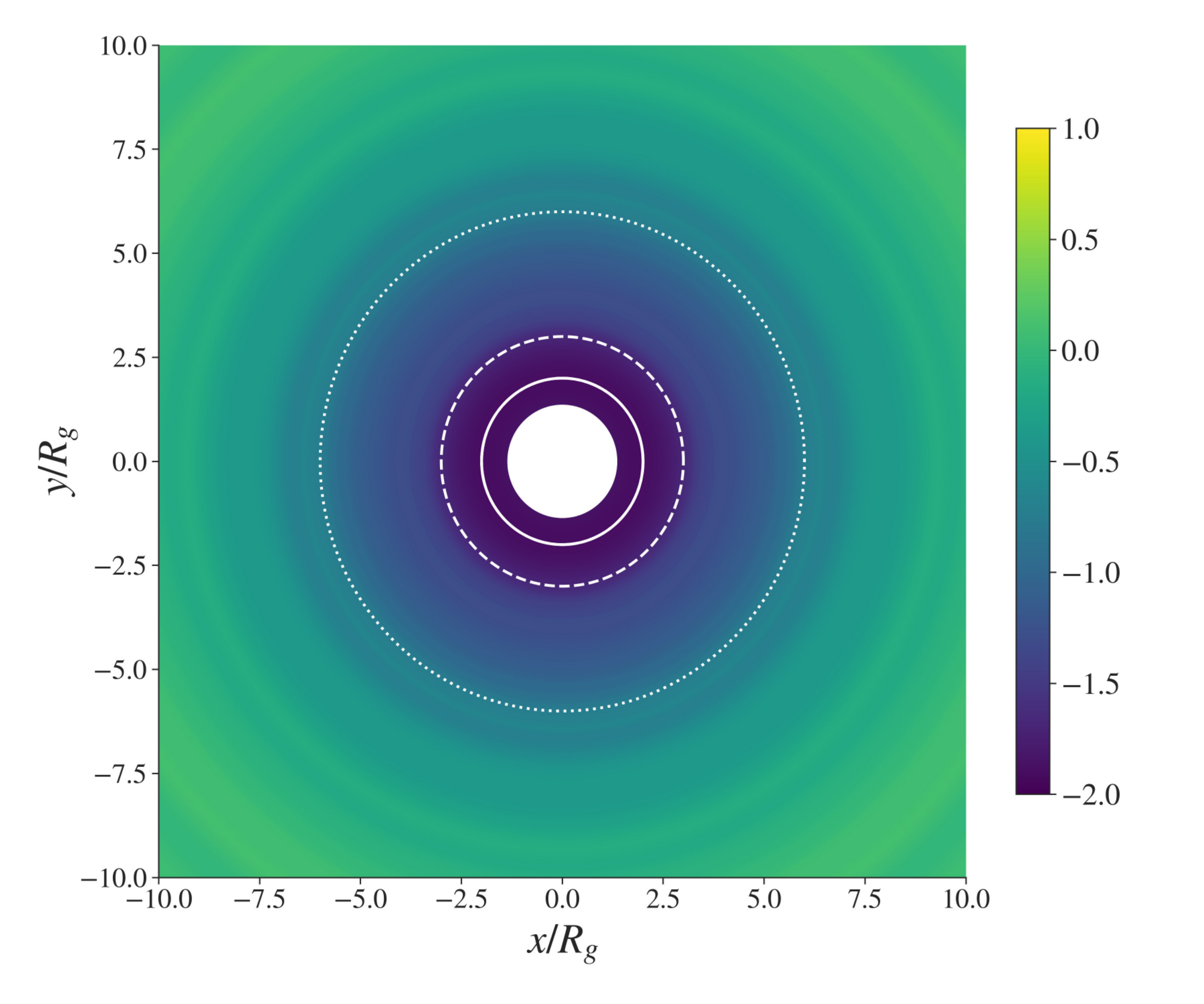}
    \includegraphics[width=0.66\columnwidth]{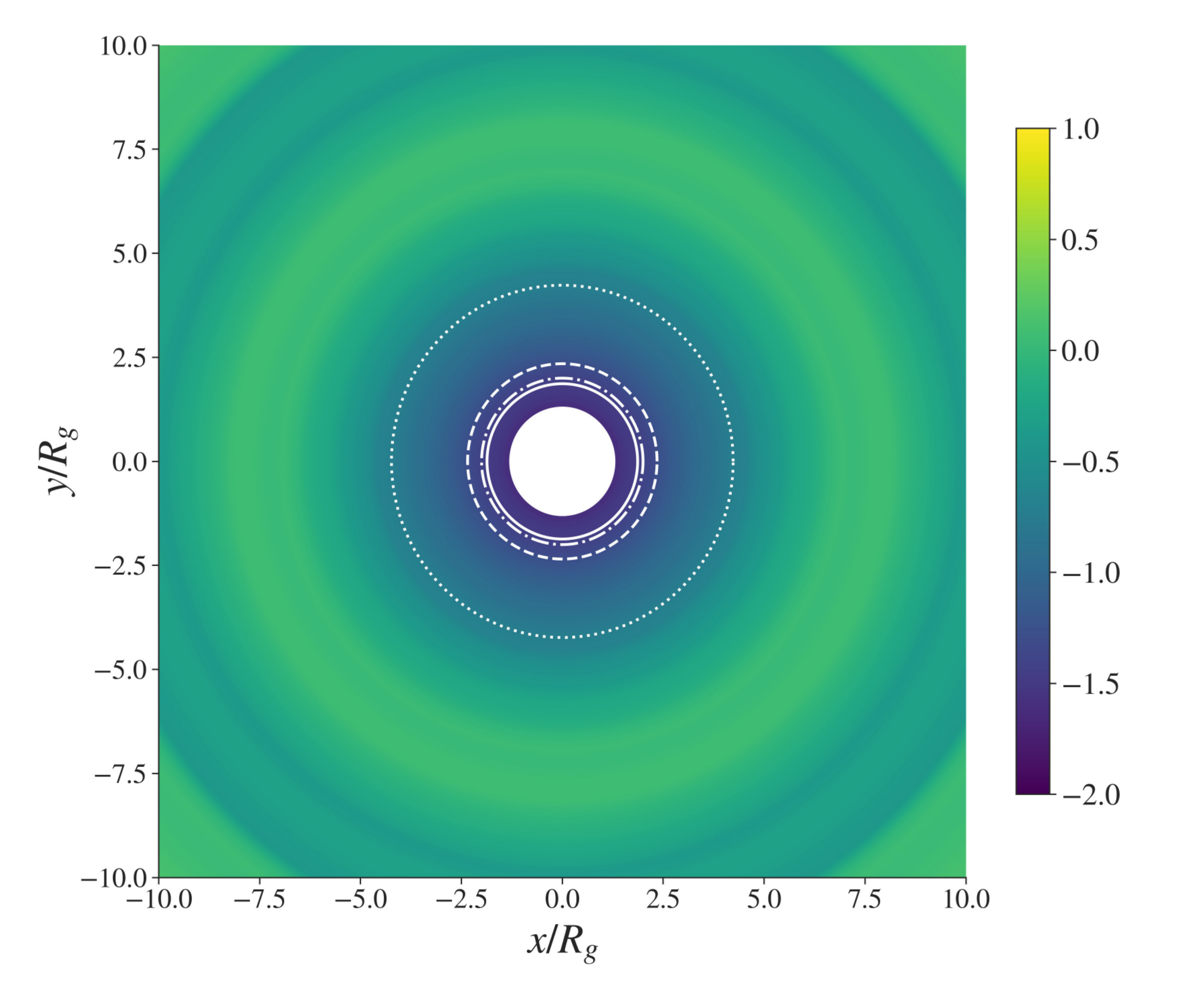}
    \includegraphics[width=0.66\columnwidth]{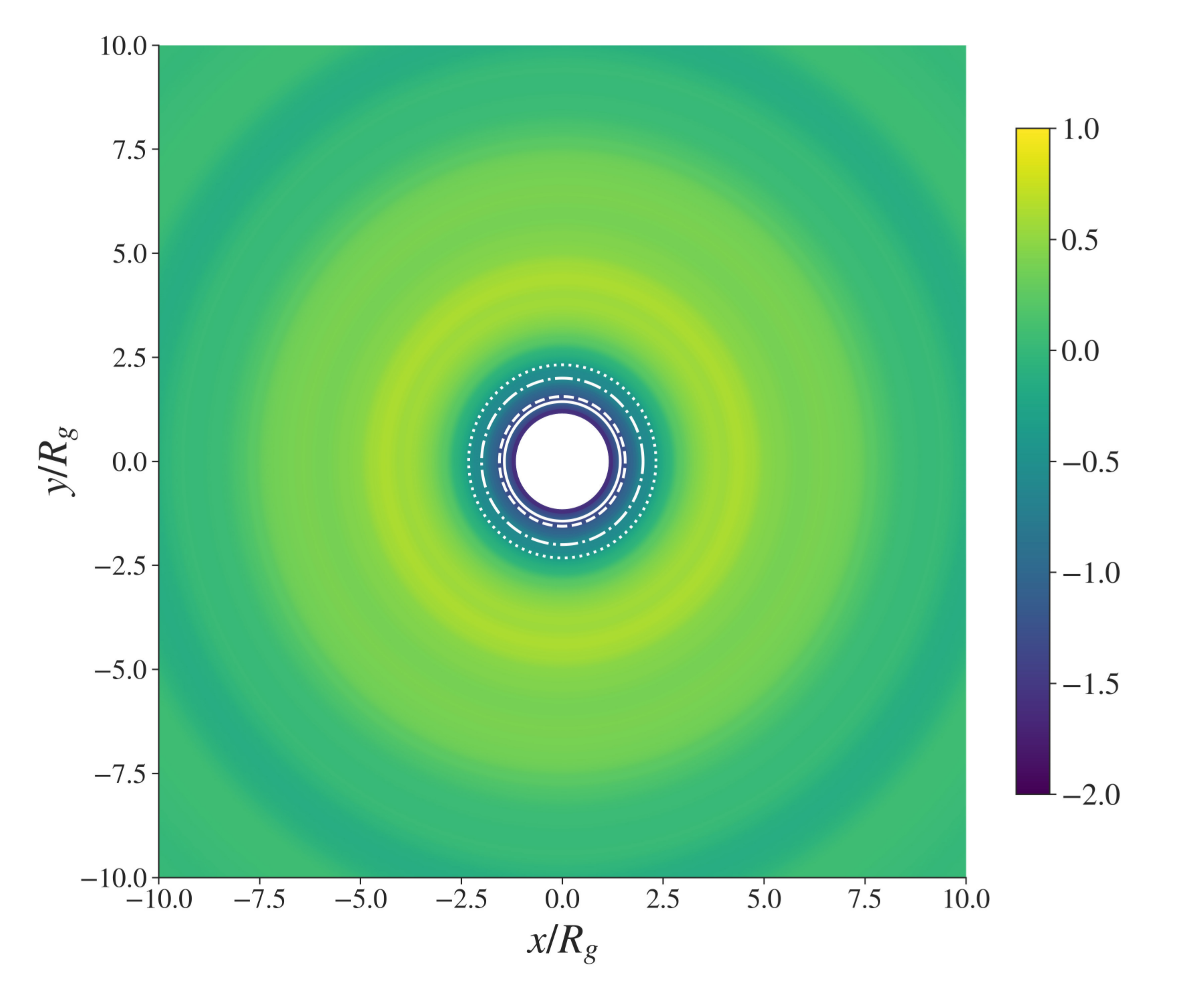}
     \caption{The surface density (log scale, in code units) at time $t=4000$ in the equatorial
     plane close to the horizon $R_{\text{g}}<10$ for three simulations with different Kerr parameters 
     ($a=0, 0.5, 0.9$ from left to right). 
    The four circles indicate the event horizon (solid line), the photon sphere (dashed line), 
    the ergosphere (dash-dotted line) and the ISCO (dotted line).}
    \label{fig:h20.3_sigma_4000}
\end{figure*}

We therefore calculate the surface density (in code units or gravitational units, denoted by the overline),
\begin{equation}
    \overline{\Sigma} = \int_{-\overline{h}(r)}^{\overline{h}(r)} \overline{\rho}(r) d\overline{s}
\end{equation}
along  the accretion stream connecting the inner disk and the horizon for our different models.
Figure~\ref{fig:h20.3_sigma_4000} shows the (axisymmetric) surface density distribution close to the horizon.
There is a clear trend towards higher surface densities (yellowish, in code units) for increasing black hole spin 
parameter when compared at a certain radius. 

For example at $r=4$ the surface density increases by a factor 100, when comparing spin parameters
of $a=0, 0.5, 0.9$, respectively (see Figure~\ref{fig:h20.3_sigma_4000}). 
The optical depth would differ by the same factor, if (!) assuming the same disk density scaling $\rho_0$ for all simulations.
Note, however, that for different spin also the metric changes and comparing physical variables at a fixed radius
is not necessarily meaningful.
For $a=0.9$ the ISCO is inside $r=4$ and the yellow structure in Figure~\ref{fig:h20.3_sigma_4000}
still resembles a rotating accretion disk, { while for a Schwarzschild black hole this radius is inside the ISCO}.

We may thus better compare the surface densities at the radius of the respective photon spheres (dashed line) 
where most of the lensed radiation originates.
Essentially, we find that $\overline{\sigma}$ is similarly small for all three spin parameters (blueish colors).
Also, the $\overline{\sigma}$ inside the the ISCO is small for all three cases shown
(see blueish colors inside the dotted circles).
That again implies that the photon orbit - and thus the photons lensed into this orbit - could be visible in all 
these cases, supposed that the optical depths are not extremely high (see below).
{ Note that here we do not consider radiation from the disk or the outflow, but only the visibility of a hypothetical
emission of photons that were lensed into the photon ring.
That photon ring would always be located within the radius of disk accretion and thus potentially in a low density area.}

{ In order to calculate the opacities of the disk-jet material, we need to know the physical densities.} 
However, since the disk mass in our simulations acts as test mass on the Kerr metric, we cannot specify the 
density in physical units without further assumptions.
Re-scaling our normalized accretion rate (in code units)\footnote{Note that the accretion rates given in the previous
figures $\propto 10^{-4}$ are normalized by the initial disk mass, in order to be able to compare different simulations} of
$\overline{\dot{M}} \simeq 0.1$ to astrophysical units assuming typical AGN accretion rates of 
\begin{equation}
    \dot{M} \equiv \rho_0  R_{\text g}^2  {\text c} \overline{\dot{M}} \simeq 0.01\,M_{\odot}\text{yr}^{-1},
\end{equation}
we can constrain the disk densities to
\begin{equation}
\rho_0 = 4\times 10^{-14}\frac{\text{g}}{\text{cm}^3}  \left[ \frac{ \dot{M} }{ 0.01\,M_{\odot}/\text{yr}} \right] 
                                    \left[ \frac{ M_{\text{BH}} }{ 5\times 10^9 M_{\odot} } \right]^{-2}
                                    \left[ \frac{ \overline{\dot{M} } }{ 0.1 } \right]^{-1}\!\!\!\!.
\end{equation}        
The re-scaled astrophysical density is $\rho = \rho_0 \bar{\rho}$ where $\bar{\rho}$ in code units follows from 
our simulations.
For comparison, for M87 the EHT collaboration derived an accretion rate of $\dot{M}\sim 2.7\times 10^{-3}\,M_{\sun}\rm yr^{-1}$ \citep{EHT2019e}
assuming a { black hole} mass of $M = 6.2\times10^9\,M_{\sun}$.
Earlier estimates suggested $\dot{M} \simeq (0.2-1)\times 10^{-3}\,M_{\sun}\rm yr^{-1}$ \citep{Feng2016}.

We may now estimate the optical depth of the disk material applying opacities for Thomson scattering
$\kappa(r) = \sigma_{\text T} n_{\text e}(r) =\sigma_{\text{T}} \rho(r) / m_{\text {p}} $.
We integrate vertically over a sufficiently large geometrical scale height $h(r)$ across the mid-plane,
and find the optical depth of the inner accretion stream
\begin{eqnarray}
\label{eq:tau}
\tau(r) & = & \int_{-h(r)}^{h(r)} \kappa(r) ds = 
              \frac{\sigma_{\text{T}}}{m_{\text {p}} \rho_0} R_{\text g} \int_{-h(r)}^{h(r)} \overline{\rho}(r) d\overline{s} \\
        & = & 0.1\, \overline{\Sigma}(r) 
              \left( \frac{ \rho_0 }{ 4\times10^{-14} \text{g\,cm}^{-3}} \right) 
              \left( \frac{M_{\text{BH}}}{5\times10^9 \text M_{\odot}}\right) \nonumber  \\
        & = & 0.1\, \bar{\Sigma}(r)
               \left( \frac{ \dot{M} }{0.01 M_{\odot}/\text{yr}} \right) 
               \left( \frac{ M_{\text{BH}} }{ 5\times10^9 \text M_{\odot} } \right)^{-1}
               \left( \frac{ \overline{\dot{M}} }{ 0.1 } \right)^{-1}, \nonumber
\end{eqnarray}
with $\overline{\Sigma}(r)$ again being the surface density in code units (see Figure~\ref{fig:h20.3_sigma_4000}).

Radiation that is lensed into the photon sphere could then be observed id the for accretion stream at this 
radius is optically thin, $\tau << 1$.
Taking the $\overline{\Sigma}\simeq 10^{-2}$ we find that for the area close to the photon orbit, the latter is 
actually the case for all three spin parameters and for the normalization used in Eq.~\ref{eq:tau}.
Considerably higher accretion rates, say $\dot{M} \simeq 10\,M_{\odot}/\text{yr}$ may lead to an opaque situation,
which is also the case for smaller black hole masses, say $M \simeq 10^8 \text M_{\odot}$.

How the image of the very central region actually looks alike, is beyond the scope of our paper as we 
do not consider the radiation from the gas in our simulations.
So, what we see as a large ''hole" in the surface density towards the center of the Schwarzschild simulation 
(see Figure~\ref{fig:h20.3_sigma_4000}, left),
may actually be bright due to radiation from hot gas falling towards the horizon.
How that emission would compare to the lensed signal at the photon-sphere we cannot tell.
Simulations by the EHT collaboration suggest that the signal from the photon-sphere is dominating.

\begin{figure}
    \centering
     \includegraphics[width=0.9\columnwidth]{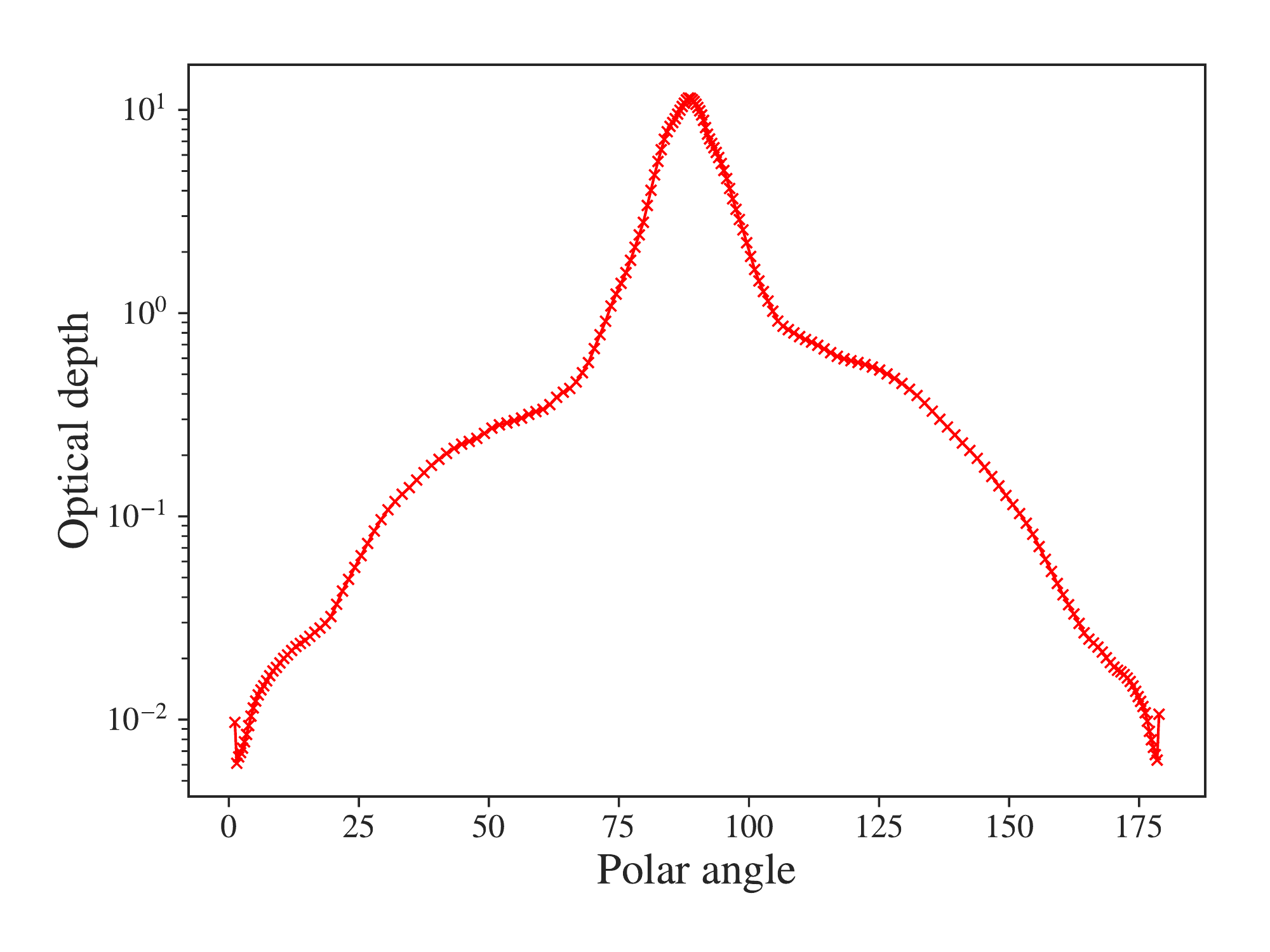}
     \caption{The optical depth Eq.~\ref{eq:tau} towards the central area of the reference simulation,
     integrated along radius vectors (assumed l.o.s.) 
     for all polar angles at time $t=4000$. The integration ranges from $r=0$ to $r=1000$.
     For the example of M87 the optical depth has to multiplied by a factor of $1/3$ considering
     the lower accretion rate and the higher central mass as compared to Eq.~\ref{eq:tau}.}
    \label{fig:h20.3_tau_4000}
\end{figure}

A remaining question is the visibility of the innermost region when observed from high inclination, meaning along
viewing angles close to the rotational axis.
{ In case of M87 the l.o.s. is about $20\degr$.}
In our models, as well as in almost all simulations in the literature, this is the area of the funnel flow of 
low (floor) density.
For our reference simulation, when integrating for the surface density along radial directions we find
the distribution shown in Figure~\ref{fig:h20.3_tau_4000}.

We find that when implying the same normalization as above, the optical depth along the line of 
sight close to rotational axis is low.
As expected, the disk material blocks the radiation along light paths close to the equatorial plane.
Interestingly, we also see that the disk wind clearly contributes to the opacity, and, may thus, depending on the 
physical density (meaning accretion or ejection rate), also block the view towards the central black hole.

Overall, we find that for reasonable disk densities, $\rho_0 = 10^{-14}{\text{g}} {\text{cm}^{-3}}$, 
the l.o.s. towards the central black hole and its shadow is not blocked by the outflow material for viewing angles 
below $25^\circ$ to the rotational axis.
For a l.o.s. inclination larger than $25^\circ$ a massive disk wind blocks the view towards the center where $\tau(\theta)>1$.

\section{Summary}
\label{sec:summary}
In this paper, we have extended the newly developed resistive GR-MHD code \texttt{rHARM} \citep{QQ1} 
to the parallel version \texttt{HARM3D} in order to apply our models of jet launching from thin accretion disks surrounding 
a black hole to longer time scales and larger spatial scales, also considering a higher numerical resolution.

In our model the disk is threaded by inclined open poloidal field lines.
In general our simulation results demonstrate how the magnetic field strength, the disk magnetic diffusivity, and 
the black hole spin influence the MHD launching of disk winds and the Blandford-Znajek jet from the black hole.
Essentially we are able to compare the strength and power of both jet components for different Kerr parameters.
In the following we summarize our results.

(1) Our implementation of resistivity is based on \citet{QQ1} following \citet{BdZ2013}.
We tested the code by simulating the decay of the magnetic field 
and comparing the evolution with the analytic solution. 
We find a perfect match for diffusivities \lax$10^{-2}$.
A further test applied a resistive version of the classical shock tube problem.

(2) As a test for the model setup we run a set of GR-MHD simulations in the mildly-relativistic limit.
Here, only the disk evolution was simulated with the inner grid boundary far from the black hole.
Strong disk outflow were found, similar to the magneto-centrifugally driven outflows observed in the non-relativistic 
simulations in the literature. 

(3) As a reference simulation we applied the code for a setup considering a black hole with Kerr parameter $a=0.9$
together with a disk magnetic diffusivity profile that follows a Gaussian distribution.
We have investigated the physics of the accretion-ejection mechanism between the disk and the launched wind while focusing
somewhat on the nature of the outflows and the development of the disk wind.

(4) We provide a detailed study of the MHD characteristics of the disk-wind structure. 
A thin disk exists until accretion and disk wind have depleted the initial mass reservoir of the disk.
We resolve the disk surface were accretion of material is turned into ejection.
The Alfv\'en surface of the disk wind is close to the disk surface - the disk wind is thus launched with sub-Alfv\'enic
speed, but quickly accelerated to super-Alfv\'enic velocities.
The counter-rotating disk seems to develop a different accretion mode with layered accretion in the upper disk levels.

(5) Two different types of disk winds were identified.
The first one arises from the inner part of the disk $r \lesssim 10 $ and is dominated by the toroidal magnetic field component,
while carrying a large part of the mass flux.
This type of disk wind has many similarities with the wind investigated by \citet{QQ2} where it was identified as a tower jet \citep{LBell1996}. 
In contrast to \citet{QQ2} we observe a second type of disk wind.
This feature is launched from the larger radii, and is dominated by the poloidal magnetic field.
So far we believe that this is mainly due to the fact that the outer disk is less evolved in comparison with the 
inner part.
The $B_{\phi}$-dominated disk wind shows higher radial mass flux even though it is not as highly magnetized.

(6) We compare the accretion rates for different black hole spin parameters.
For the same level of magnetic diffusivity ($\eta_0 = 0.001$), we find for increasing spin the accretion rate decreases 
close to the horizon.
At the same time, the accretion rate increases, and with it the mass flux of the launched disk wind in both the 
polar direction (launching) and radial direction (acceleration) increase as well.
This result comes in contrast with previous works \citep{QQ2} where the connection between accretion and disk wind was
much stronger.

(7) We compare the accretion rates for different levels of magnetic diffusivity.
For the same black hole spin we find that increasing diffusivity lowers the accretion rate, and results in a decrease in the mass flux of the disk wind launched from the disk surface.
The radial mass fluxes show only small differences that do not allow us to say beyond any doubt if they are affected by 
the changes in diffusivity.
Definitely, a weaker coupling between matter and magnetic field, induced by the increase in magnetic diffusivity, affects both accretion rate 
and mass loading of the wind in a similar way.

(8) The electromagnetic energy flux that is carried by different parts of the outflow is dominated by the flux of the jet funnel.
This flux in the jet funnel is highly affected by the black hole rotation as this part of the outflow driven by the BZ mechanism.
We find that the disk and the Poynting-dominated outflows are strongly connected as the level of magnetic diffusivity does
affect the electromagnetic flux in the jet as well -- in spite of the fact that the diffusivity close to the horizon is negligible.
Similar to the peak in the mass fluxes for the disk wind, the Poynting flux reaches a peak value for $\eta_0 = 10^{-3}$.
We believe that this critical level for the resistivity is a result of the interplay between re-connection decreasing the 
magnetic flux launching the outflow and magnetic diffusion and ohmic heating, both increasing the mass flux. 

(9) The simulation of a counter-rotating black hole revealed an interesting feature.
The retrograde rotation induces additional field reversals in the toroidal component of the magnetic field in the inner disk area.
In this case, the accretion is supported mainly from the surface material of the inner disk area (where $B_{\phi}=0$), 
though without significantly affecting the accretion rate itself.

(10) Motivated by the recent discovery of the M87 black hole shadow, we calculate the optical depth of the inner-most accretion flow 
and the outflow structure around it.
We find that for high accretion rates, $\dot{M} \simeq 10\,M_{\odot}/\text{yr}$ for a black hole mass of 
$M \simeq 10^9 \text M_{\odot}$, the innermost accretion stream may be opaque for the lensed signal, while the 
jet-outflow launched from disk and black hole will most probably remain optically thin.

In summary, we have compared the efficiency of GR-MHD jet launching for a sample of combinations of the accretion disk magnetic
diffusivity and the black hole spin.
We find a substantial mass loading of the disk wind that accelerates up to $0.1\,c$.
The low-density high-velocity jet funnel generated by the BZ mechanism can be affected by the resistive, turbulent environment
of the accretion disk.
The two components of the disk wind follow the same trend even though its strength is not suppressed by a high disk 
diffusivity but continuously supported by mass loading. 
Besides magnetic diffusivity, also re-connection and ohmic heating govern the strength of the disk wind.

\acknowledgements
{We acknowledge a number of helpful suggestions by an anonymous referee.
C.V. is thankful for financial support by the International Max Planck Research School for Astronomy and Cosmic Physics
at the University of Heidelberg (IMPRS-HD).}

\appendix

\section{Test simulations considering magnetic diffusivity}
\label{sec:tests}
In order to verify the implementation of magnetic diffusivity into \texttt{HARM3D},
we have performed two test simulations.
Our tests are similar to those applied by \citet{QQ1}.
In the first test we follow the diffusion of a parallel magnetic field in a rectangular box through 
for different strength of the magnetic diffusivity and compare it with the time-dependent analytic 
solution to the diffusion equation. 
The second test problem is a classic shock tube that allows us in addition to check how magnetic diffusivity 
affects the shock capturing abilities of the code.

\subsection{Diffusive decay of a vertical field}
\label{sec:box}
The setup for the simulations treating the diffusive decay of a vertical field considers a hydro-static gas 
distribution located in an almost rectangular box that is threaded by a weak magnetic field. 
A uniform magnetic diffusivity is applied for the whole box and is the only parameter affecting the 
magnetic field evolution. 
Applying different levels of magnetic diffusivity we compere the simulated evolution of the magnetic field with 
the analytic solution.
As in \citet{QQ1} we find an almost perfect match.

\subsubsection{Numerical setup}
The box simulations are performed in a $256^2$ grid in a small sector of our domain space, along the equatorial plane, 
extending by $\Delta r$ in radius and $\Delta \theta$ in latitude.
By choosing a large enough radius $r_0$ to place the box ($\Delta r << r_0$), we establish that its shape is as 
close as possible to a perfect square, with a side length $r\in [r_{0}-\Delta r/2, r_{0}+\Delta r/2]$, and a 
latitude $\Delta \theta$ that corresponds to a z-direction side $\Delta z$, where $z = r \sin(\pi/2 - \theta)$ and
$\theta \in [\pi/2-\Delta \theta/2, \pi/2+\Delta \theta /2]$.

A relativistic gas is applied in the area of the box with a polytropic index of $\gamma = 4/3$. 
The gas is in hydro-static equilibrium with radial profiles of density,
$\rho (r) = C \; r^{\alpha}$
and pressure
$p(r) = \beta \; \rho^{\gamma_{\rm G}}$,
where $\alpha=1/(1-\gamma_{\rm G})$, $\beta=1/(1-\alpha)$ and $C$ denotes a proper normalization constant. 
This profile balances the gravitational force at large distances from the black hole (where GR effects are negligible). 
The magnetic field is uniform in $\theta$-direction.
In $r$-direction it follows a (time dependant) Gaussian profile, 
\begin{equation}
B_{\theta}(r,\tilde{t})=\frac{1}{\sqrt{\,\tilde{t}\,}} \exp\left(-\frac{ (r-r_{0})^{2} }{ 4 \eta \tilde{t} }\right).
\label{eq:B_theta}
\end{equation}
We apply a very high plasma-$\beta$, $\beta_{HM} = 10^8$ in order to establish a weak magnetic field
that does not initiate any advection of magnetic flux. 
The time variable $\tilde{t}=t_{0}+t$ is connected with the code running time $t$ with the parameter $t_0$ which 
basically normalizes the Gaussian profile. 
Finally, we are using outflow boundary conditions in all four boundaries of the box.

\subsubsection{Simulation runs}
\begin{figure*}
    \centering
    \includegraphics[width=8.8cm]{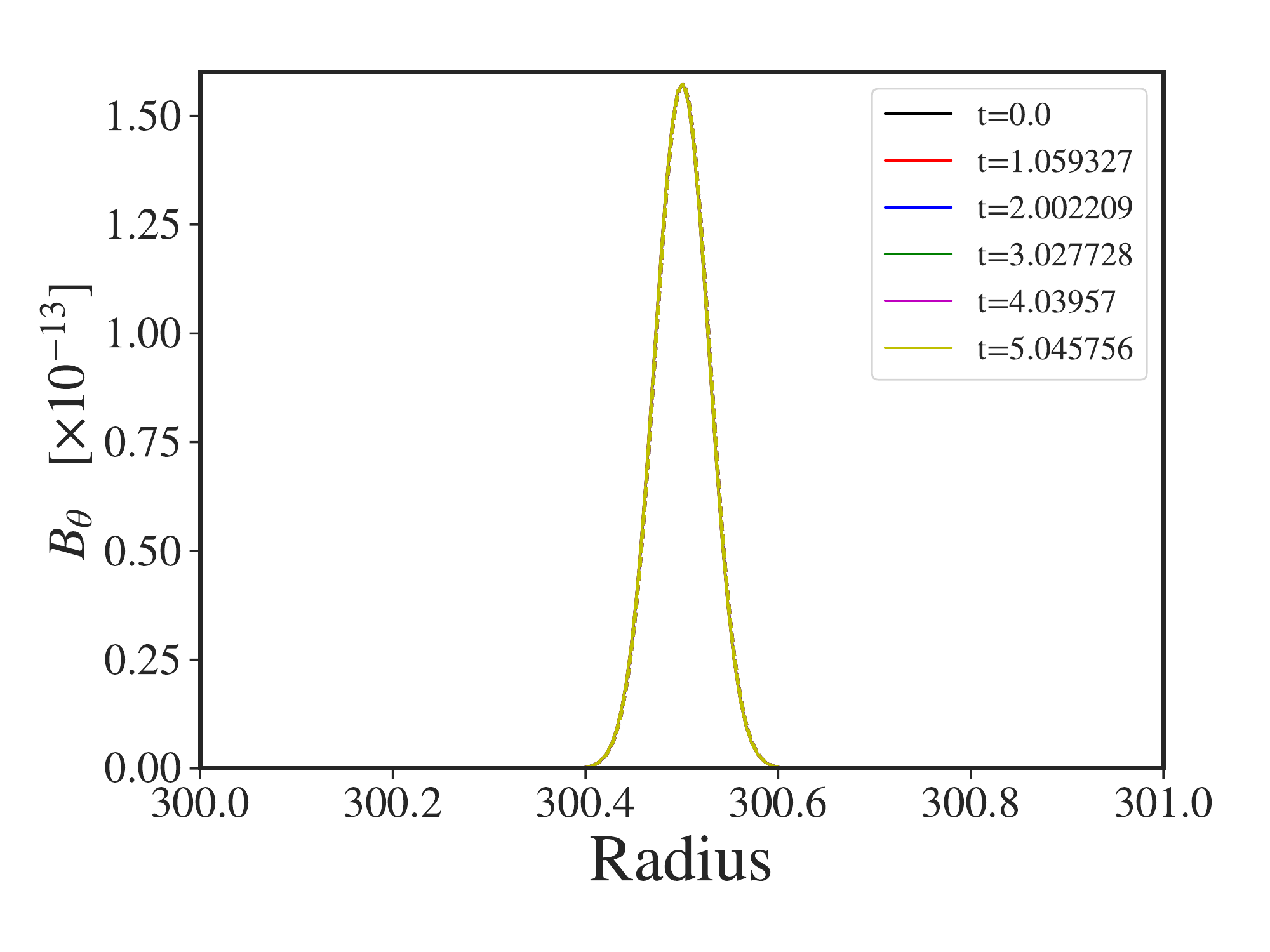}
    \includegraphics[width=8.8cm]{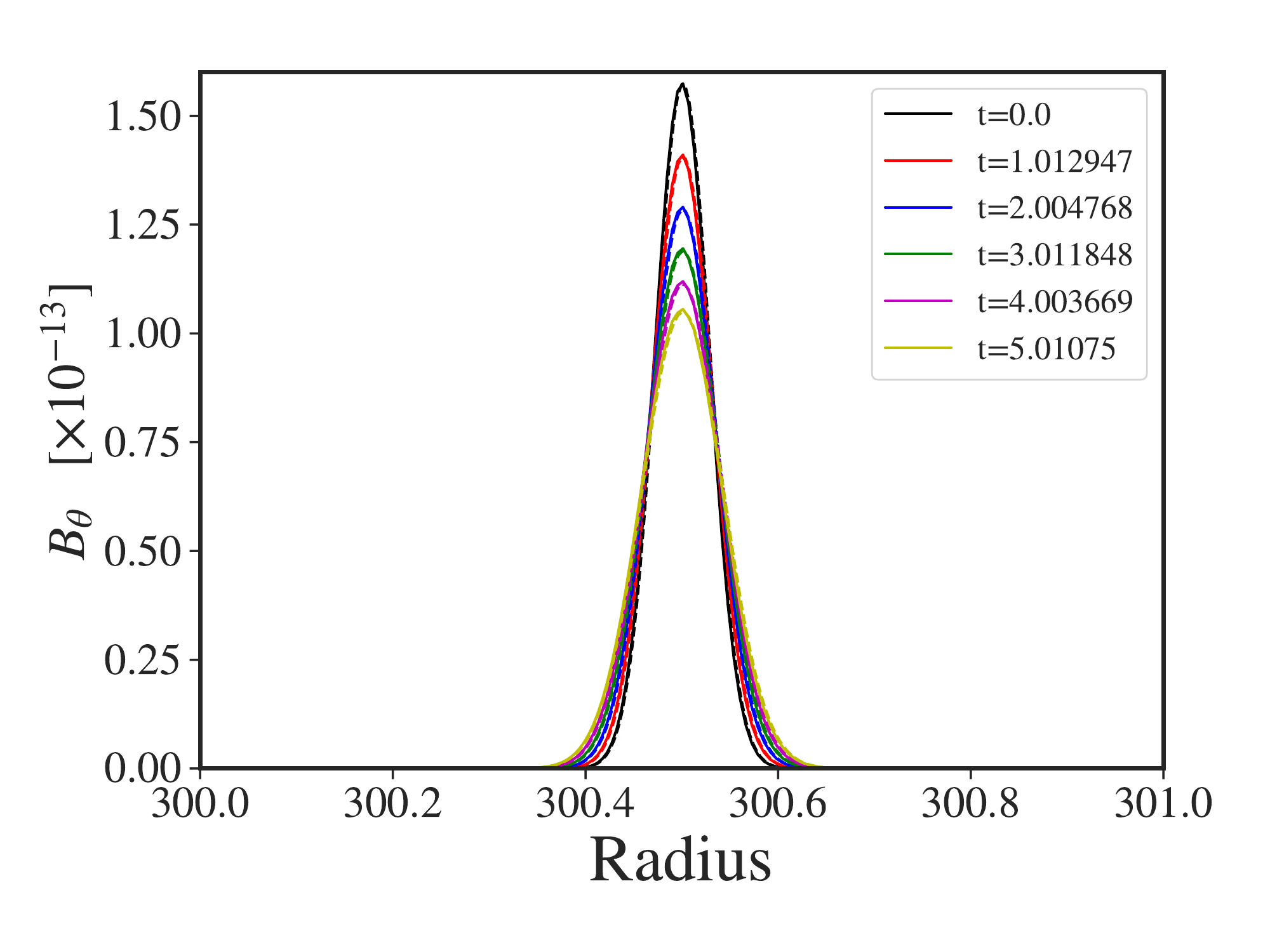}
    \includegraphics[width=8.8cm]{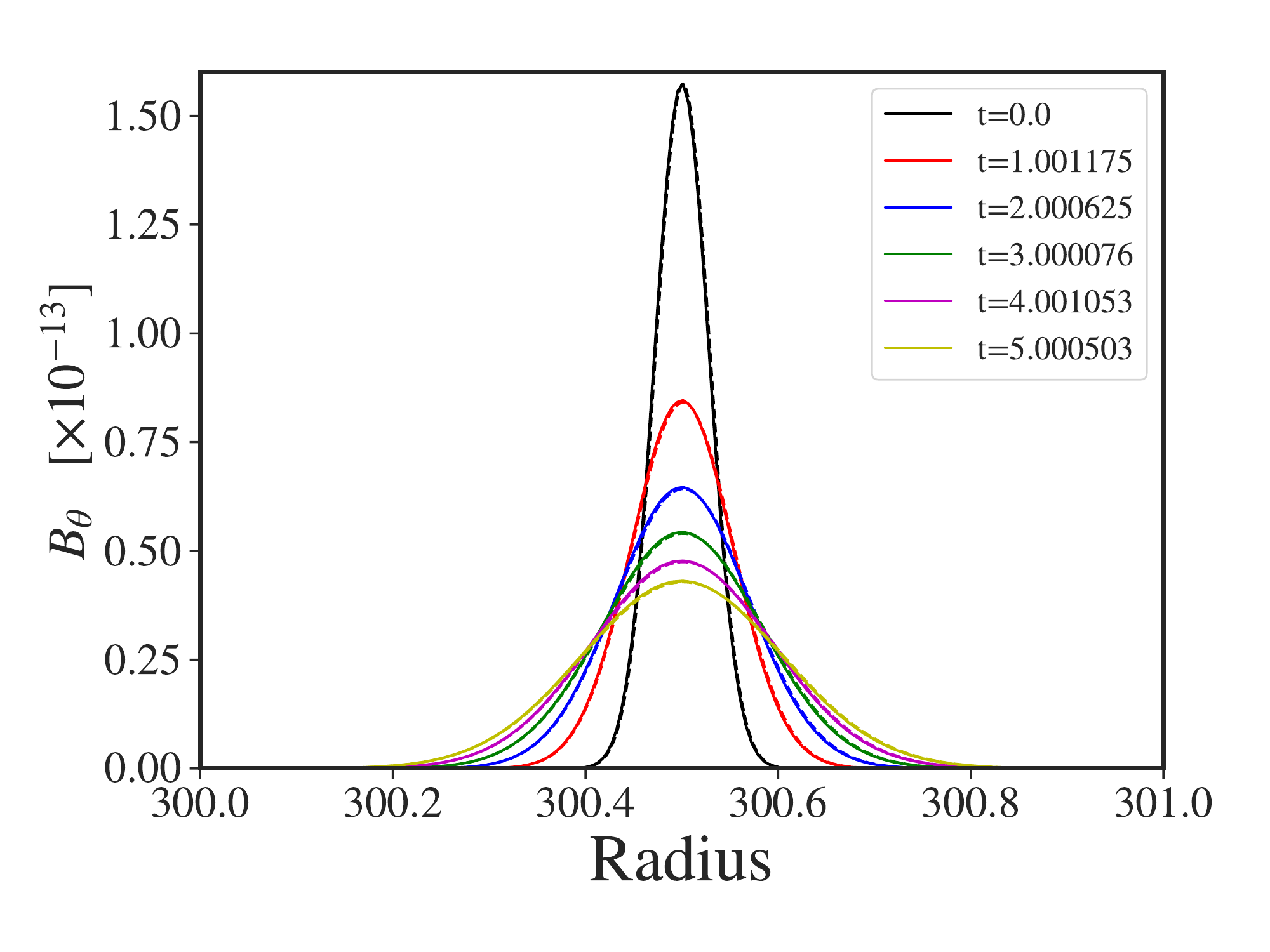}
    \includegraphics[width=8.8cm]{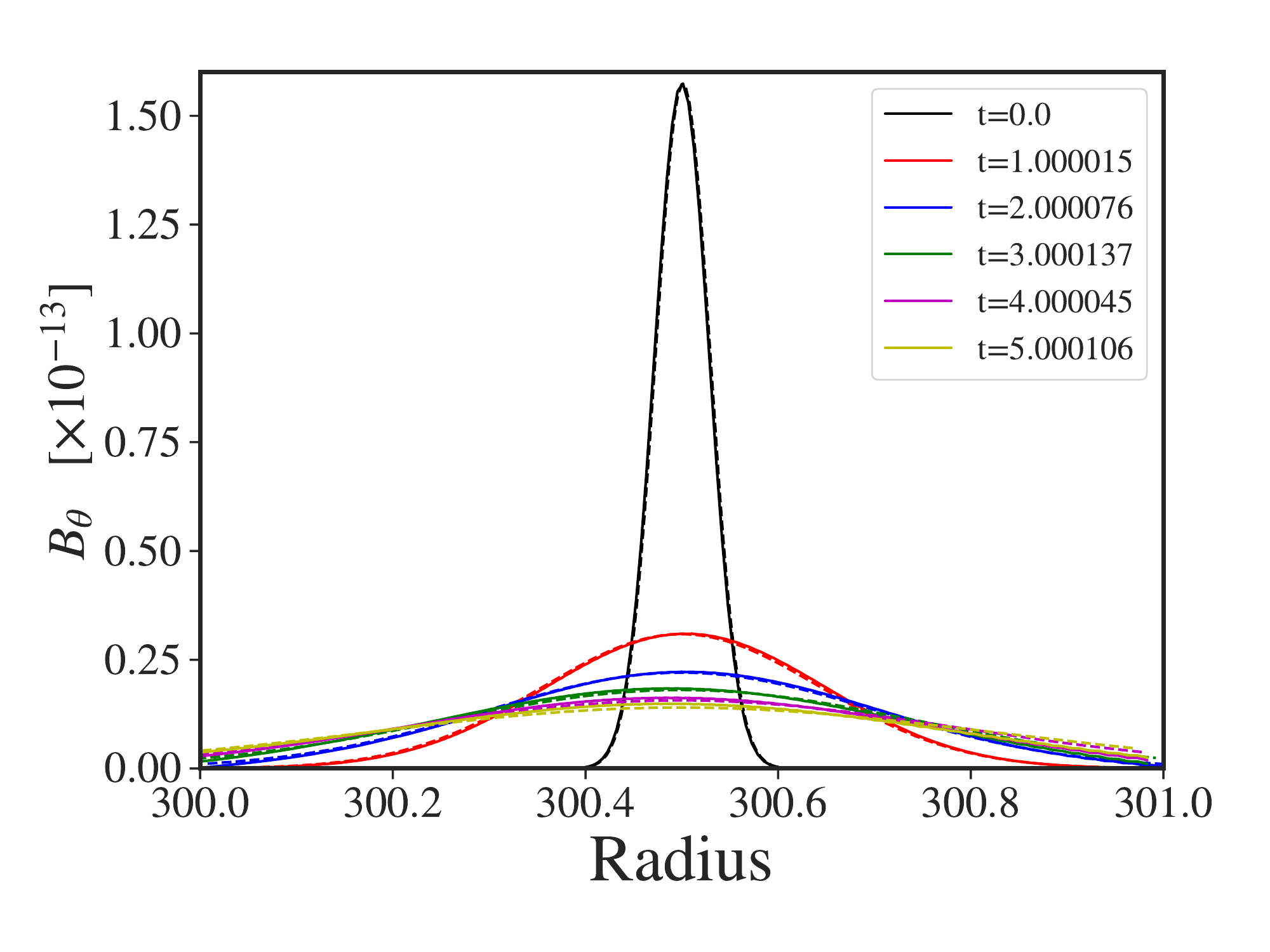}
    \caption{Diffusive decay of a vertical magnetic field.
    Evolution of the $\theta$-component of the magnetic for simulation runs applying four 
    values for the magnetic diffusivity,
    $\eta=10^{-12}$ (upper left), 
    $\eta=10^{-4}$ (upper right), 
    $\eta=10^{-3}$ (lower left), and 
    {\em box2} with 
    $\eta=10^{-2}$ (lower right). 
    Each color represents a different time step $t$ in the simulation. 
    Solid lines show simulation results while dashed lines show the analytical solution.  }
    \label{fig:box1}
\end{figure*}

We have placed the simulation box far from the black hole at a radius $r_0 = 300.5$ with a 
side length of $\Delta r = 1 \approx \Delta z $.
At this distance the shape of our box is quite close to square as $\Delta r << r_0$.
We follow the magnetic field evolution as given initially by Equation \ref{eq:B_theta} along the equatorial plane.
We run a series of simulations for different strength of magnetic diffusivity. 
In Figure \ref{fig:box1} we compare the simulation results (solid lines) with the analytic solution (dashed lines). 

For simulation {\em box12} with $\eta=10^{-12}$ there is barely any change in the magnetic field distribution 
and the simulation perfectly matches the ideal MHD limit.
Note that for the very high resolution applied in these simulations, also the numerical diffusivity
is low\footnote{See \citet{QQ1} for an assessment of the numerical diffusivity of \texttt{HARM-2D}}.
As we increase the magnetic diffusivity ({\em box4, box3, box2}) to the values of $\eta=10^{-2}$, the magnetic field 
decays - faster for higher diffusivity.
Overall, the initial field distribution decays following exactly the analytical solution.

However, for high levels of the magnetic diffusivity, $\eta>0.1$, the code fails. 

In this case the magnetic field has completely lost it's initial Gaussian distribution which poses a limit in the values of diffusivity we are allowed to use in our simulation.

\subsection{Diffusive shock tube test}
\label{sec:shocktube}
Following \citet{QQ1} we perform a series of tests with our resistive code based on the classical 1D shock tube test 
that demonstrates the shock capturing capability of the code. 
We employ a computational domain that extends for $x\in [298.75, 302.25]$ in the limit of Minkowski space-time using Cartesian
coordinates with 4000 cells to reduce the effect of numerical diffusion. 
The initial condition of the test follows the setup of \citet{DumbserZanotti2009} and \citet{BdZ2013}. 
We implement a discontinuity in the density of the gas, in the gas velocities and in the magnetic field, thus

\begin{equation}
\begin{split}
    (\rho, &p, v^x, v^y, v^z, B^x, B^y, B^z) = \\
            &(1.08, 0.95, 0.4, 0.3, 0.2, 2.0, 0.3, 0.3)
\end{split}
\label{tube:init1}
\end{equation}
for $x<300.5$, and
\begin{equation}
\begin{split}
    (\rho, &p, v^x, v^y, v^z, B^x, B^y, B^z) = \\
            &(1.0, 1.0, -0.45, -0.2, 0.2, 2.0, -0.7, 0.5)
\end{split}
\label{tube:init2}
\end{equation}
for $x>300.5$,
The initial electric field is set to the ideal MHD value. 
The boundary condition at the ends of the tube is fixed to the initial values (Dirichlet boundary conditions). 
For the equation of state we choose a polytropic index $\gamma = 5/3$. 

\begin{figure*}
    \centering
    \includegraphics[width=7cm]{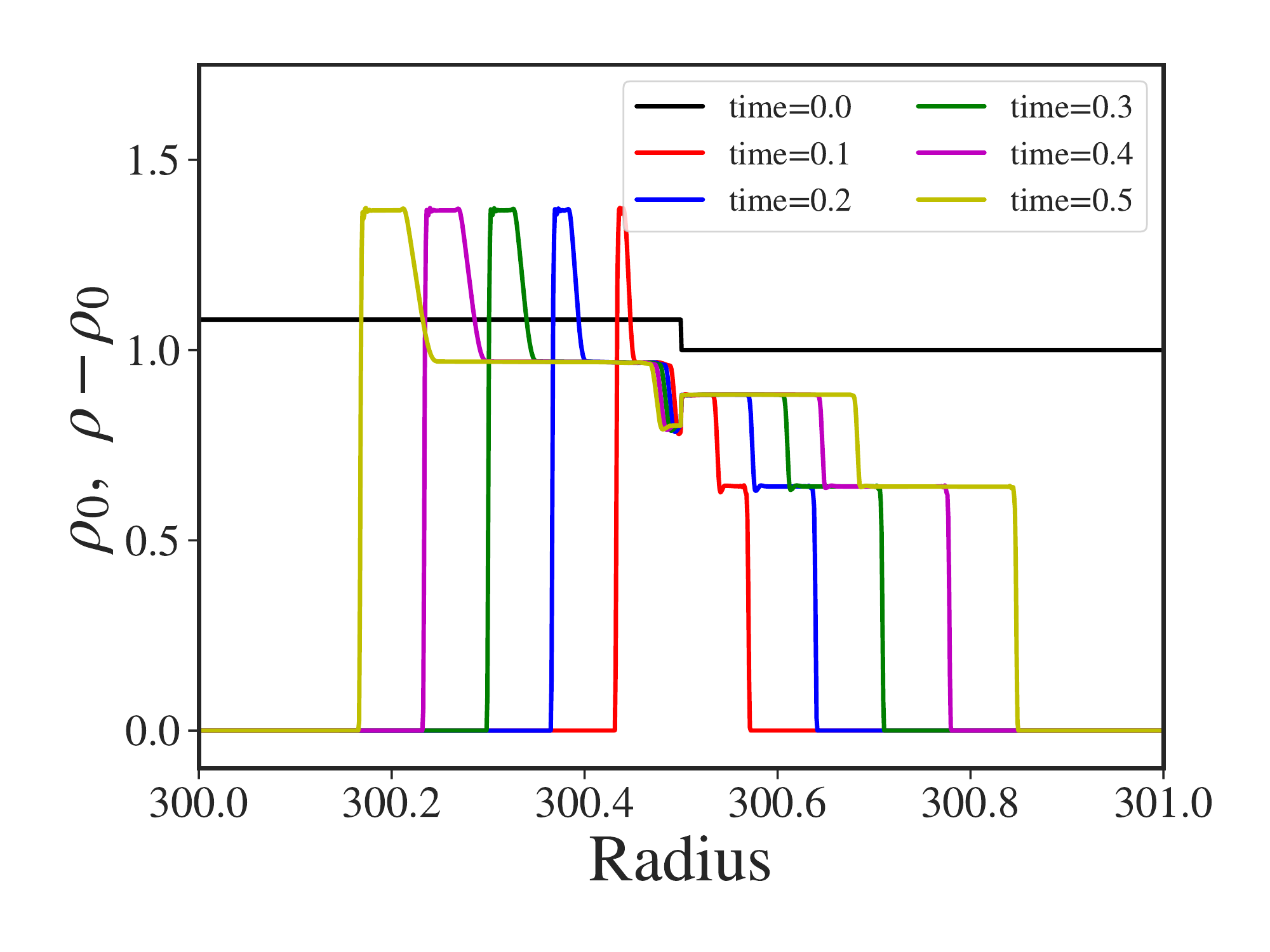}
    \includegraphics[width=7cm]{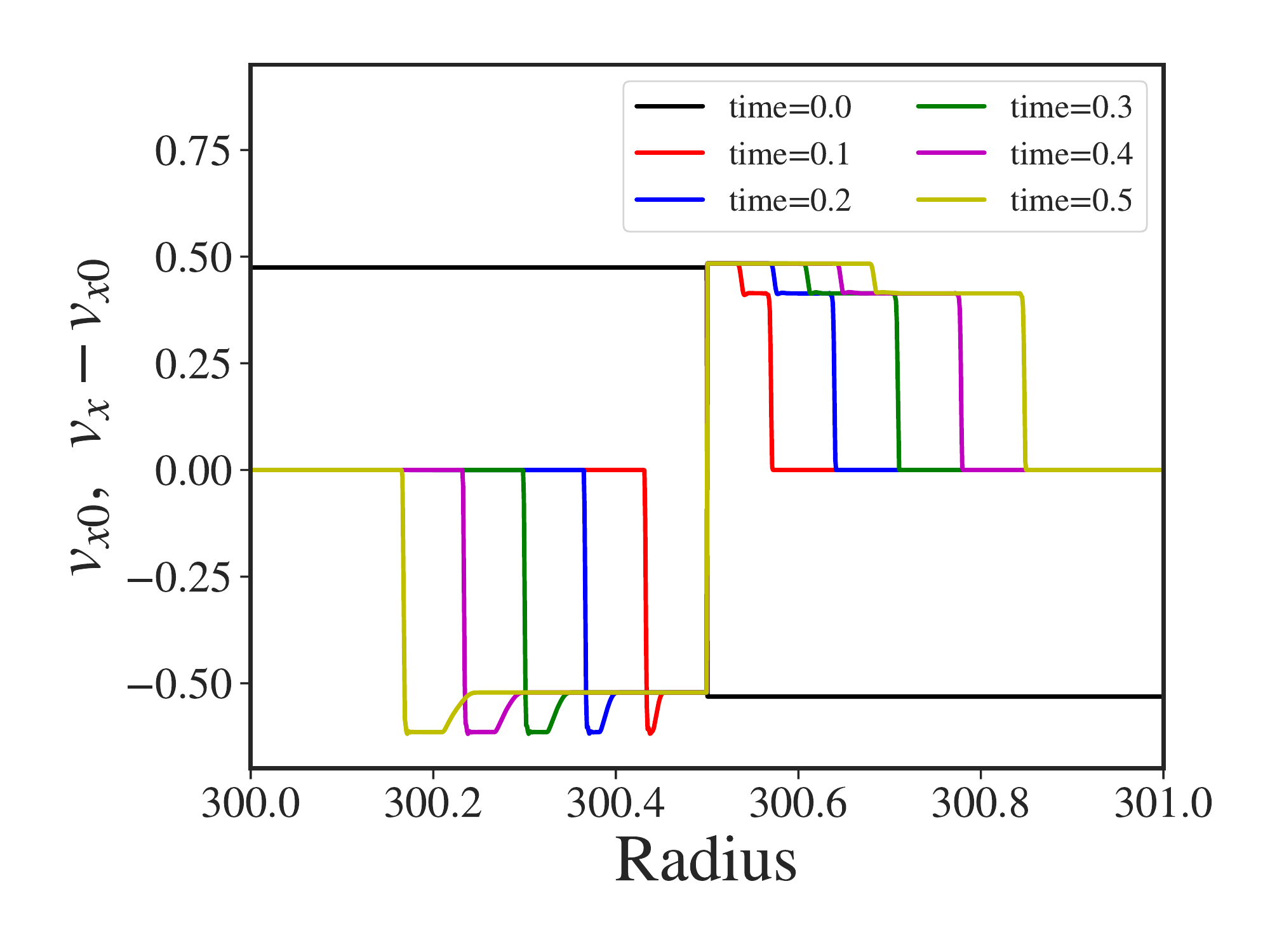}
    \includegraphics[width=7cm]{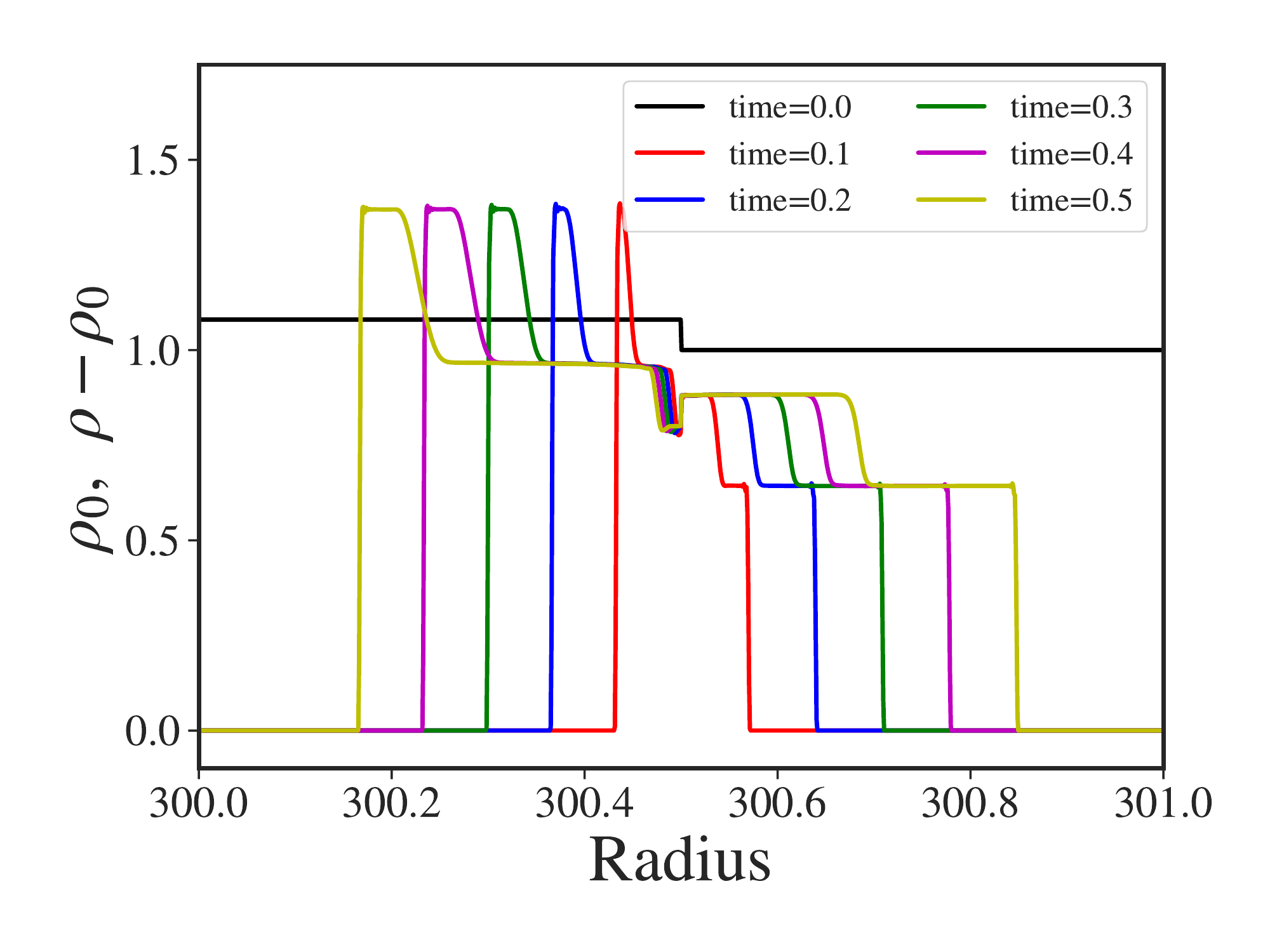}
    \includegraphics[width=7cm]{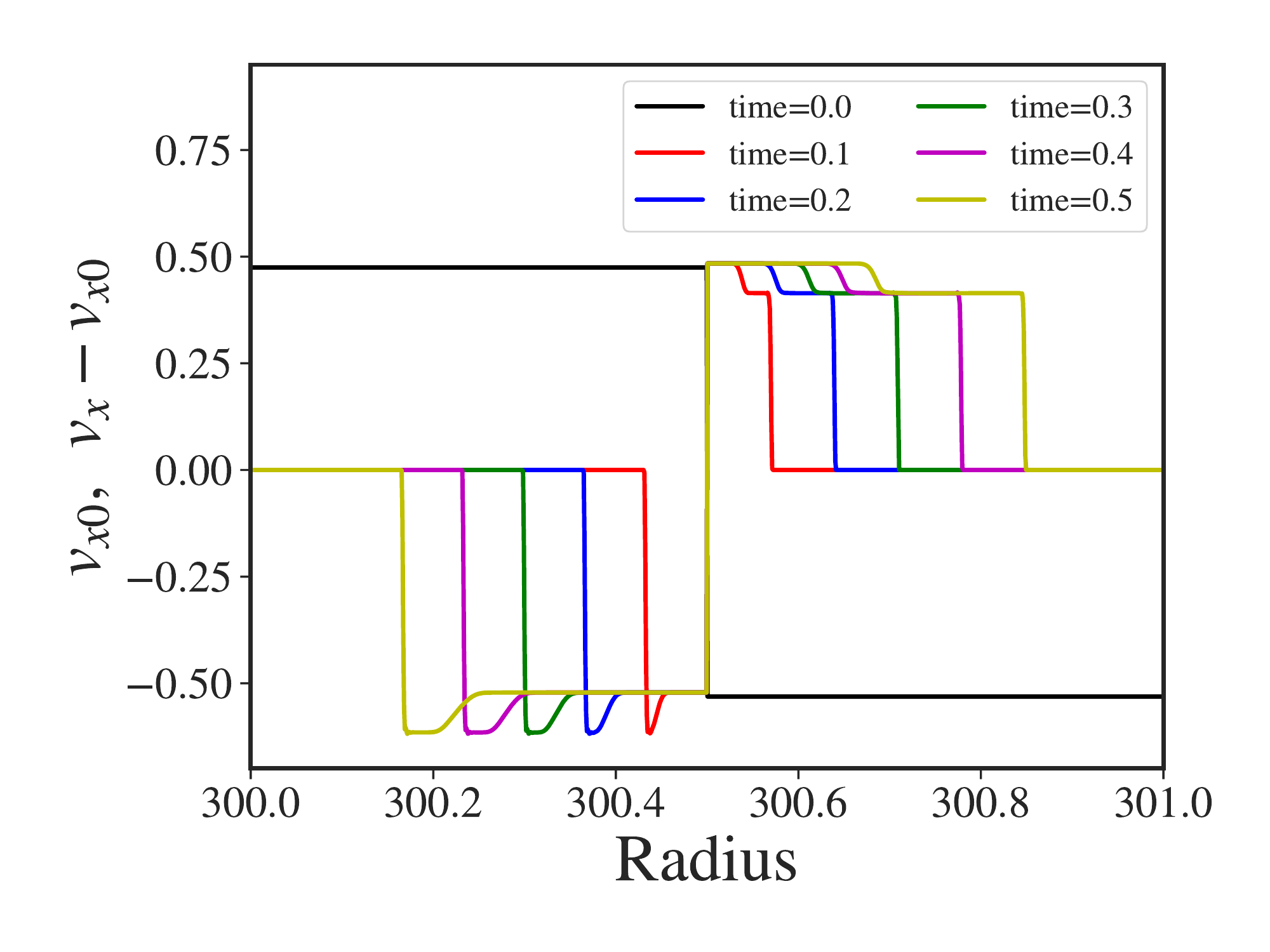}
    \includegraphics[width=7cm]{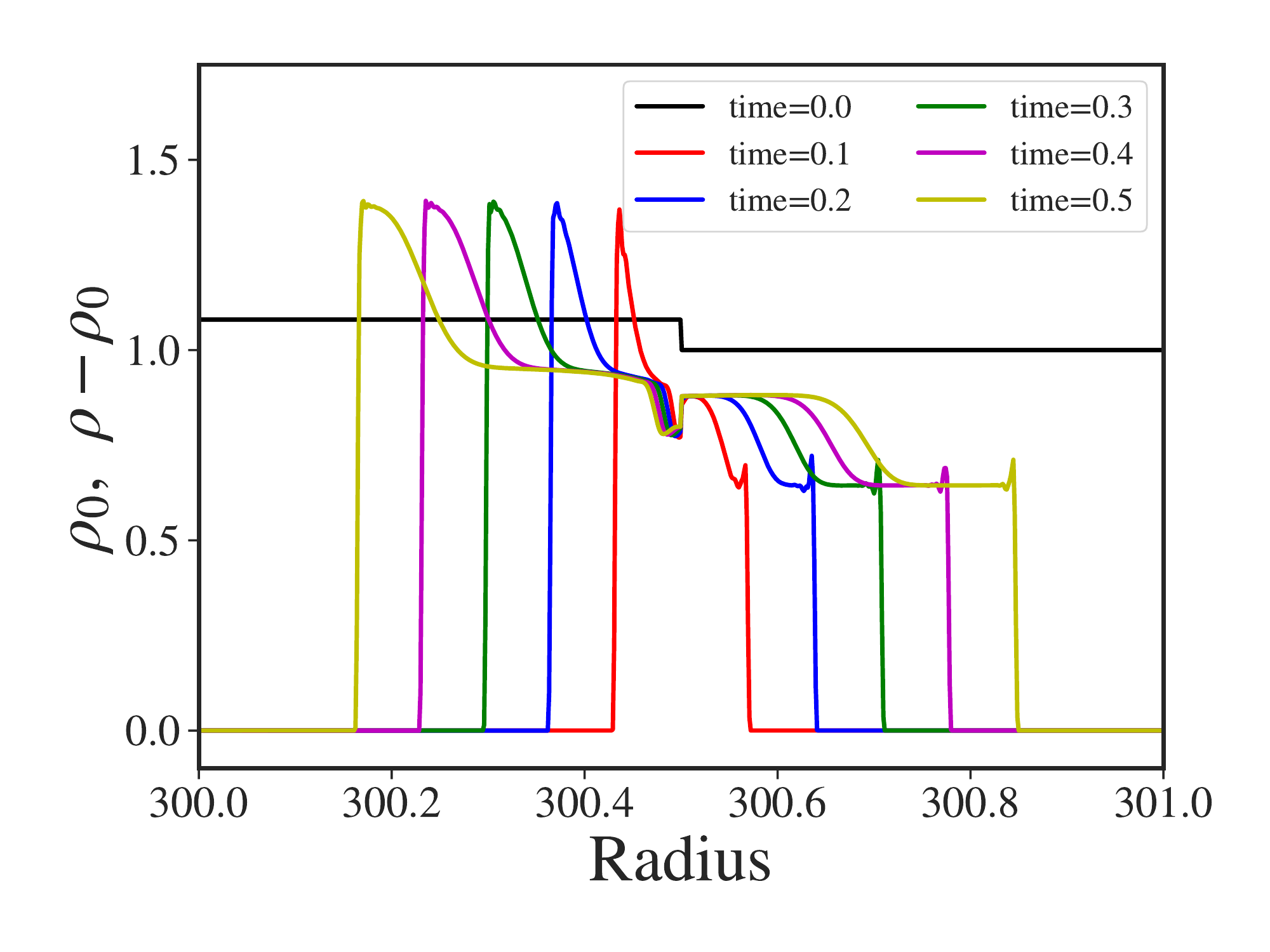}
    \includegraphics[width=7cm]{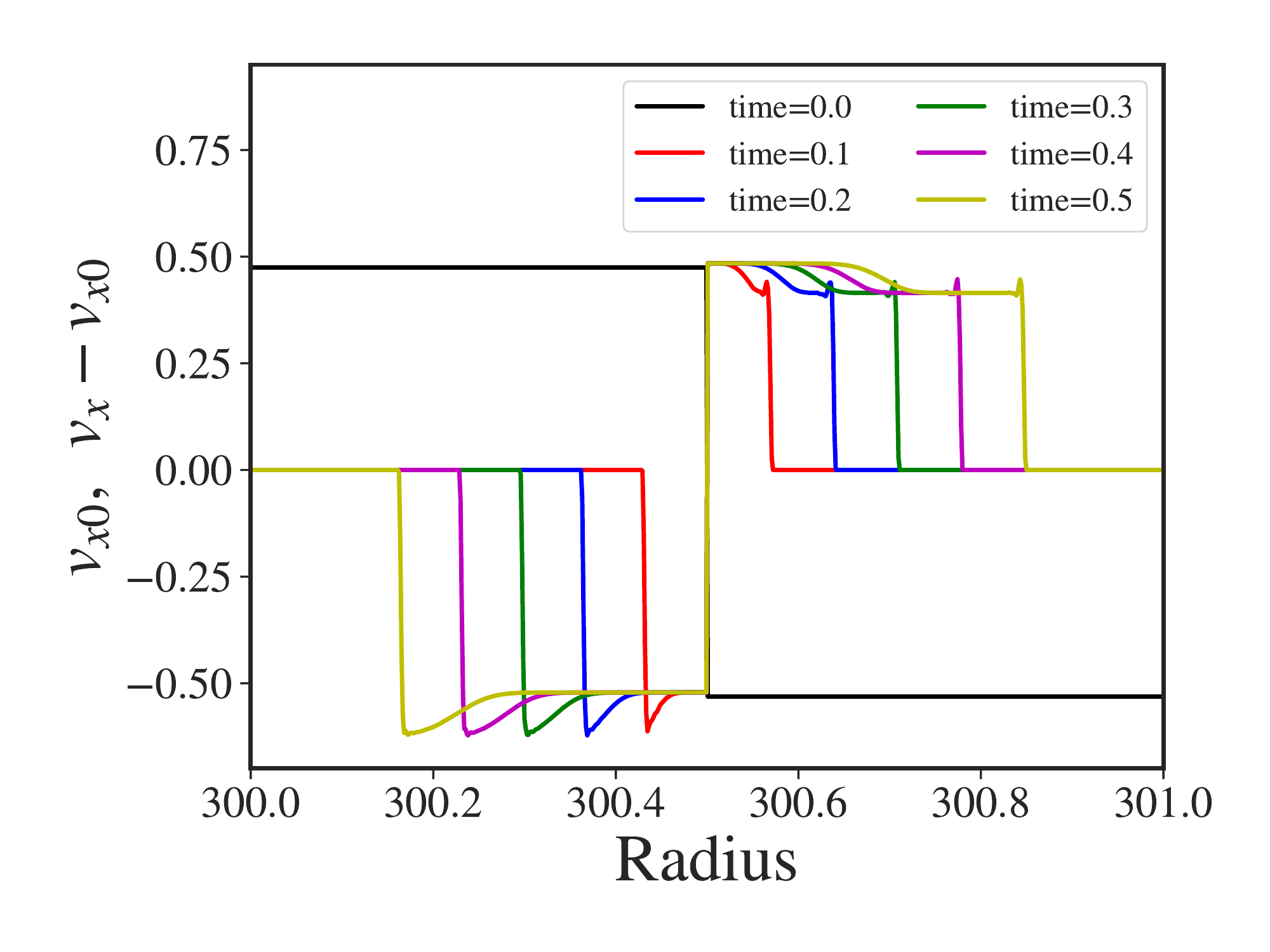}
    \includegraphics[width=7cm]{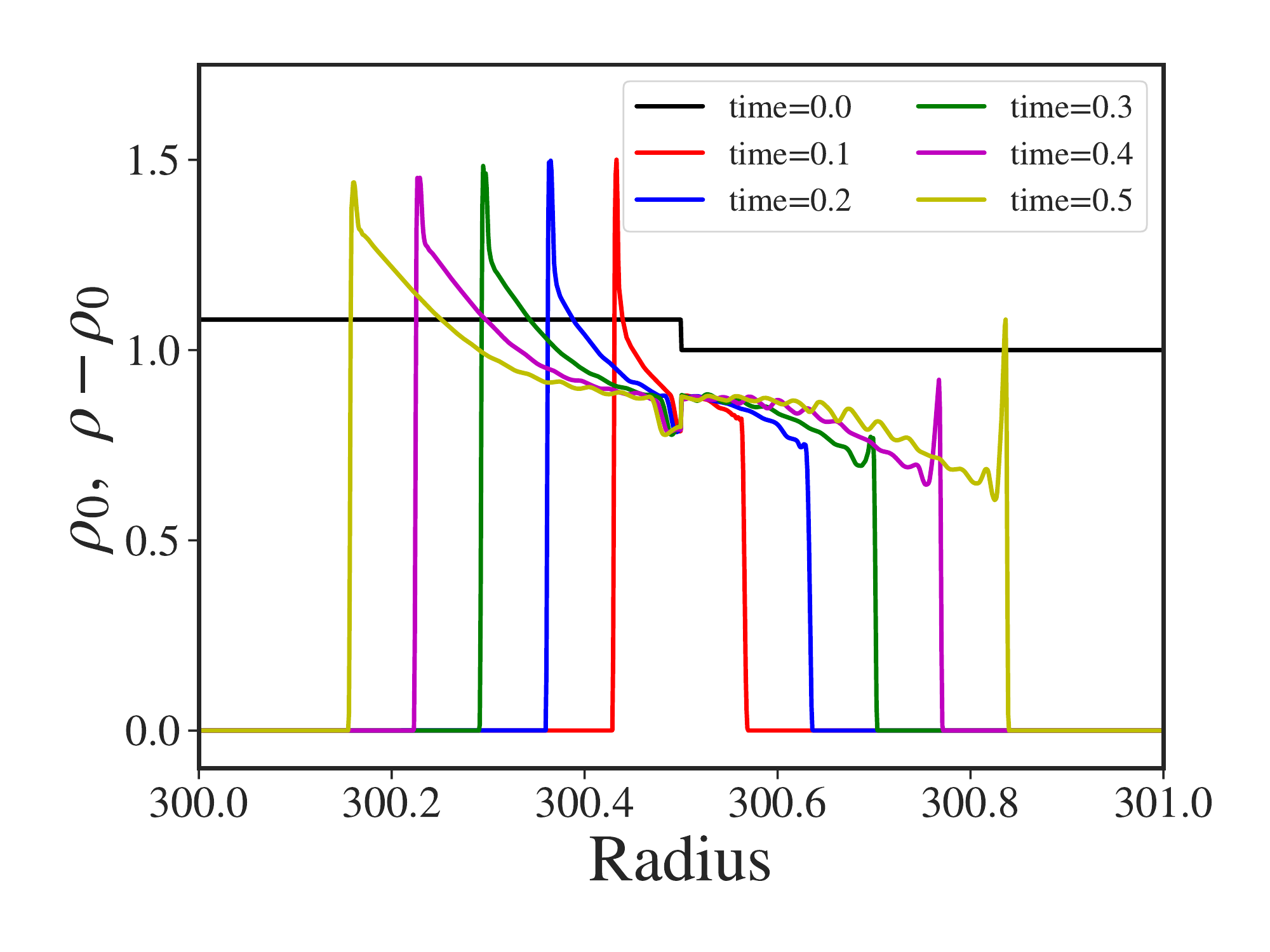}
    \includegraphics[width=7cm]{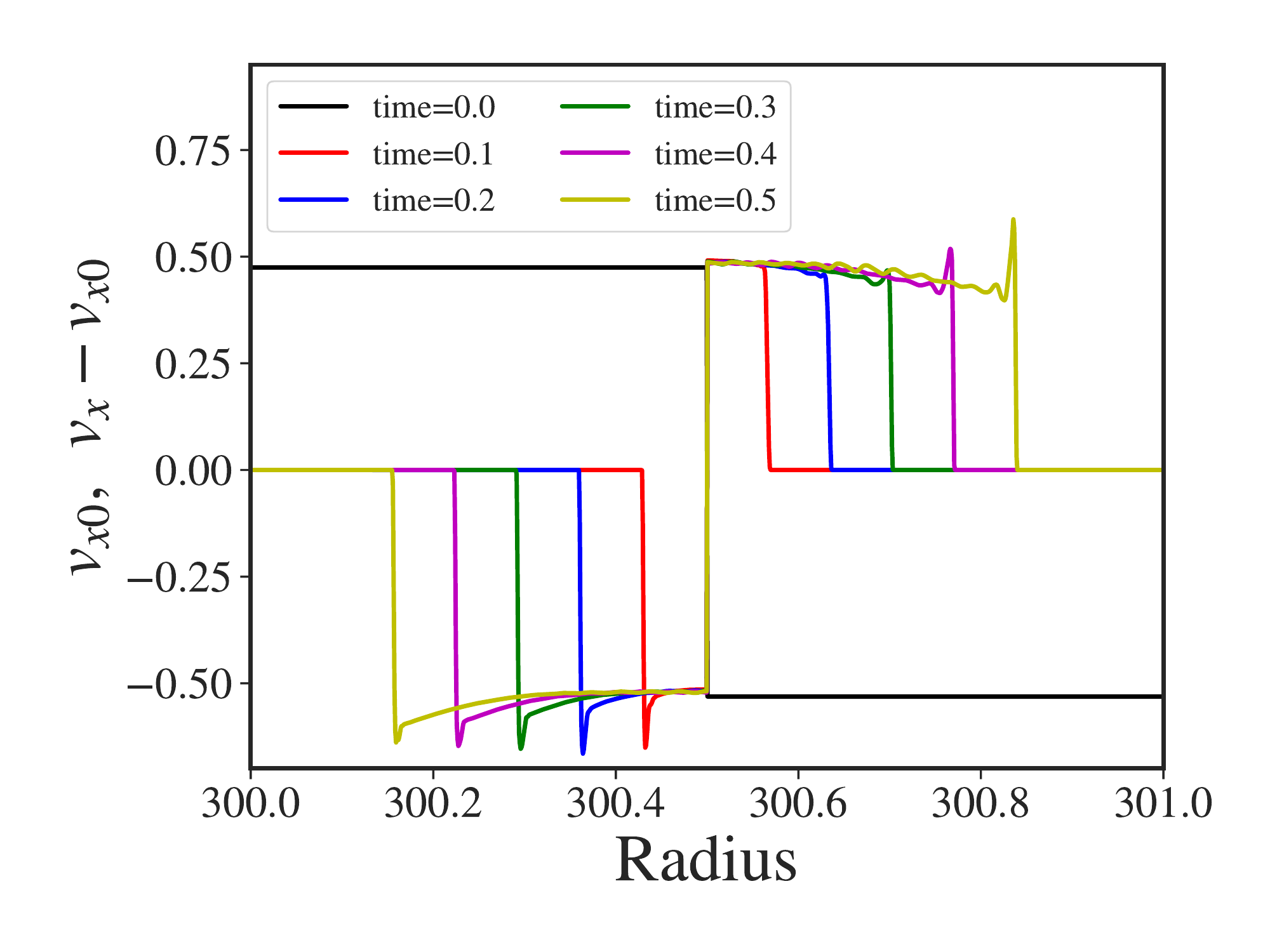}    
\caption{Time evolution of the classic 1D shock tube test.
  Shown are gas density (left) and horizontal velocity (right) for different levels of magnetic diffusivity,
  $\eta = 10^{-12}, 10^{-3}, 10^{-2}, 10^{-1}$ (from top to bottom).
  The initial conditions for $\rho_0$ and $V_{x0}$ are denoted by black lines, 
  while colored lines denote the evolution for five consecutive times steps.
  }
    \label{fig:tube1}
\end{figure*}

In Figure \ref{fig:tube1} we show the evolution of the discontinuity in gas density and horizontal velocity for 
different values of magnetic diffusivity. 
We note the in all cases we see the distinct features that result from the breaking of the initial discontinuity 
and the velocity values describe accurately the behavior of the gas density. 
The left-going rarefaction wave has a negative velocity and moves faster than the compound wave that follows it. 
The contact discontinuity propagates with the same speed as the compound wave which also appears in the density 
distribution, while the discontinuity moves slowly away from it's initial position at $x=0$. 
The discontinuity is followed by a slowly moving shock front and a fast moving rarefaction wave, both with positive 
velocities.

\begin{figure}
    \centering
    \includegraphics[width=8cm]{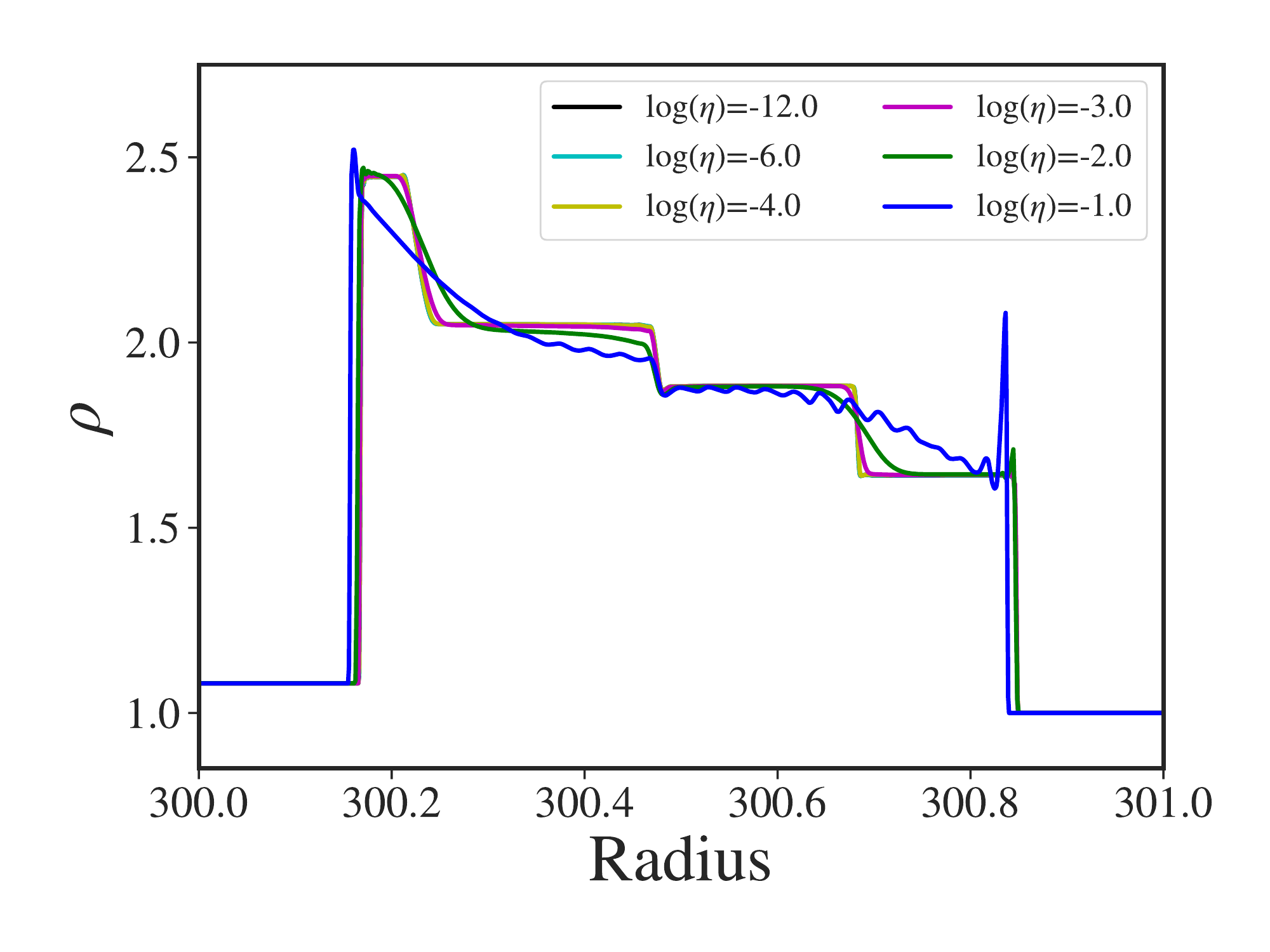}
    \includegraphics[width=8cm]{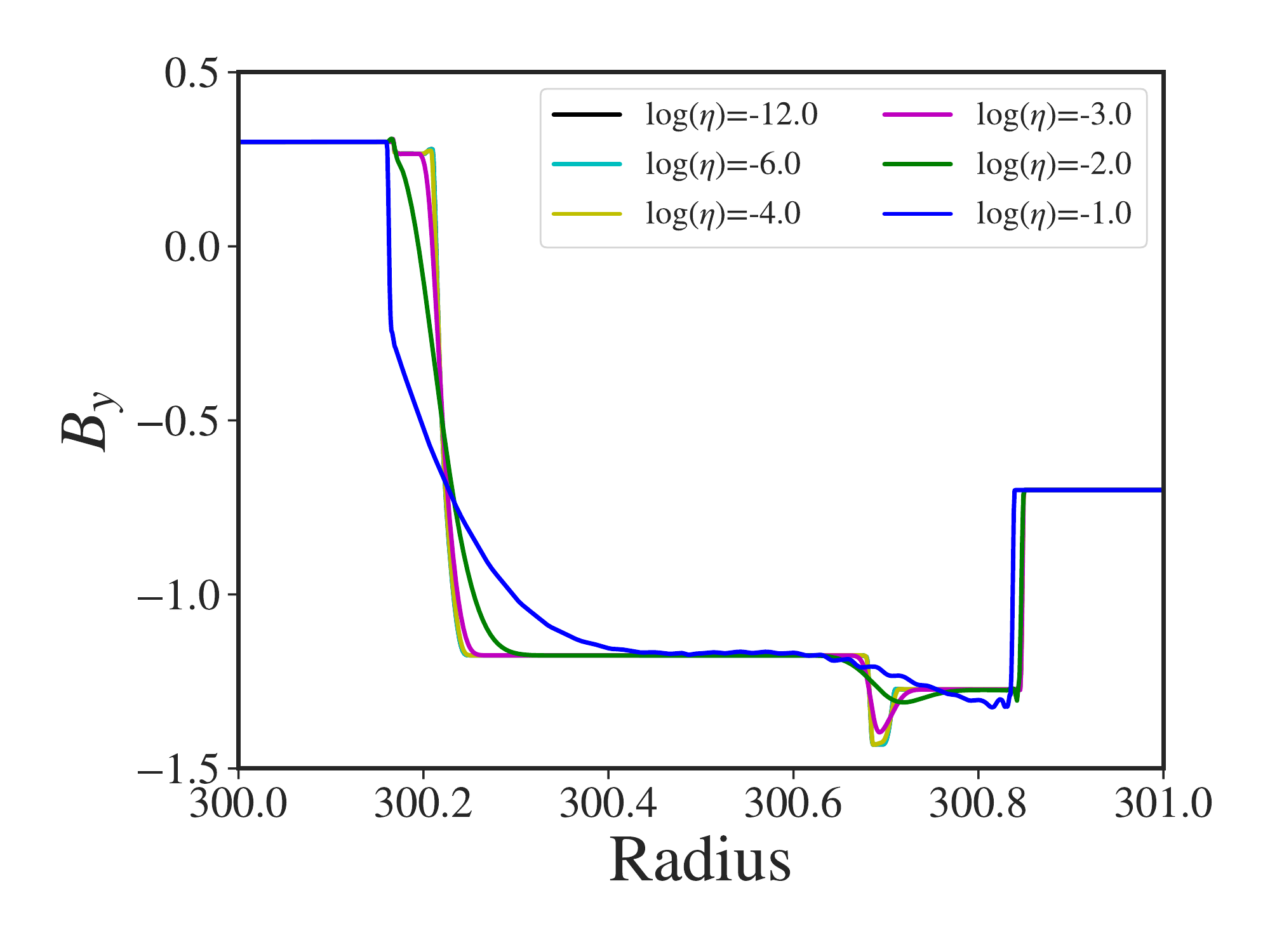}
\caption{Diffusive shock tube test.
      Shock longitudinal structure in density (left) and perpendicular magnetic field (right) at time $t=0.5$ for
      different levels of magnetic diffusivity,
      $\eta=10^{-12}$ (black line, ideal MHD),
      $\eta=10^{-6}$ (cyan),
      $\eta=10^{-4}$ (yellow), 
      $\eta=10^{-3}$ (magenta),
      $\eta=10^{-2}$ (green) and
      $\eta=10^{-1}$ (blue).
      Note that for $\eta < 10^{-4}$ the lines are overlapping.
}
    \label{fig:tube2}
\end{figure}

In Figure \ref{fig:tube2} we compare the distribution of gas density and vertical magnetic field for different levels
of magnetic diffusivity. 
We see that for $\eta < 10^{-4}$ there is little difference between the simulations.

\section{Mildly-relativistic limit}
\label{sec:limit}
Non-relativistic jet-launching simulations of the disk-outflow transition have detected strong outflows
from the innermost disk area
\citep{CK2002, Zanni2007, MurphyFerreiraZanni2010, Somayeh2012, 
Stepanovs2, Stepanovs3},
numerically confirming analytical derivations in steady state by \citet{BP1982} and \citet{Ferreira1997}.

In order to test our general relativistic simulation setup it is therefore interesting to do a comparison simulation
towards the non-relativistic limit,
thereby placing the inner grid boundary and the inner disk boundary at a radius further out. 
This excludes almost all general relativistic effects from the simulation, in particular any influence from a
central black hole.
For this simulation the inner disk radius has been placed at the location of the inner boundary 
at $r = 40$, well outside the marginally stable orbit.
The other simulation parameters were:
$\eta-0 - 10^{-5}$, $a=0$, $\beta = 50$; $K = 10^{-3}$; $\Gamma = 5/3$.

\begin{figure*}
      \centering
      \includegraphics[width=5.7cm]{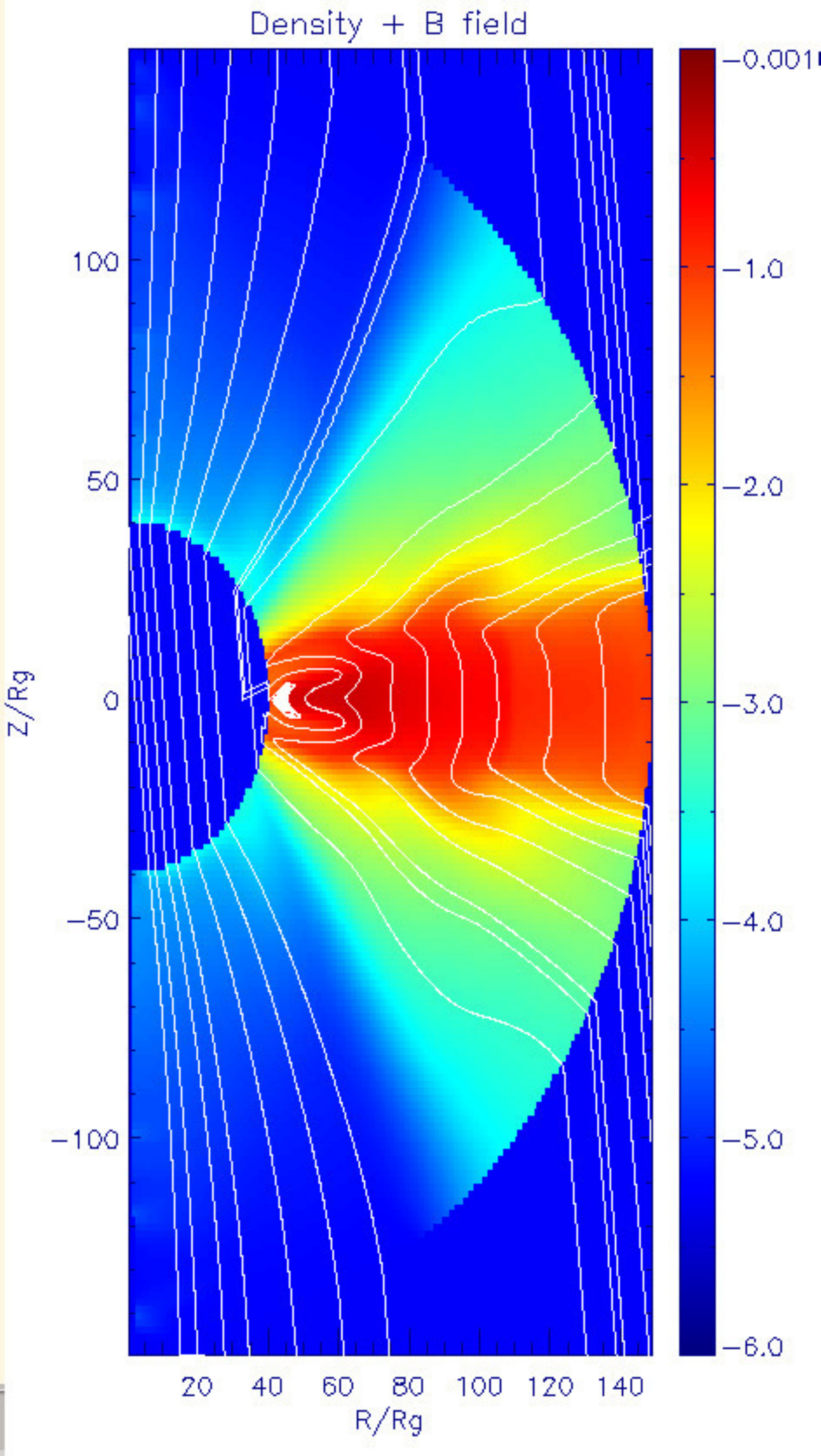}
      \includegraphics[width=5.7cm]{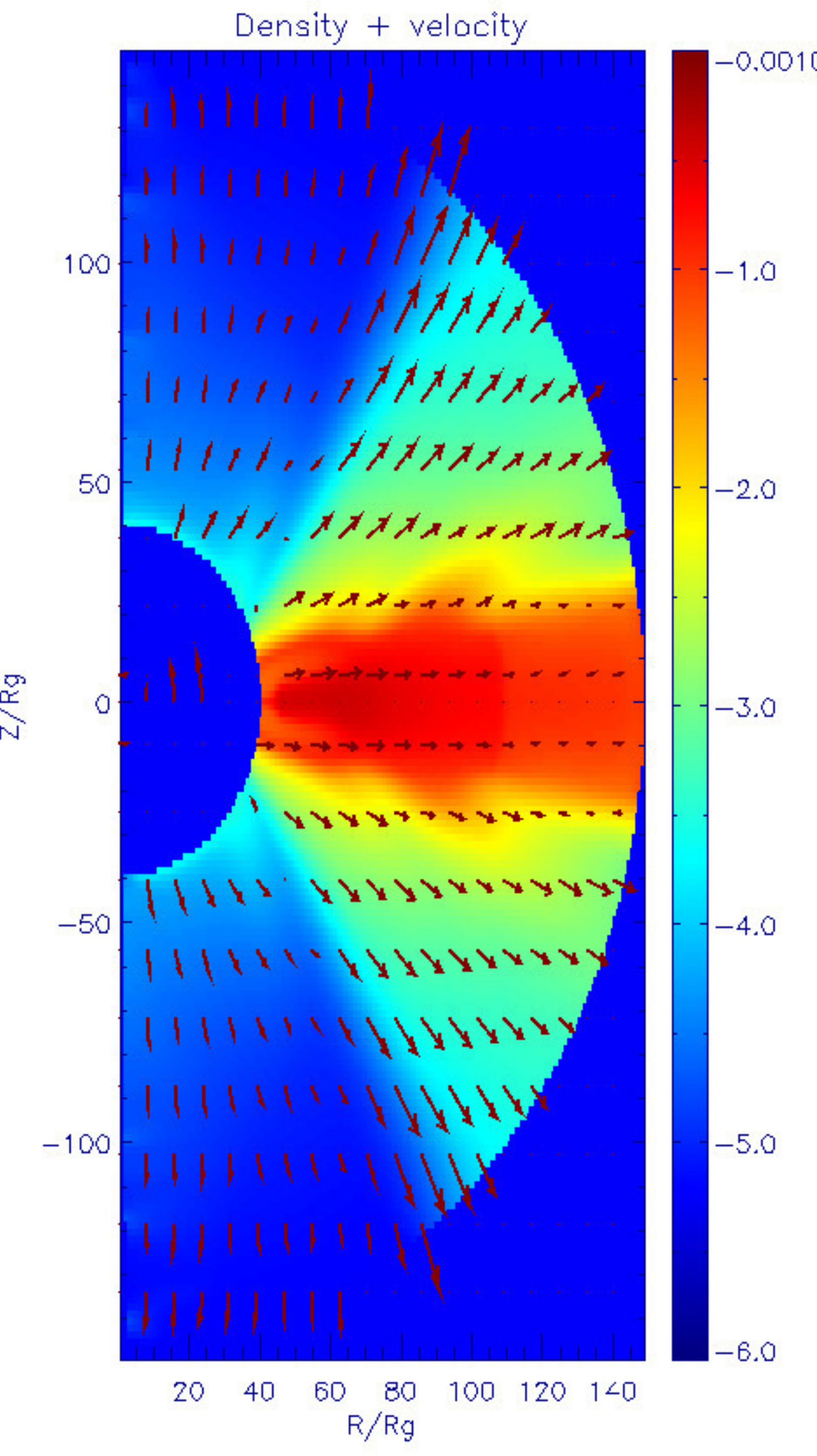}
      \includegraphics[width=5.7cm]{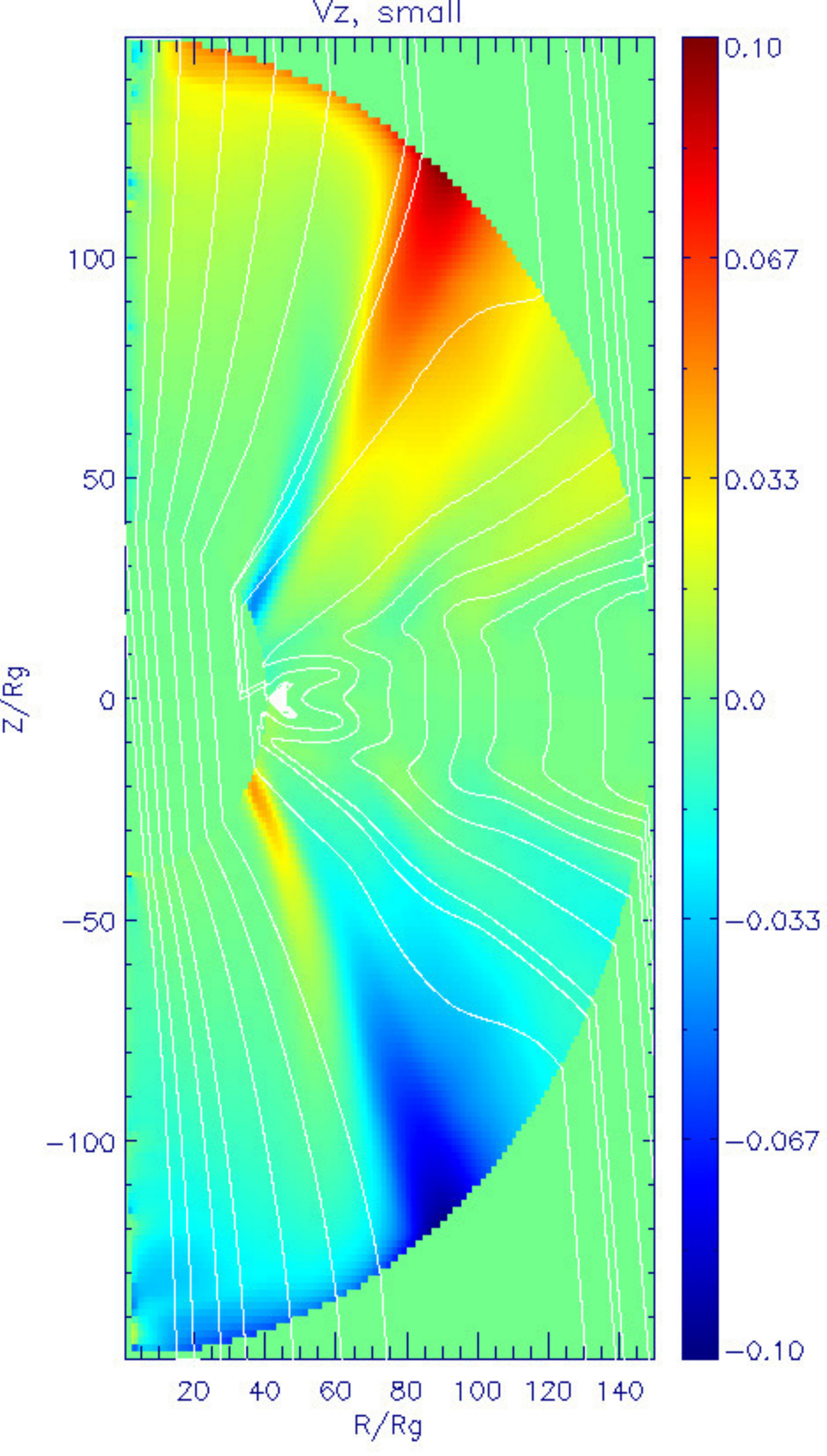}
      \caption{The test case of a mildly-relativistic simulation with a inner disk radius located at the 
      inner boundary at $r=40 r_{\rm g}$.
      Shown is the mass density distribution overlayed with poloidal magnetic field lines, 
               the mass density distribution overlayed with poloidal velocity vectors, 
               and the vertical velocity distribution overlayed with poloidal magnetic field lines (from left to right),
               all at time $t=6000$.
               }
   \label{fig:sim16}
  \end{figure*}

Figure~\ref{fig:sim16} shows the evolutionary state of such a mildly-relativistic simulation at time $t=6000$.
We see a clear disk outflow along the poloidal field lines anchored in the disk.
The maximum speed reached by this configuration is somewhat larger than 0.1\,c at a distance of $r = 150$
from the launching point.
This is corresponding to the orbital speed at the launching radius of $r=40$ and, thus, what we expect from
the classical Blandford-Payne magneto-centrifugal acceleration in the non-relativistic limit (see publications
cited just above, see in particular Figure.~1 in \citealt{Zanni2007} ).

Note that in comparison to the non-relativistic simulations cited above the evolutionary time step is rather low.
Non-relativistic simulations have been performed till several 100,000 rotational periods of the inner disk
\citep{Stepanovs2}.
For our mildly-relativistic simulation the time unit is the light crossing time over the gravitational radius $t_{\rm g}$
and thus $t=6000$ correspond to about 50 inner disk orbits only.
However, already a substantial disk wind is launched, as the disk evolution time is only several orbits and also 
the outflow kinematic time scale is much shorter than the disk evolutionary time scale.

Note also the axial flow of low density along the rotational axis.
The density distribution follows mainly from our floor model, applying also a relatively high
internal energy, that leads to a pressure driven axial flow.
While this seems kind of artificial, it can be motivated by the existence of a central wind (stellar wind
or Blandford-Znajek-driven jet) and helps to stabilize the central area against collapse..\

With these simulations we have therefore proven the applicability of our setup for magneto-centrifugally driven
disk winds in a mildly-relativistic setup.
While we expected to see similar trends also for the general relativistic simulations, those simulations show in fact 
a much more violent and variable characteristics for the disk wind (see discussion in our paper above).

\bibliographystyle{apj}

\begin{thebibliography}{33}
\expandafter\ifx\csname natexlab\endcsname\relax\def\natexlab#1{#1}\fi


\bibitem[Asada \& Nakamura(2012)]{AsadaNakamura2012} Asada, K., \& Nakamura, M.\ 2012, \apjl, 745, 28

\bibitem[Baiotti et al.(2005)]{Whisky2005}
         Baiotti, L., Hawke, I.,Montero, P.~J., et al.
         \ 2005, \prd, 71, 024035 

\bibitem[Balbus \& Hawley(1991)]{BH1991} Balbus, S.~A., \& Hawley, J.~F.\ 1991, \apj, 376, 214 

\bibitem[Balbus \& Hawley(1998)]{BH1998} Balbus, S.~A., \& Hawley, J.~F.\ 1998, RvMP, 70, 1 

\bibitem[Biretta et al.(1995)]{Biretta1995} Biretta, J.A., Zhou, F., Owen, F.N. \ 1995, \apj, 447, 582

\bibitem[Blandford \& Payne(1982)]{BP1982} Blandford, R.~D., \& Payne, D.~G.\ 1982, \mnras, 199, 883 

\bibitem[Blandford \& Znajek(1977)]{BZ1977} Blandford, R.~D., \& Znajek, R.~L.\ 1977, \mnras, 179, 433 

\bibitem[Boccardi et al.(2016)]{Boccardi2016} 
         Boccardi, B., Krichbaum, T.~P., Bach, U., et al. 
         \ 2016, \aap, 585, 33 

\bibitem[Britzen et al.(2017)]{Britzen2017} 
         Britzen, S., Qian, S.-J., Steffen, W., et al. 
         \ 2017, \aap, 602, 29 

\bibitem[Bucciantini \& Del Zanna(2013)]{BdZ2013} Bucciantini, N., \& Del Zanna, L.\ 2013, \mnras, 428, 71 

\bibitem[Bugli et al.(2014)]{Bugli2014} Bugli, M., Del Zanna, L., \& Bucciantini, N.\ 2014, \mnras, 440, L41

\bibitem[Casse \& Keppens(2002)]{CK2002} Casse, F., \& Keppens, R.\ 2002, \apj, 581, 988 

\bibitem[De Gouveia dal Pino \& Lazarian(2005)]{DeGouveia2005} de Gouveia Dal Pino, E.~M., Lazarian, A.,\ 2005, \aap, 411, 845

\bibitem[De Gouveia dal Pino et al.(2010)]{DeGouveia2010} de Gouveia Dal Pino, E.~M., Piovezan, P.~P., Kadowaki, L.~H.~S.\ 2010, \aap, 518, 5

\bibitem[Del Zanna et al.(2007)]{DelZanna2007} 
         Del Zanna, L., Zanotti, O., Bucciantini, N., \& Londrillo, P.\ 2007, \aap, 473, 11 


\bibitem[De Villiers \& Hawley(2003)]{DeVilliers2003a} De Villiers, J.-P., \& Hawley, J.~F.\ 2003, \apj, 589, 458 

\bibitem[De Villiers \& Hawley(2003)]{DeVilliers2003b} De Villiers, J.-P., \& Hawley, J.~F.\ 2003, \apj, 592, 1060 

\bibitem[De Villiers et al.(2005)]{DeVilliers2005} De Villiers, J.-P., Hawley, J.~F.,  Krolik, J.~H.,  Hirose, S.\ 2005, \apj, 620, 878 

\bibitem[Dionysopoulou et al.(2013)]{Dionysopoulou2013} Dionysopoulou, K., Alic, D., Palenzuela, C., Rezzolla, L., \& Giacomazzo, B.\ 2013, \prd, 88, 044020 

\bibitem[Dionysopoulou et al.(2015)]{Dionysopoulou2015} Dionysopoulou, K., Alic, D.,  Rezzolla, L.\ 2015, \prd, 92, 084064

\bibitem[Doeleman et al.(2012)]{Doeleman2012} Doeleman S.S., et al. \ 2012, Science, 338, 355

\bibitem[Dumbser \& Zanotti (2009)]{DumbserZanotti2009} Dumbser, M., \& Zanotti, O. \ 2009, JCoPh, 228, 6991

\bibitem[Event Horizon Telescope Collaboration (2019a)]{EHT2019a} Akiyama, K., Alberdi, A., Alef, W., et al. \ 2019, \apjl, 875, L1

\bibitem[Event Horizon Telescope Collaboration (2019b)]{EHT2019e} Akiyama, K., Alberdi, A., Alef, W., et al., \ 2019, \apjl, 875, L5

\bibitem[Einstein A.(1915)] {Einstein1915} Einstein, A. \ 1915, \ Sitzungsberichte der K\"oniglich-Preu\ss ischen Akademie der Wissenschaften, 844

\bibitem[Feng et al.(2016)]{Feng2016} Feng, J., Wu, Q., Lu, R.\ 2003, \apj, 830, 6

\bibitem[Ferreira(1997)]{Ferreira1997} Ferreira, J.\ 1997, \aap, 319, 340 

\bibitem[Fishbone \& Moncrief(1976)]{FishboneMoncrief1976} Fishbone, L.~G., \& Moncrief, V.\ 1976, \apj, 207, 962 

\bibitem[Fleming et al.(2000)]{Fleming2000} Fleming, T.~P., Stone, J.~M., \& Hawley, J.~F.\ 2000, \apj, 530, 464 

\bibitem[Gammie et al.(2003)]{Gammie2003} Gammie, C.~F., McKinney, J.~C., \& T{\'o}th, G.\ 2003, \apj, 589, 444 

\bibitem[Giovannini et al.(2018)]{Giovannini2018} Giovannini G., \ 2003, Nat. Astr., 2, 472 

\bibitem[Gressel(2010)]{Gressel2010} Gressel, O. \ 2010, \mnras, 405, 41

\bibitem[Hirose et al.(2004)]{Hirose2004} Hirose, S., Krolik, J.~H., De Villiers, J.-P., Hawley, J.~F.\ 2004, \apj, 606, 1083

\bibitem[Hawley \& Balbus(1992)]{HawleyBalbus1992} Hawley, J.~F., \& Balbus, S.~A.\ 1992, \apj, 400, 595 

\bibitem[Koide et al.(1999)]{Koide1999} Koide, S., Shibata, K., \& Kudoh, T.\ 1999, \apj, 522, 727 

\bibitem[Komissarov(2007)]{Komissarov2007} Komissarov, S.~S.\ 2007, \mnras, 382, 995 

\bibitem[Longaretti \& Lesur(2010)]{LongarettiLesur2010} Longaretti, P.-Y., \& Lesur, G.\ 2010, \aap, 516, A51 

\bibitem[Lynden-Bell(1996)]{LBell1996} Lynden-Bell, D.\ 1996, \mnras, 279, 389 

\bibitem[Maxwell J.C.(1865)]{Maxwell1865} Maxwell, James Clerk, \ 1865, Philosophical Transactions of the Royal Society of London, 155, 459 

\bibitem[McKinney \& Gammie(2004)]{McKinney2004} McKinney, J.~C., \& Gammie, C.~F.\ 2004, \apj, 611, 977 

\bibitem[McKinney et al.(2012)]{McKinneyTchekh2012} McKinney, J.~C., Tchekhovskoy, A., \& Blandford, R.~D.\ 2012, \mnras, 423, 3083 

\bibitem[McKinney et al.(2013)]{McKinneyTchekh2013} McKinney, J.~C., Tchekhovskoy, A., \& Blandford, R.~D.\ 2013, Science, 339, 49

\bibitem[McKinney et al.(2014)]{McKinneyTchekh2014} McKinney, J.~C., Tchekhovskoy, A., Sadowski, A., \& Narayan, R. \ 2014, \mnras, 441, 3177 


\bibitem[Misner et al.(1973)]{MTW1973} 
         Misner, C.~W., Thorne, K.~S., \& Wheeler, J.~A.\ 1973, San Francisco: W.H.~Freeman and Co. (1973)

\bibitem[Murphy et al.(2010)]{MurphyFerreiraZanni2010} 
       {Murphy}, G.~C. and {Ferreira}, J. \& {Zanni}, C. 2010, A\&A, 512, 82

\bibitem[Narayan et al.(2003)]{Narayan2003} 
         Narayan, R., Igumenshchev, I.~V., \& Abramowicz, M.~A.\ 2003, \pasj, 55, L69 

\bibitem[Nakahara et al.(2018)]{Nakahara2018} 
         Nakahara S., Doi A.,Murata Y., Hada K., Nakamura M., Asada K. \ 2018, \apj, 854, 148

\bibitem[Nakamura et al.(2018)]{Nakamura2018} 
         Nakamura, M., Asada, K., Hada, K., et al. \ 2018, \apj, 868, 146

\bibitem[Noble et al.(2006)]{Noble2006} Noble, S.~C., Gammie, C.~F., McKinney, J.~C., \& Del Zanna, L.\ 2006, \apj, 641, 626 

\bibitem[Noble et al.(2009)]{Noble2009} {Noble}, S.~C., {Krolik}, J.~H., {Hawley}, J.~F.\ 2009, \apj, 692, 411

\bibitem[Noble et al.(2010)]{Noble2010} {Noble}, S.~C., {Krolik}, J.~H., {Hawley}, J.~F.\ 2010, \apj, 711, 959

\bibitem[Noble et al.(2011)]{Noble2011} {Noble}, S.~C., {Krolik}, J.~H., {Schnittman}, J.~D.,  {Hawley}, J.~F.\ 2009, \apj, 743, 115

\bibitem[Novikov \& Thorne(1973)]{NT1973} 
        {Novikov}, I.~D. and {Thorne}, K.~S. \ 1973, Astrophysics of black holes, in: Black Holes (Les Astres Occlus), ed: C.~DeWitt and B.~DeWitt, 
        (Gordon and Breach, N.Y.), p.343 


\bibitem[Paczy{\'n}sky \& Wiita(1980)]{PW} Paczy{\'n}sky, B., \& Wiita, P.~J.\ 1980, \aap, 88, 23 

\bibitem[Palenzuela et al.(2009)]{Palenzuela2009} Palenzuela, C., Lehner, L., Reula, O., \& Rezzolla, L.\ 2009, \mnras, 394, 1727 

\bibitem[Palenzuela (2013)]{Palenzuela2013} Palenzuela, C.\ 2013, \mnras, 431, 1853

\bibitem[Penna et al.(2010)]{Penna2010} Penna, R.~F., McKinney, J.~C.,Narayan, R., Tchekhovskoy, A., Shafee, R., McClintock, J.~E.\ 2010, \mnras, 408, 752


\bibitem[Porth \& Fendt(2010)]{Porth2010} Porth, O., \& Fendt, C.\ 2010, \apj, 709, 1100 


\bibitem[Porth et al.(2017)]{Porth2017} Porth, O., Olivares, H., Mizuno, Y., t al.
         \ 2017, ComAC, 4, 1 

\bibitem[Porth et al.(2019)]{Porth2019} Porth, O. et. al \ 2019, arXiv e-prints,  	arXiv:1904.04923



\bibitem[Qian et al.(2017)]{QQ1} Qian, Q., \& Fendt, C., \& Noble, S.~C., \& Bugli, M., \ 2017, \apj, 834, 29

\bibitem[Qian et al.(2018)]{QQ2} Qian, Q., \& Fendt, C., \& Vourellis, C. \ 2018, \apj, 859, 28

\bibitem[Ripperda et al.(2019)]{Ripperda2019} Ripperda, B. et al. \ 2019, arXiv e-prints, arXiv: 1907.07197


\bibitem[S{\c a}dowski et al.(2012)]{SadowskiNarayan2014} S{\c a}dowski, A., Narayan, R., McKinney, J.~C., \& Tchekhovskoy, A. \ 2014, \mnras, 439, 503



\bibitem[Shakura \& Sunyaev(1973)]{SS1973} Shakura, N.~I., \& Sunyaev, R.~A.\ 1973, \aap, 24, 337 

\bibitem[Sheikhnezami et al.(2012)]{Somayeh2012} Sheikhnezami, S., Fendt, C., Porth, O., Vaidya, B., \& Ghanbari, J.\ 2012, \apj, 757, 65 

\bibitem[Stepanovs \& Fendt(2014)]{Stepanovs1} Stepanovs, D., \& Fendt, C.\ 2014, \apj, 793, 31 

\bibitem[Stepanovs \& Fendt(2014)]{Stepanovs2} Stepanovs, D., \& Fendt, C., Sheikhnezami, S. \ 2014, \apj, 796, 29

\bibitem[Stepanovs \& Fendt(2016)]{Stepanovs3} Stepanovs, D., \& Fendt, C.\ 2016, \ 2016, \apj, 825, 14  

\bibitem[Tchekhovskoy et al.(2010)]{TchekhMcKinney2009} Tchekhovskoy, A., McKinney, J. C., Narayan, R.\ 2009, \apj, 699, 1789

\bibitem[Tchekhovskoy et al.(2010)]{TchekhNarayan2010} Tchekhovskoy, A., Narayan, R., McKinney, J. C.\ 2010, \apj, 711, 50

\bibitem[Tchekhovskoy et al.(2011)]{TchekhNarayan2011} Tchekhovskoy, A., Narayan, R., McKinney, J. C.\ 2011, \mnras, 418, 79 

\bibitem[Tchekhovskoy et al.(2012)]{TchekhMcKinney2012} Tchekhovskoy, A., McKinney, J. C.\ 2012, \mnras, 423, 55

\bibitem[Ustyugova et al.(1995)]{Ustyugova1995} 
         Ustyugova, G.~V., Koldoba, A.~V., Romanova, M.~M.,	Chechetkin, V.~M., Lovelace, R.~V.~E.\ 1995, \apj, 439, L39



\bibitem[Zanni et al.(2007)]{Zanni2007} {Zanni}, C. and {Ferrari}, A., {Rosner}, R., {Bodo}, G. \& {Massaglia}, S. 2007, A\&A, 469, 811

\bibitem[Zanotti et al.(2015)]{ZanottiDumbser2015} Zanotti, O., Dumbser, M., \ 2015, Computer Physics Communications , 188, 110



\end{thebibliography}

\end{document}